\documentclass[acmlarge]{acmart}
\usepackage{soul}
\usepackage{graphicx}
\usepackage{subfigure}
\usepackage{float}
\usepackage{color}
\usepackage[ruled,vlined,linesnumbered,lined,boxed,commentsnumbered]{algorithm2e}

\usepackage[font=small,labelfont=sf]{caption}

\AtBeginDocument{%
  \providecommand\BibTeX{{%
    \normalfont B\kern-0.5em{\scshape i\kern-0.25em b}\kern-0.8em\TeX}}}

\setcopyright{acmcopyright}
\acmJournal{IMWUT}
\acmYear{2020}\acmVolume{4}\acmNumber{2}\acmArticle{70}
\acmMonth{6}\acmPrice{15.00}\acmDOI{10.1145/3397321}

\begin{document}

\newcommand{\systemname}{V\textsuperscript{2}iFi}

\title{\texorpdfstring{V\textsuperscript{2}iFi}{V2iFi}: in-Vehicle Vital Sign Monitoring via Compact RF Sensing} 
\author{Tianyue Zheng}
\authornote{Both authors contributed equally to this research.}
\email{tianyue002@e.ntu.edu.sg}
\author{Zhe Chen}
\authornotemark[1]
\email{chen.zhe@ntu.edu.sg}
\affiliation{%
  \institution{Nanyang Technological University}
  \country{Singapore}}

\author{Chao Cai}
\email{chris.cai@ntu.edu.sg}
\affiliation{%
  \institution{Nanyang Technological University}
  \country{Singapore}}
  
\author{Jun Luo}
\email{junluo@ntu.edu.sg}
\affiliation{%
  \institution{Nanyang Technological University}
  \country{Singapore}}
  
\author{Xu Zhang}
\email{zhangxu@uchicago.edu}
\affiliation{%
  \institution{University of Chicago}
  \city{Chicago}
  \country{United States}}

\renewcommand{\shortauthors}{Zheng and Chen, et al.}

\begin{abstract}
Given the significant amount of time people spend in vehicles, health issues 
under driving condition have become a major concern. Such issues may vary from fatigue, asthma, stroke, to even heart attack, yet they can be adequately indicated by vital signs and abnormal activities. Therefore, in-vehicle vital sign monitoring can help us predict and hence prevent these issues.
Whereas existing sensor-based (including camera) methods could be used to detect these indicators, privacy concern and system complexity
both call for a convenient yet effective and robust alternative. This paper aims to develop \systemname, 
an intelligent system performing monitoring tasks using a COTS impulse radio mounted on the windshield. \systemname\ is capable of reliably detecting driver's vital signs under driving condition and with the presence of passengers, thus allowing for potentially inferring corresponding health issues. 
Compared with prior work based on Wi-Fi CSI, \systemname\ is able to
distinguish reflected signals from multiple users, and hence provide
finer-grained measurements under more realistic settings.
We evaluate \systemname\ both in lab environments and during real-life road tests; %
the results demonstrate that respiratory rate, heart rate, and heart rate variability can all be estimated accurately. %
Based on these estimation results,
we further discuss how machine learning models can be applied on top of \systemname\ so as to improve both physiological and psychological wellbeing in driving environments.
  
\end{abstract}



\maketitle
\section{Introduction}
With the rapid development of artificial intelligence, vehicles are becoming increasingly ``smart'' nowadays:
they not only learn to drive by itself, but also attempt to better understand and interact with the driver and passengers. In particular, health monitoring is one of the essential components of intelligent vehicles;
it aims to figure out the health status of people under driving condition, especially after long hours of daily commutes and frustrating traffic jams~\cite{b1:critical}.
For example, drowsy driving has been reported to have caused $25\%$ fatal vehicle crashes~\cite{flatley2004sleep}, and vital signs such as respiration and heartbeat are important indicators for early detection of drowsiness~\cite{b3:hrv_drowsy,solaz2016drowsiness}. 
Consequently, an in-vehicle health monitoring system that keeps watching these indicators offers both the vehicle and driver better perception and decision-making capabilities, thus leading to a healthy and safe driving experiences. 

To enable healthy and safe driving, automotive companies have built various monitoring system, including ATTENTION ASSIST\textsuperscript{\textregistered} by Mercedes-Benz~\cite{benz}, Driver Alert Control (DAC) by Volvo~\cite{volvo} and SmartSenior by BMW ~\cite{bmw}. In the meantime, research groups from around the world have also been attracted to this topic \cite{BreathListener-MobiSys19, Vitamon-Sensys, Smartwatch-IEEESens, monitoring-Accid, awais2017hybrid, multisensor-ICRA}.
By far, two main categories of such systems work on monitoring person (driver in particular) directly, namely
camera-based~\cite{volvo, Vitamon-Sensys, multisensor-ICRA} and wearable sensor-based~\cite{awais2017hybrid, Smartwatch-IEEESens, monitoring-Accid} (e.g., EEG and ECG). Although these approaches work well technically, they may not be exactly practical in real life, because of well-known weaknesses including privacy concern, poor performance in low-light, as well as the use of intrusive sensors (causing uncomfortable user experience). 

Fortunately, recent progress in wireless sensing has brought us new hope to overcome the aforementioned weaknesses~\cite{adib2015smart, AcousticResp-Ubicomp, Fullbreathe-UbiComp, Indotrack-UbiComp}. Such a
system always transmits wireless (e.g., radio and acoustic) signals and captures
their reflections. As the reflections %
off persons carry their respective vital signs,\footnote{As vital signs are actually micro-scale body activities, the ability to monitor the former implies the same to the latter. Therefore, our technical discussions focus only on vital sign monitoring for the sake of conciseness.}
one may extract these vital signs from the reflections without intrusive sensors touching human bodies, while respecting the privacy of users and avoiding the interference of ambient light.
However, different from the state-of-the-art wireless based vital sign monitoring, in-vehicle monitoring is much more complicated because of the confined and noisy nature of driving environments. To tackle these challenges, there have been a few existing studies on in-vehicle monitoring of either general health or specific vital signs via acoustic~\cite{BreathListener-MobiSys19} or Radio Frequency (RF)~\cite{WiFind-IEEEToBD} signals. Although acoustic signals can deliver vital sign monitoring ubiquitously via smart devices such as smartphones~\cite{BreathListener-MobiSys19}, they are prone to be
contaminated by background sounds under driving condition. Moreover, babies and pets may feel uncomfortable as they are sensitive to the high frequency acoustic signals used by acoustic sensing. 

\begin{figure}[hb]
	\centering
	\subfigure[Wi-Fi monitoring for driver alone.]
	{
		\centering
		\includegraphics[width=0.31\textwidth]{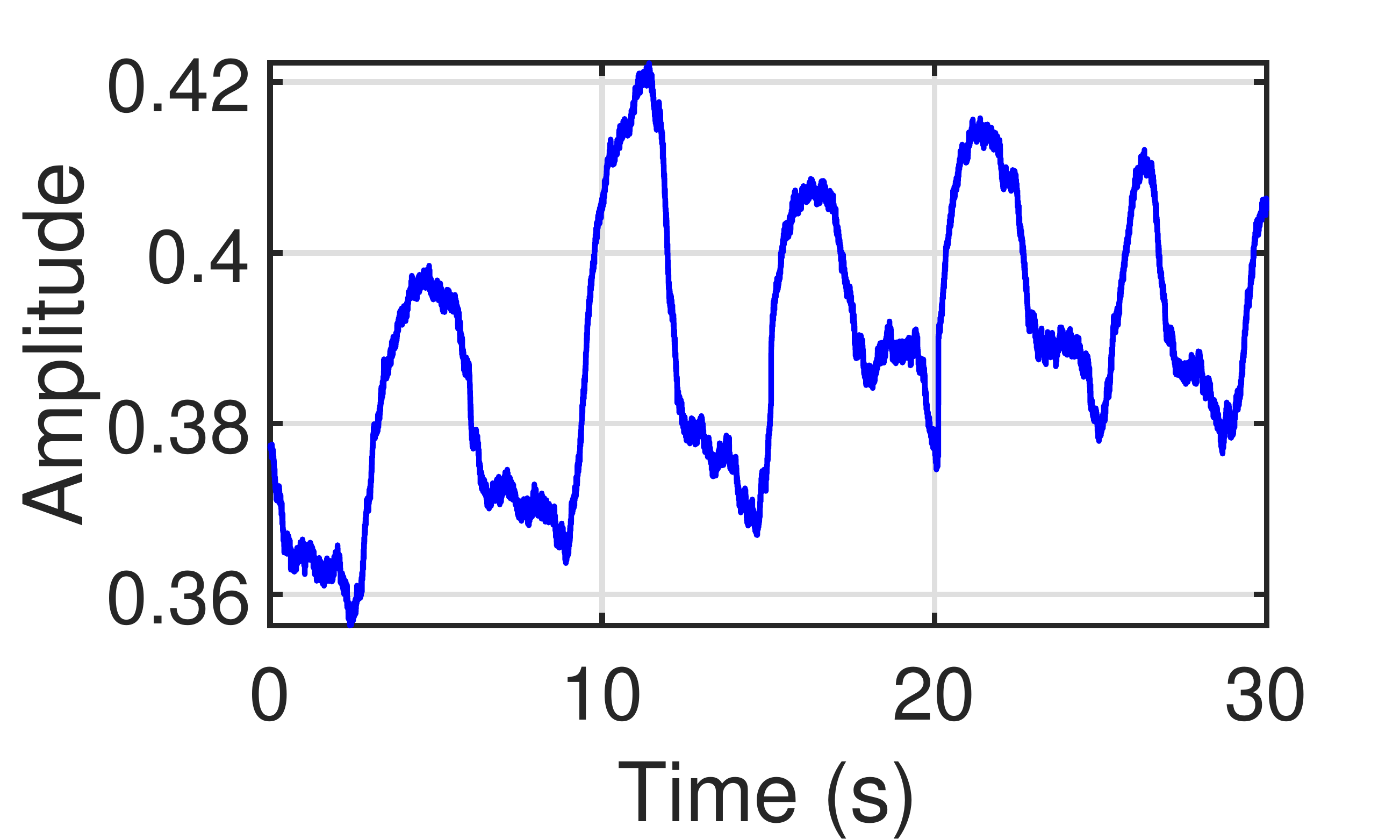}
		\label{fig:wifi_person1}		
	}
	\subfigure[Wi-Fi monitoring with multiple persons.]
	{
		\centering
		\includegraphics[width=0.31\textwidth]{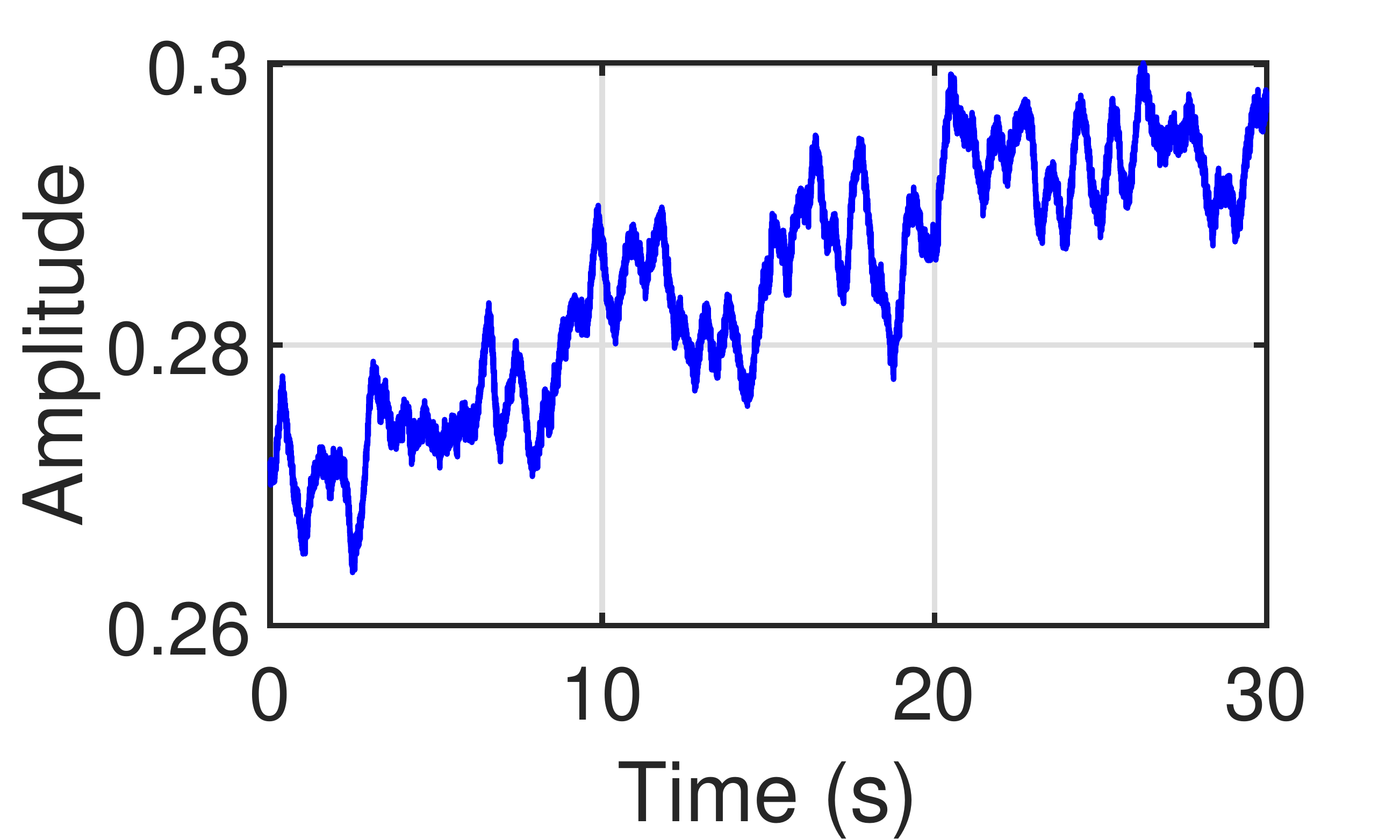}
		\label{fig:wifi_person3}		
	}
	\subfigure[Impulse radio with multiple persons.]
	{
		\centering
		\includegraphics[width=0.31\textwidth]{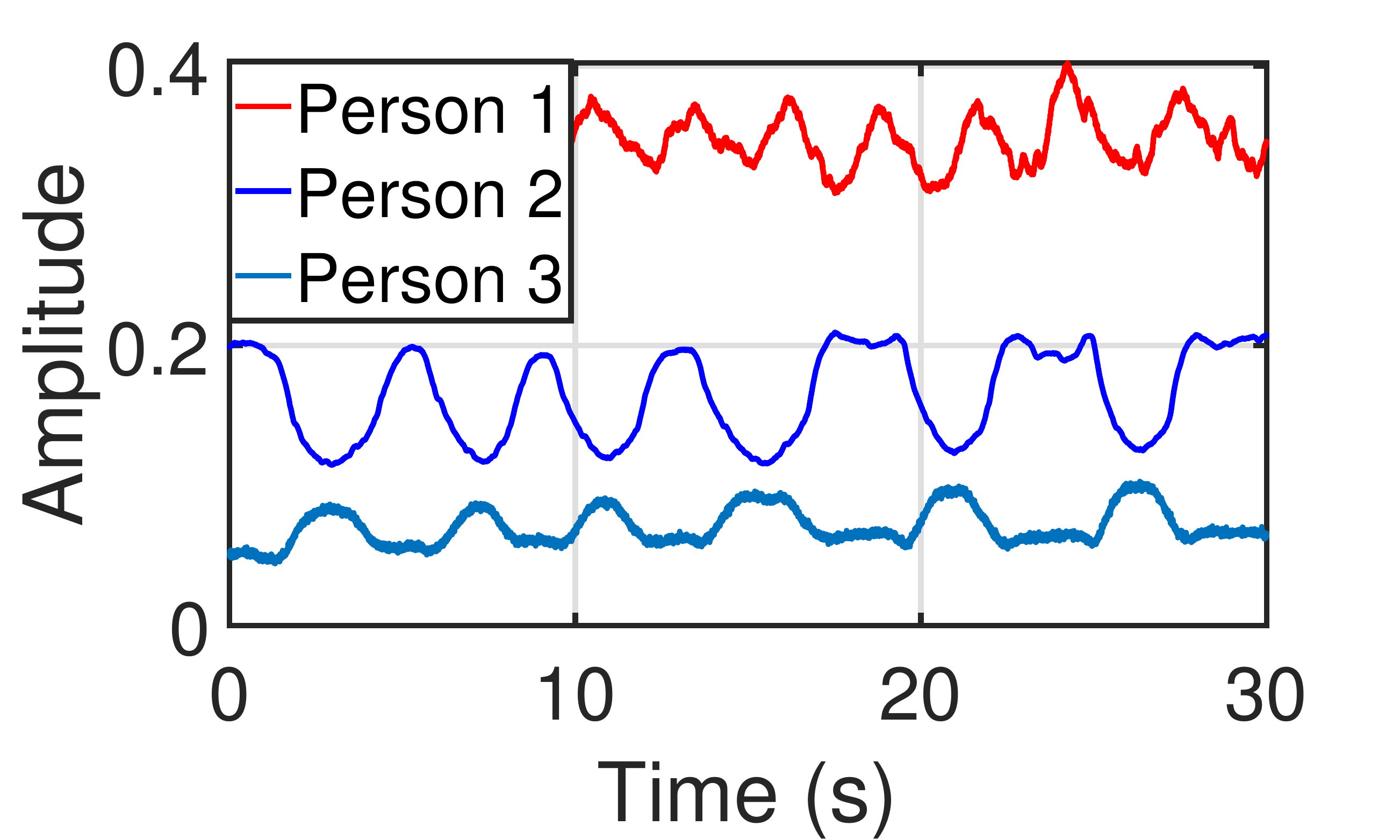}
		\label{fig:uwb_persons}		
	}
	\caption{Drivers' respiratory waveform can be detected without any other person using Wi-Fi CSI shown in (a). However, (b) illustrates that multiple persons degrade the respiratory waveform of driver due to narrow band Wi-Fi in the rich multipath in-vehicle environment. Instead, \systemname\ leverages a COTS impulse radio to distinguish multiple persons, so as to further filter out multiple persons' interference as demonstrated in (c).}
\end{figure}

Compared with acoustic signals, RF signals are even less sensitive to ambient noises such as heat and sounds, hence it poses a more promising solution for in-vehicle monitoring.
WiFind~\cite{WiFind-IEEEToBD} leverages Channel State Information (CSI) of Wi-Fi devices  to detect driver motions and respiratory rate, based on which fatigue state could potentially be inferred. However, it only offers coarse-grained measurements
due to its narrow Wi-Fi bandwidth. To make things worse, the system may fail in the presence of multiple persons in the vehicle. In order to better illustrate this latter issue, we use Wi-Fi card Intel 5300~\cite{b5:csitool} to detect respiratory waveform in a vehicle. Whereas the respiratory waveform of a driver alone can be clearly identified in Fig.~\ref{fig:wifi_person1}, the periodic feature of this waveform is corrupted and cannot be easily distinguished when multiple passengers are present, as shown in Fig.~\ref{fig:wifi_person3}. The reason is that the narrow Wi-Fi bandwidth cannot support high temporal resolution in such an extreme multipath environment. Note that differentiating multiple persons via their respective respiratory rates does not help, as identifying the critical monitoring target (e.g., the driver) is still impossible without position information, and the respiratory rates can be similar even for different persons. 
As a teaser to our work, Fig.~\ref{fig:uwb_persons} demonstrates that, when properly used, an impulse radio with ultra-wideband offers a much higher temporal resolution, and hence can readily differentiate multiple persons based on their relative distances, 
\captionsetup[figure]{labelfont=sf}
and then focus on monitoring the driver. Last but not least, using standard Wi-Fi radio for sensing purpose may affect its normal data communication function, and the resulted setting is \textit{bistatic} (i.e., transmitter and receiver are separated~\cite{raemer1996radar}); these all make it difficult to test and deploy a system. 

To summarize, in order to further improve the performance of driver vital sign monitoring system
based on RF sensing, we face the following challenges that largely confine the capability of existing systems:
\begin{itemize}
    \item \textbf{Deployment complexity.} In-vehicle vital sign monitoring based on Wi-Fi CSI requires an Access Point~(AP) and a STAtion~(STA)~\cite{WiFind-IEEEToBD}. Such a bistatic system is too complicated to be adopted as an intelligent component of smart vehicles in practice.
    \item \textbf{Passenger interference.} Previous research works based on Wi-Fi CSI are not robust to the cases where multiple persons are in a vehicle. Respiratory movements caused by other passengers in the cabin can distort the CSI and worsen the performance.  
    \item \textbf{Excessive noise.} Previous vital sign monitoring systems are mostly designed for a home setting~\cite{adib2015smart, Fullbreathe-UbiComp}. Performance of in-vehicle vital sign monitoring can deteriorate with the presence of excessive noise caused by the running vehicle and moving human body parts.
    \item \textbf{Low sensitivity.} Technologies based on Wi-Fi CSI cannot ensure sufficient temporal
    resolution because of the narrow bandwidth. Therefore, measurements of vital signs using Wi-Fi CSI cannot be very accurate, especially given the vibration background created by engines and road conditions. 

\end{itemize}
\begin{figure}[b]
	\centering
	\includegraphics[width=1\linewidth]{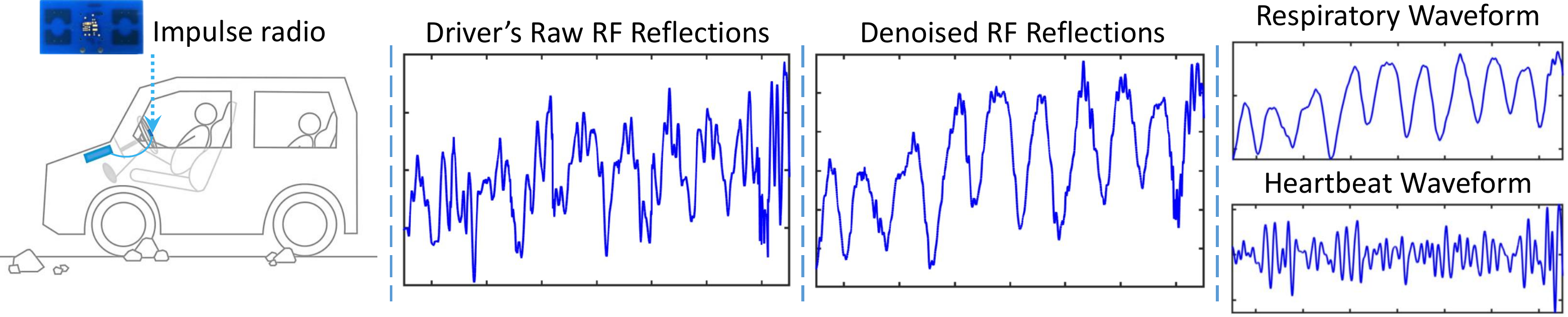}
	\caption{A vital signs monitoring scenario of driving vehicle. The driver's RF reflections always contains his or her vital signs with noise introduced by vehicle's vibrations.  \systemname\ separates such vibrations to reconstruct signals of vital signs.}
	\label{fig:case_eg}
\end{figure}
To answer these challenges, we propose \systemname\ as a  compact in-\textbf{V}ehicle \textbf{Vi}tal sign monitoring system endowed with fine-grained sensitivity. It consists of low-cost Commercial Off-The-Shelf (COTS) components, and is hence readily deployable, as illustrated in Fig.~\ref{fig:case_eg}. \systemname\ can filter the interference from passengers and vibrations caused by moving vehicles while focusing on driver's vital sign monitoring. The reconstructed RF signals induced by vital signs can then be used to estimate respiratory rate, heart rate, and even finer-grained heart rate variability. %
As \systemname\ offers continuous monitoring on driver's vital signs, we will be able to perform real-time and comprehensive diagnosis of the health status of the driver.
Particularly, we should be able to detect driver drowsiness in its early stage and hence prevent vehicle accidents.

Specifically, \systemname\ employs a COTS impulse radio to collect RF reflections from people. The COTS impulse radio is compact and \textit{monostatic} (i.e., the transmitter and receiver are co-located~\cite{raemer1996radar}); it can be readily integrated with embedded devices such as Raspberry Pi~\cite{rpi} adopted by us to form an edge computing node, rendering a straightforward in-vehicle deployment shown in Fig.~\ref{fig:case_eg} (leftmost).
The radio's ultra-wideband enables us to develop algorithms that accurately sorts out signals reflected from different people under complex multipath conditions. This further allows us to zoom in and inspect the RF reflections from the driver and hence extract his/her vital signs. In addition, the competent sensitivity of \systemname\ delivers abundant information for combating other background noises. Leveraging this capability, our new Multi-Sequence Variational Mode Decomposition (MS-VMD) algorithm decomposes a set of related time series into multiple band-separated intrinsic  modes. This enables \systemname\ to filter out noises such as  vibrations of vehicle, reconstruct both respiratory and heartbeat waveforms, and in turn estimate respiratory rate, heart rate, and even heart rate variability accordingly. 
In summary, our major contributions in this paper are:
\begin{itemize}

\item  Our hardware design for \systemname\ innovatively combines a COTS impulse radio with a compact edge computing platform. Leveraging the high temporal resolution offered by an impulse radio, \systemname\ is able to distinguish signals reflected from various users and hence focus on monitoring the driver. Moreover, compared with existing wireless solutions (e.g.,~\cite{WiFind-IEEEToBD}), \systemname\ is convenient to deploy thanks to its low system complexity, making it a practical system for complicated driving environments.

\item We emphasize on the robustness in vital sign monitoring to overcome noise in driving environments. To prepare the signal for vital sign extraction, \systemname\ uses a combination of filters to remove noise and improve signal quality, it then detects and discards unstable data frames corrupted by noise. As a result, \systemname\ can extract vital sign information from noisy data frames to the largest extent.

\item We develop a novel MS-VMD algorithm to fully leverage the wide bandwidth of the UWB signal and perform signal separation via optimization techniques. The MS-VMD algorithm allows \systemname\ to jointly extract vital signs from multiple sequences of signals. This enables \systemname\ to accurately estimate driver's vital signs carried by reflected signals, including even the subtle heart rate variability, which, to the best of our knowledge, has never been extracted by wireless sensing under driving condition.

\item We conduct extensive experiments to evaluate the performance of \systemname\ under various conditions. All the results demonstrate that \systemname\ is robust to various unfavorable factors including bumpy road and heavy clothing of the driver, proving its outstanding effectiveness in practice.

\end{itemize}

The rest of the paper is organized as follows. Sec.~\ref{sec:vs_driver} explains the vital signs to be measured by \systemname. Sec.~\ref{sec:systemdesign} presents the system design of \systemname. Sec.~\ref{sec:eval} reports the evaluation results. Limitations and potentials are discussed in Sec.~\ref{sec:limit_fw}.  Sec.~\ref{sec:rw} gives a brief review of related work. Sec.~\ref{sec:con} concludes this paper.

\section{Vital Signs as Health Indicators in Driving Environments} \label{sec:vs_driver}
Before proceeding to present our system details, we first discuss the measurable vital signs as health indicators in driving environments. 
The essential idea is to provide a necessary background and strong motivation for the development of \systemname, in the sense of which vital signs can be quantitatively measured and how they are relevant to personal health status.
Basically, we propose adopting the following dimensions: i) respiratory rate: the number of breaths a person takes per minute, ii) heart rate: the number of times a person's heart contracts and relaxes per minute, and iii) heart rate variability (HRV): the physiological phenomenon of variation in time intervals between heartbeats. 
All these can be measured by \systemname's impulse radio, and they are closely related to personal health status.

\subsection{Respiratory Rate and Heart Rate}
Respiratory rate 
is one of the crucial indicators of a person's health issues. Abnormal respiratory rate under 12 breaths-per-minute (bpm)
or over 25~\!bpm has been shown to be an important predictor of health issues, such as asthma~\cite{catterall1982irregular}, anxiety~\cite{giardino2007anxiety}, pneumonia~\cite{strauss2014prognostic}, lung disease~\cite{clark1990clinical}, heart disease~\cite{cretikos2008respiratory,fieselmann1993respiratory} and drug overdose. In particular, Solaz et al.~\cite{solaz2016drowsiness} observe abnormal change in respiratory rates and use it to infer abnormal health issues like driver drowsiness. Lisper et al.~\cite{lisper1973effects} report that, during a long-time continuous driving, respiratory rate of the driver and passengers are found to have slowed down gradually. %
Similarly, irregular heart rate is also a crucial indicator of serious health issues such as cardiac disease~\cite{cysarz2007regular, nabi2010psychological}. Also, Sharma et al.~\cite{sharma2011effect} show that irregular heart rate is a sign of electrolyte imbalance. In terms of driving environments, Jo et al.~\cite{korean2019hr} demonstrate that heart rate significantly decreases during sleepy driving. Banning et al.~\cite{banning2012driving} report that people with irregular heartbeat are more likely to cause a traffic accident.

The aforementioned discussions have clearly shown the relevance of both respiratory rate and heart rate as indicators for driving-specific health status (e.g., driver fatigue, drunk driving), so they should be chosen as key vital signs for in-vehicle monitoring, yet we would need to avoid using intrusive sensors that cause uncomfortable user experience. 
Recently progress in RF sensing provide us with an alternative solution for remotely extracting these vital signs. 
Zeng et al.~\cite{Fullbreathe-UbiComp} design a Wi-Fi based system for respiration detection, and 
Adib et al.~\cite{adib2015smart} demonstrate that respiratory rate and heart rate can both be measured remotely by leveraging an FMCW radio. In particular, for the driving environments, Peng et al.~\cite{WiFind-IEEEToBD} has migrated the Wi-Fi based for estimating respiratory rate, while Park et al.~\cite{park2019noncontact} design a special RF sensor for monitoring both respiratory rate and heart rate of the driver. However, it is still not clear if these proposals may effectively handle realistic scenarios with multiple passenger present.

\subsection{Heart Rate Variability}
It is well known that a healthy heart does not beat at a regular interval~\cite{shaffer2014healthy}. 
In other words, the rhythms of a healthy heart are not as stable as heart rate; they are instead non-linear and offer a higher-dimensional information. %
Therefore, it is a more useful indicator of a personal health status than respiratory and heart rates. 
Variations in heart beat intervals are controlled by the unconscious part of human nervous system, i.e., the Automatic Nervous System (ANS)~\cite{ernst2017heart}. This part of the nervous system regulates heart beat, respiration, pupillary response, urination, sexual arousal and digestive process. The ANS can be further divided into two sub-systems,
i.e., sympathetic and parasympathetic nervous system. Human heart without external input will beat at constant rate of approximately 100~\!bpm. However, the two nervous subsystems compete with each other, influencing nerves in heart tissues and causing variations in heart beat intervals. Therefore, HRV clearly reflects the general state of the nervous system and thus personal health status.

Specifically, HRV has been used to detect multiple health issues. In~\cite{sessa2018heart}, Sessa et al. report abnormal HRV precedes sudden cardiac attack. In~\cite{fujiwara2018heart}, Fujiwara et al. demonstrate that changes in sleep patterns affect nervous system, and HRV changes as a result. In another study, Benichou et al.~\cite{benichou2018heart} show that type 2 diabetes cause significant decrease in HRV. HRV is also shown to be linked to asthma~\cite{lutfi2015patterns}, anxiety~\cite{chalmers2014anxiety} and drug overdose~\cite{waring2008impaired}. 
Similar to respiratory rate and heart rate, HRV can also be obtained from reflected RF signals. Although HRV signal is weak and hard to capture, Zhao et al.~\cite{zhao2016emotion} demonstrated that ECG-like HRV signal can be extracted from mixed signal of heartbeat and respiration using numerical differentiation. Whereas their findings have confirmed the existence of HRV signal in the reflected RF signal, their method may not be robust to noise and interference under driving conditions, so we require a novel estimation algorithm for \systemname. 

In brief, HRV is particularly useful for indicating health status in any environments, and it can be obtained from RF signals, so we choose it to be one of the key vital signs for \systemname.
However, there are multiple indices that describe HRV, and they can be categorized into categories  summarized in Table~\ref{tab:sys_eval}
~\cite{shaffer2017overview}.
\captionsetup[table]{labelfont=sf}
\begin{table}[ht] 
	\centering	
	\caption{A summary of HRV measurements.} \label{tab:sys_eval}
	\begin{tabular} {p{4cm} p{10cm}}
		\hline
		Category  & Parameters \\
		\hline
		Time Domain &  Interbeat Interval (IBI), SDNN, SDRR, SDANN, SDNN index (SDNNI), pNN50, HR Max - HR Min, RMSSD, HRV triangular index  \\
		\hline
		Frequency Domain &  ULF power, VLF power, LF peak, LF power, HF peak, HF power, LF/HF   \\
		\hline
		Non-linear &  SD1, SD2, SD1/SD2, ApEn, SampEN, DFA$\alpha$1, DFA$\alpha$2, $D_2$   \\
		\hline
	\end{tabular}
\end{table}
Some of these indices require complex computation and processing (e.g., ECG-like waveform). 
In our work, as RF signal only carries temporal information about heartbeats. We choose Interbeat Interval (IBI) as the HRV index. In fact, some other indicators, such as SDNN, SDRR and LF/HF, can be derived from IBI.

\section{System Design of \texorpdfstring{\systemname}{V2iFi}} \label{sec:systemdesign}
\systemname\ is composed of two hardware components, a single-board computer and an impulse radio, as well as backend algorithms for signal processing. It utilizes RF signals transmitted by the impulse radio for vital sign monitoring. Upon receiving the reflected signals, delicate procedures are taken to distill the targeted vital signs out of them, notably by our novel MS-VMD algorithm.
In the following, we first provide a software system overview in Sec.~\ref{ssec:overview}, while leaving the explanations for hardware components to Sec.~\ref{ssec:expset}. Then we present the RF channel model in Sec.~\ref{ssec:rf_channel} and discuss algorithms for signal prepossessing, signal separation, and vital sign extraction in Sec.~\ref{ssec:prep}, \ref{ssec:vmd}, and \ref{ssec:vs}, respectively.

\begin{figure}[ht]
	\centering
	\includegraphics[width=0.8\linewidth]{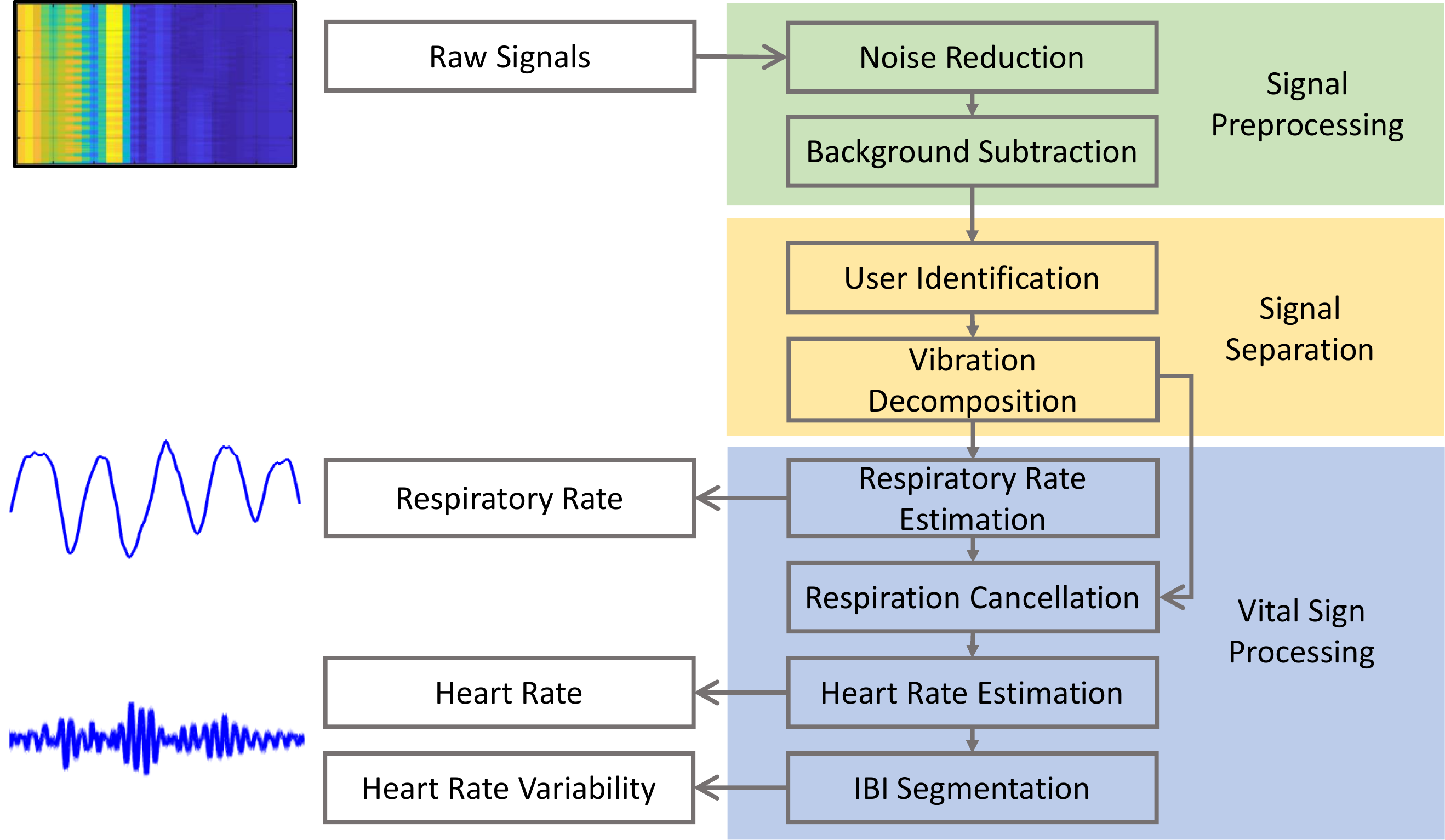}
	\caption{Software system design of \systemname. }
	\label{fig:systemdesign}
\end{figure}

\subsection{Software System Overview} \label{ssec:overview}
\systemname's software architecture is illustrated in Fig.~\ref{fig:systemdesign}. The whole system consists of three components: signal preprocessing, signal separation, and vital sign processing.

\begin{itemize}
    \item \textit{RF signal preprocessing component.} This component includes  noise reduction and background subtraction. After the reflected signal is received by the radio, the data is forwarded to the embedded computer for processing. Due to low power of the wideband Guassian pulse~( $\mathrm{-41\!~dBm/MHz}$ )~\cite{xethru_comparison}, we need to improve the SNR of signals. Basically, we employ smooth filter to reduce noise in the signals. Then, we utilize a loopback filter to remove the static reflections from clutters for background subtraction. 

	\item \textit{Signal separation component.} 
	The greatest difficulty of in-vehicle vital sign estimation is how to remove the noise caused by vibrations of vehicle from the reflectors. We perform two steps to eliminate the vibration. First, we locate the driver, and extract driver-specific signals. Second, we propose a novel MS-VMD algorithm to decompose the mixed signals containing vibration noises and vital signs via leveraging the reflections from multiple body parts (i.e., signals from different fast-time bins).

	\item \textit{Vital sign processing component.}	
	After decomposing the signals, \systemname\ identifies respiration and heartbeat signals from them. Next, \systemname\ can estimate vital signs like respiratory rate,  heart rate, and IBI.  Together with all of them can be used to determine health issues of the driver.
	
\end{itemize}
We will model RF channel first, and elaborate on all components of \systemname\ in the next few sections. 

\subsection{Modeling RF Channel} \label{ssec:rf_channel}
\begin{figure}[ht]
	\centering
	\includegraphics[width=0.6\linewidth]{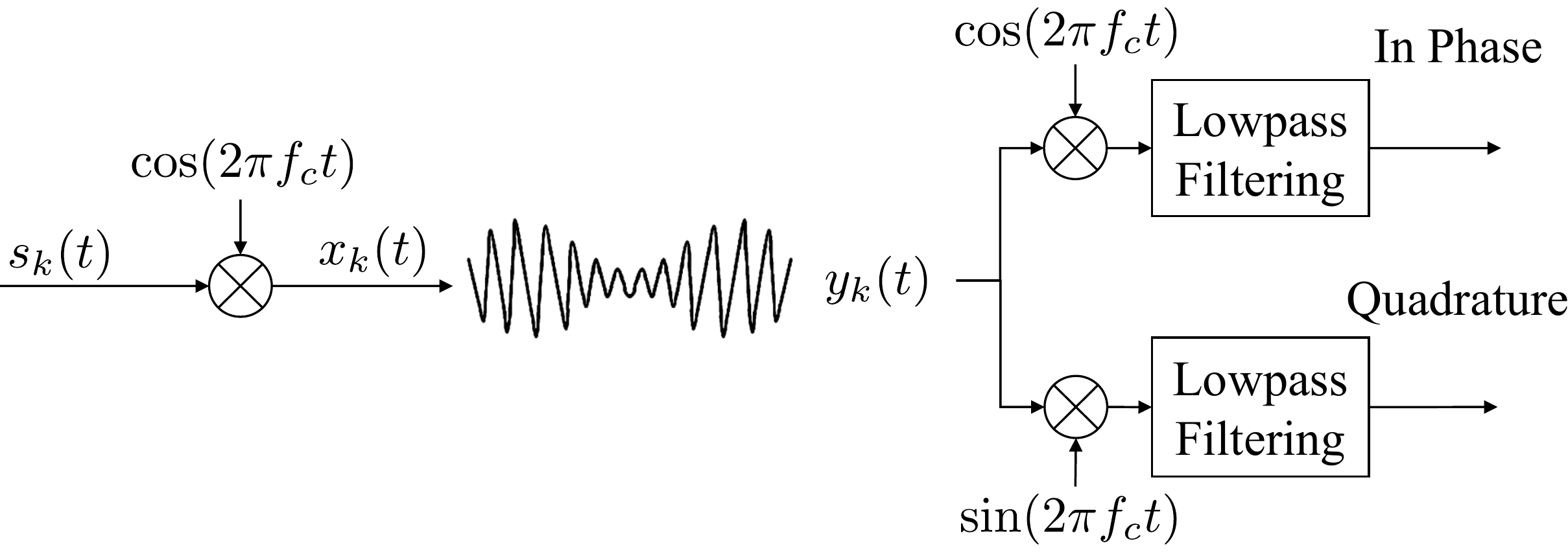}
	\caption{System diagram from the baseband transmitted signal $s_k(t)$ to the baseband received signal $y_k^b(t)$.}
	\label{fig:radarsysdiagram}
\end{figure}

In this section, we introduce the theory of RF modeling of the impulse radio. \systemname\ utilizes a system-on-chip impulse radio for transmitting and receiving wireless pulses. The system diagram of \systemname\ is illustrated in Fig.~\ref{fig:radarsysdiagram}. The transmitted signal is $s_k(t)$, the modulated signal is $x_k(t)$, the signal after passing the channel is $y_k(t)$, and the demodulated signal is $y^b_k(t)$. The architecture of the impulse radio is slightly different from traditional architecture~\cite{xethru}: it only employs an in-phase single-carrier frequency $\mathrm{cos}(2\pi f_{c} t)$ for upconversion, but In-phase and Quadrature~(IQ) sampling ($\mathrm{cos}(2\pi f_{c} t)$ and $\mathrm{sin}(2\pi f_{c} t)$)  at the receiver side for downconversion.  

\begin{figure}[ht]
	\begin{minipage}[ht]{0.44\linewidth}
		\centering
		\includegraphics[width=\textwidth]{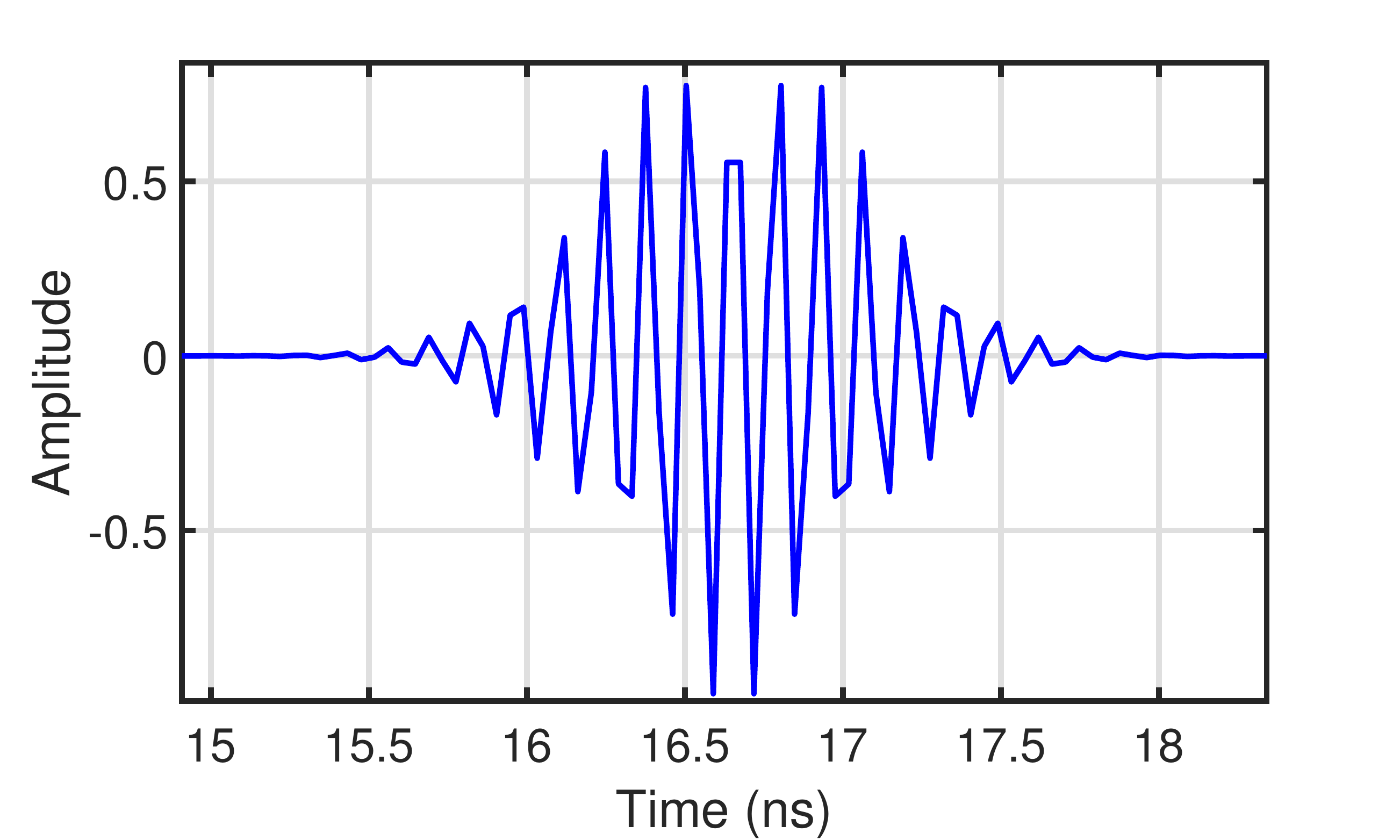}
		\caption{Transmitted $x_k(t)$ in time domain.}
		\label{fig:uwbsig}
	\end{minipage}
	\begin{minipage}[ht]{0.44\linewidth}
		\centering
		\includegraphics[width=\textwidth]{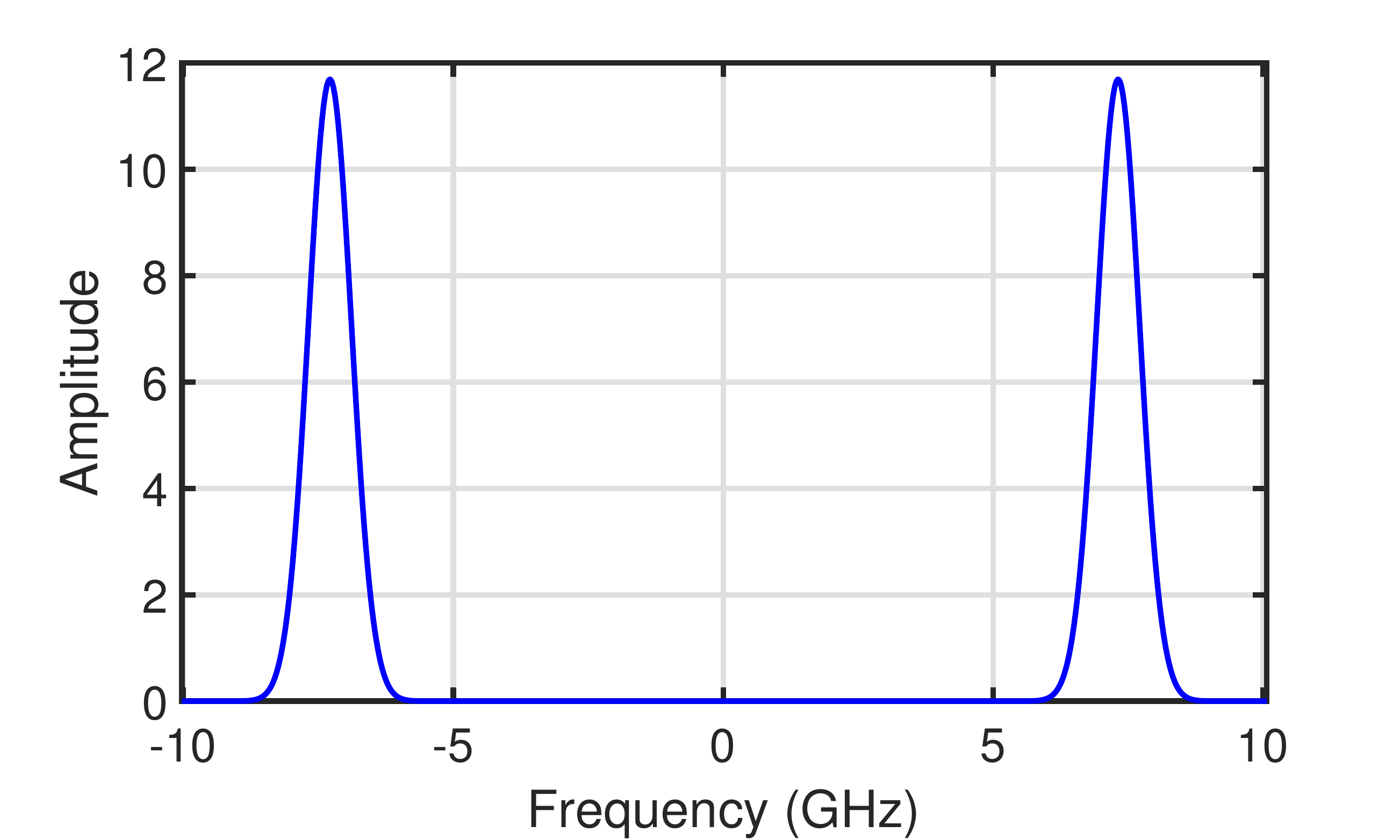}
		\caption{Transmitted $x_k(t)$ in frequency domain.}
		\label{fig:uwbsigfre}
	\end{minipage}
\end{figure}

The Gaussian pulse transmitted by \systemname\ can be expressed by the following equation:
$ s(t) = V_{tx}  \exp\left(- \frac{ (t - \frac{T_p}{2})^2 }{ 2 \sigma^2_p }\right) $
where the amplitude of the pulse is $V_{tx}$, $T_p$ is the duration of the signal, and $\sigma_p^2$ is the variance corresponding to the -10 dB bandwidth, actually we have $ \sigma_p = \frac{ 1 }{ 2 \pi B_{-10dB} (\log_{10}(e))^{1/2} }$.  
After upconversion, the transmitted signal in time domain is given by:
\begin{align} \label{eq:trans_sig}
x_k (t) = s(t - k T_s) \cdot \cos (2 \pi  f_c (t- kT_s )),
\end{align}
where $f_c$ is carrier frequency, $T_s = \frac{1}{f_p}$ is duration of frame where $f_p$ is the pulse repetition frequency,  $k$ denotes the $k$-th frame. Because the impulse radio transmits a sequence of identical pulses, we have $s(t - k T_s) = s(t)$. For simplicity, we can denote $t = t^{'} + k T_s$ with $t^{'} \in [0, T_s] $, and E.q.~\eqref{eq:trans_sig} can be written as
$x_k (t) = s(t) \cdot \cos (2 \pi  f_c t).$ 
The transmitted signal $x_k(t)$ is illustrated in Fig.~\ref{fig:uwbsig}, and its representation in the frequency domain is shown in Fig.~\ref{fig:uwbsigfre}. It can be seen that the carrier frequency is $\mathrm{7.3~\!GHz}$, and bandwidth is $\mathrm{1.4~\!GHz}$. 

The impulse response $h_k (t)$ of the in-vehicle environment is given by:
$h_k(t) = \sum_{p=1}^{P}  \alpha_p  \delta \left (t -  \tau_p    -  \tau_p^{D}(kT_s)  \right )  $ 
where $ \alpha_p $ is the channel gain of the $p^\mathrm{th}$ reflection path signal in the vehicle,  $\tau_p$ is the time delay of the $p$-th path, $\tau_p^{D}(k T_s)$ is the time delay caused by Doppler frequency shift of the $p$-th path. Moreover, for the impulse radio used in \systemname\ ,  $\tau_p = \frac{2 R_p}{c}  $ and $\tau_p^{D}(kT_s) = \frac{ 2 v_p k T_s}{c}$, where $R_p$ is the distance between the target to be detected and the UWB radio, $c$ is the speed of light, $v_p$ is the speed of the moving target. The range resolution of \systemname\ is given by
$\Delta r = \frac{c}{2B} $ 
where $B$ is the bandwidth of the impulse radio. Therefore, the time resolution of \systemname\ can be calculated as  $\Delta \tau = \frac{1}{2B}$.  

In summary, the received signal of the impulse radio is:
\begin{align} \label{eq:radar_multipath_1}
y_k (t)  = h_k(t) * x_k(t) = \sum_{p=1}^{P}  \alpha_p   \cos(2 \pi f_c (t - kT_s -  \tau_p    -  \tau_p^{D}(k T_s)  )  \cdot s(t -  kT_s -  \tau_p    -  \tau_p^{D}(k T_s) )  + n(t),
\end{align}
where $n(t)$ is Gaussian noise with a variance of $\epsilon^2$, and the symbol $*$ denotes convolutional operation. 

In a radar system, Pulse Repetition Interval (PRI) is the time between two consecutive pulses. We place the reflected signal of the same pulse in the same row, for example, the reflected signal of the first pulse in the first row, the reflected signal of the second pulse in the second row. We define the dimension of the row as the "fast-time" dimension, that is, the dimension of time slots composing a single PRI, and define the dimension of the column as the "slow-time" dimension, which updates every PRI~\cite{fast-Radarcon}.

\begin{figure}[ht]
	\begin{minipage}[ht]{0.463\linewidth}
		\centering
		\includegraphics[width=\textwidth]{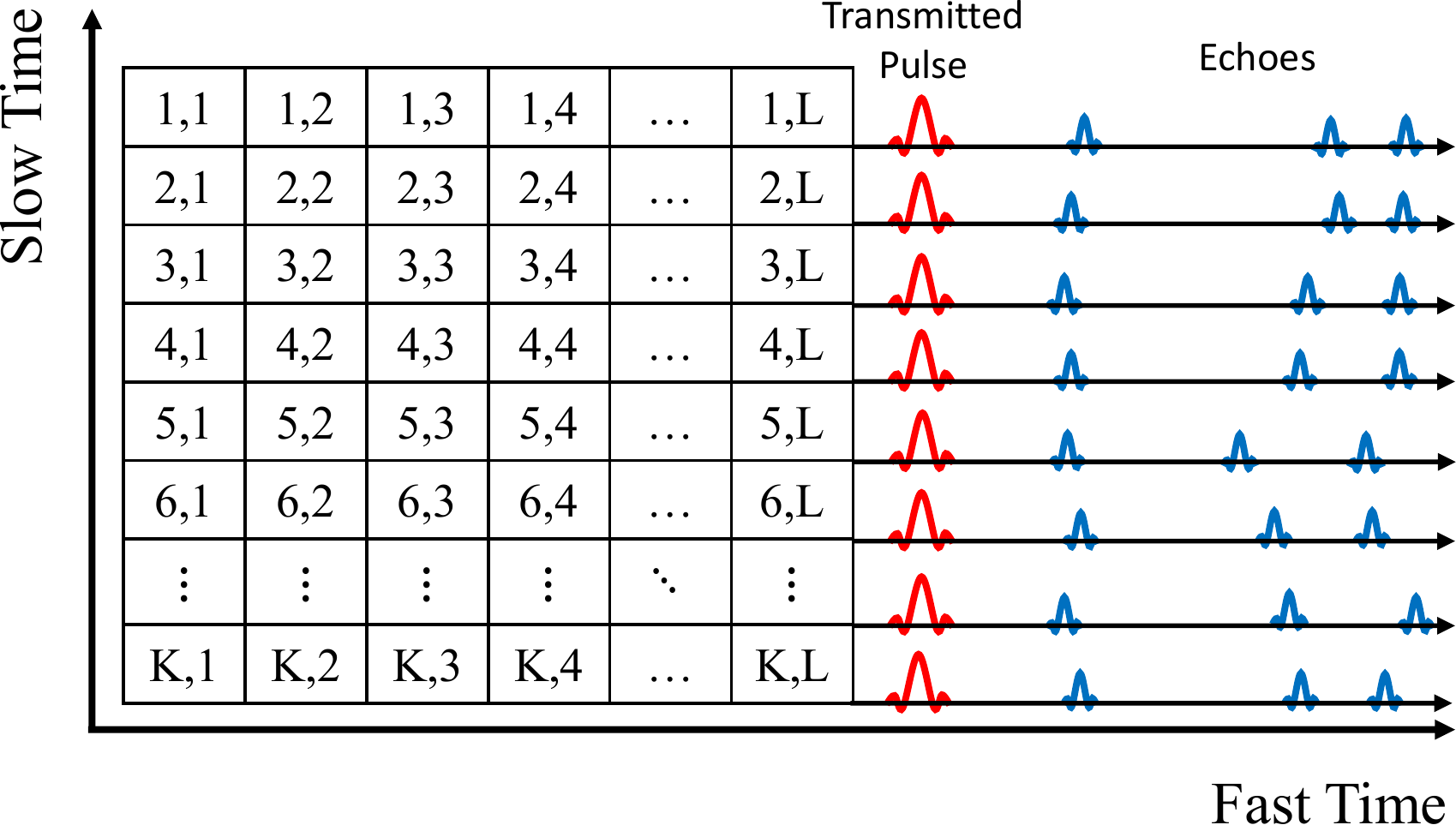}
		\caption{ The matrix of received baseband signals.}
		\label{fig:fast_slow_time}
	\end{minipage}
	\hfill
	\begin{minipage}[ht]{0.44\linewidth}
		\centering
		\includegraphics[width=\textwidth]{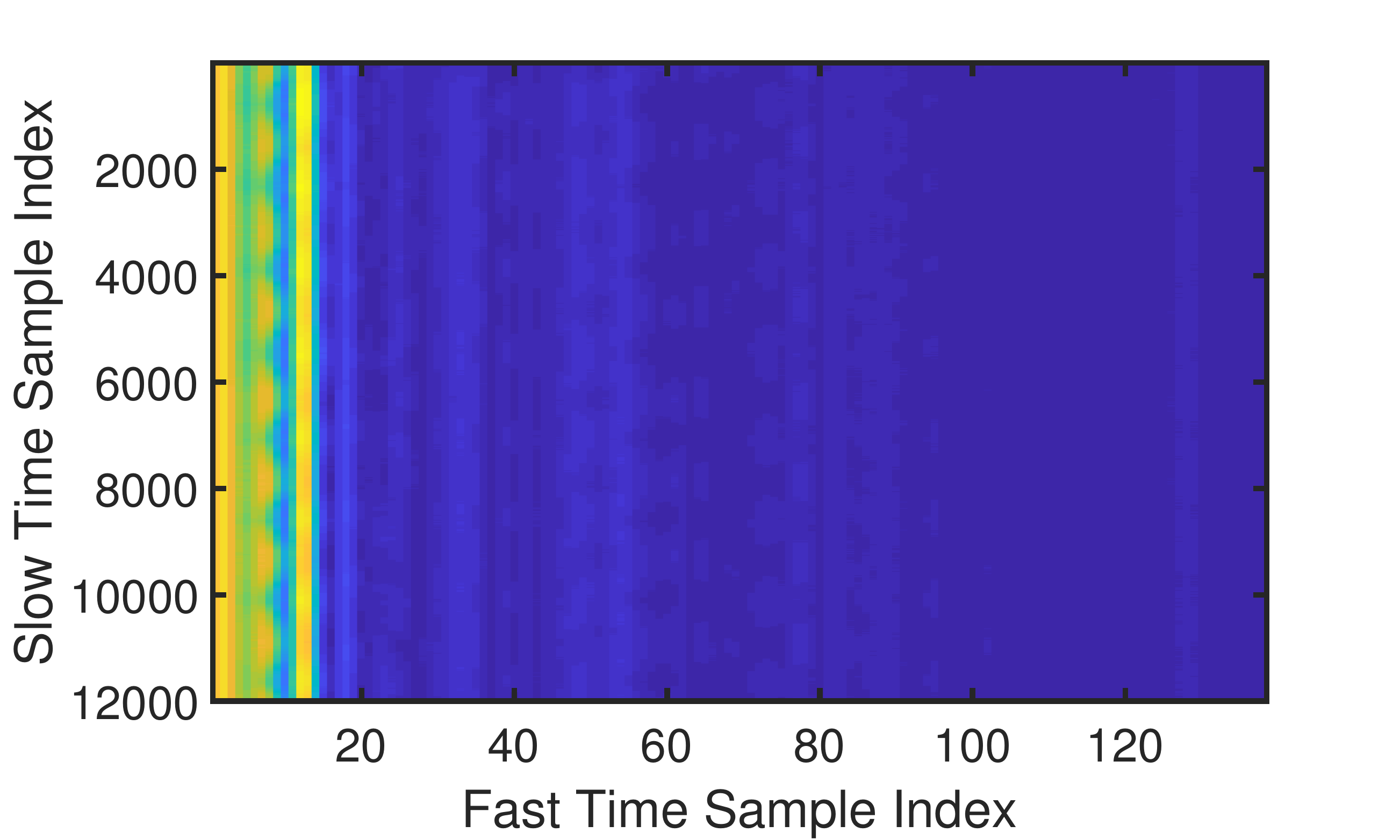}
		\caption{An example of received baseband signals. }
		\label{fig:walk_hm}
	\end{minipage}
\end{figure}

By transforming the time sequence of the signal into a matrix, the idea of "fast-time" and "slow-time" can be visualized, as shown in Fig.~\ref{fig:fast_slow_time}, and an example of real received signals from UWB radio are illustrated in Fig.~\ref{fig:walk_hm}. Generally speaking, the fast-time dimension denotes time delays of range distance, and the slow-time axis is used to estimate Doppler effect by observing over a long time span. Hereafter, we abuse the terminology by using \textit{sequence} to denote the vector of a ranging distance in the fast-time, such as a column vector in Fig.~\ref{fig:fast_slow_time}.

Received baseband signals $y_k^b (t) $ are obtained after applying IQ downconversion, we have:

\begin{align} \label{eq:radar_baseband}
y_k^b (t)  =  \sum_{p=1}^{P}   \alpha_p  e^{ 2 \pi f_c (\tau_p  +  \tau_p^{D}(k T_s)} \cdot s(t -  kT_s -  \tau_p    -  \tau_p^{D}(k T_s)   )  + n(t).
\end{align}

Let $t = lT_{n}$ represent $l$-th discrete samples obtained from ADC where $T_n$ is the sampling interval. Then the discrete baseband signals can be represented by $y_k^b (lT_n) $.  E.q.~\eqref{eq:radar_baseband} indicated that different minute movements caused by vital signs in the vehicle have different $\tau_p$ and $ \tau_p^{D}(k T_s)$ in  $y_k^b(t)$. By utilizing the differences, different vital signs can be distinguished.

\subsection{RF Signal Preprocessing} \label{ssec:prep}
Before extracting information from RF signals,  the effects from hardware or environment should be removed to guarantee signal quality. The RF signals preprocessing has two main steps: i) noise reduction and ii) background subtraction.

\subsubsection{Noise Reduction}

\begin{figure} [ht]
	\begin{minipage}[ht]{0.44\linewidth}
		\centering
		\includegraphics[width=\textwidth]{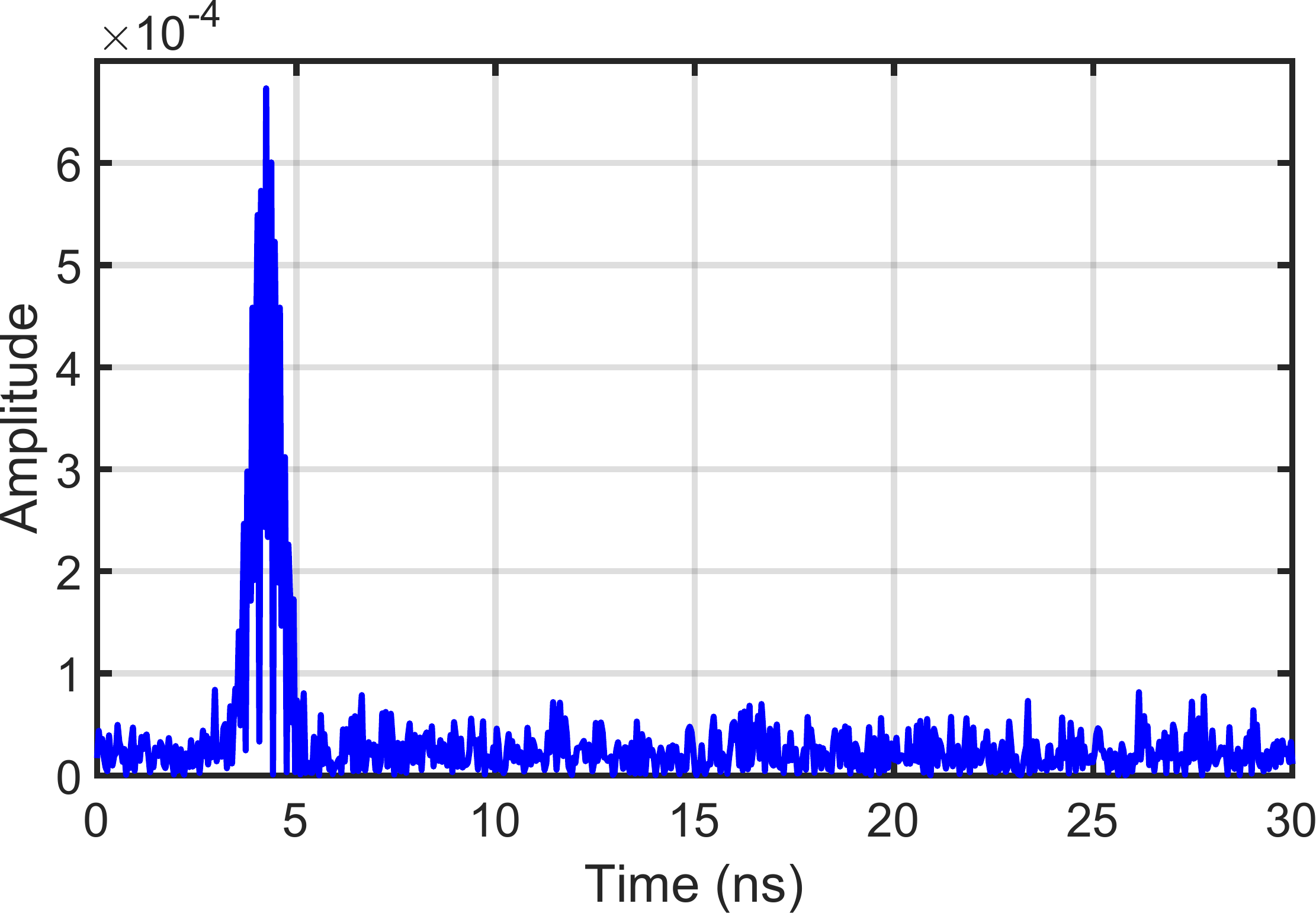}
		\caption{Received signal without SNR enhancement.}
		\label{fig:nonfilter}
	\end{minipage}
	\begin{minipage}[ht]{0.44\linewidth}
		\centering
		\includegraphics[width=\textwidth]{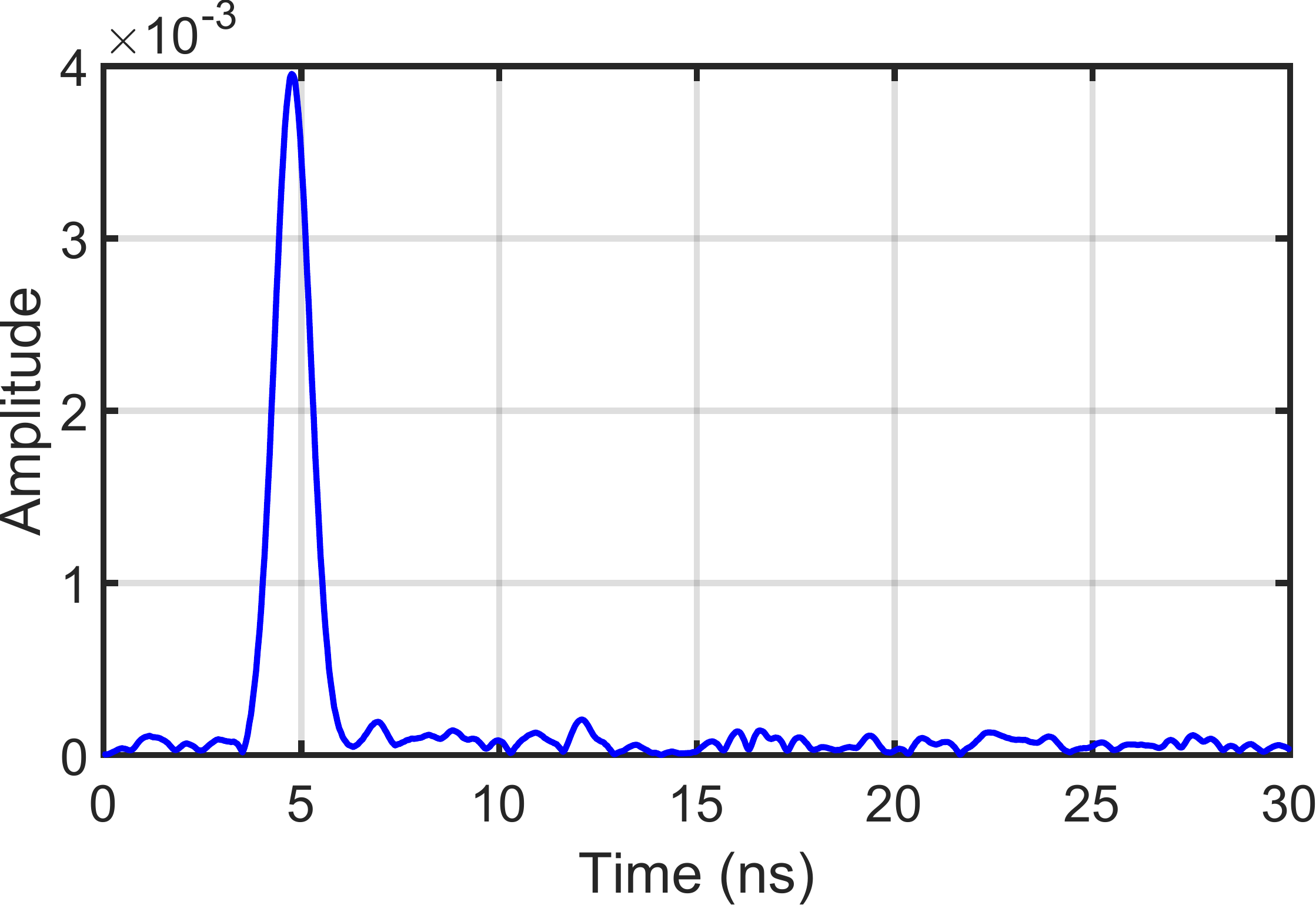}
		\caption{Received signal with SNR enhancement.}
		\label{fig:filter}
	\end{minipage}
\end{figure}

The received baseband signals are polluted with noise, as shown in Fig.~\ref{fig:nonfilter}. Noise will prevent the following vital signs extraction modules to work properly. Especially, vital signs will be immersed in noise. Therefore, a cascading filter comprised of a low-pass Finite Impulse Response (FIR) filter and a smoothing filter are utilized to enhance the SNR of the signals. The order of the designed FIR filter is 26 and Hamming window is used. The smooth filter with a window size of 50 points is used to further smooth the output signal of the FIR filter. Fig.~\ref{fig:filter}. illustrates the output of the cascading filter. It can be seen that noise is suppressed. 

\subsubsection{Background Subtraction}
\begin{figure} [ht]
    \begin{minipage}[ht]{0.44\linewidth}
    	\centering
    	\includegraphics[width=\linewidth]{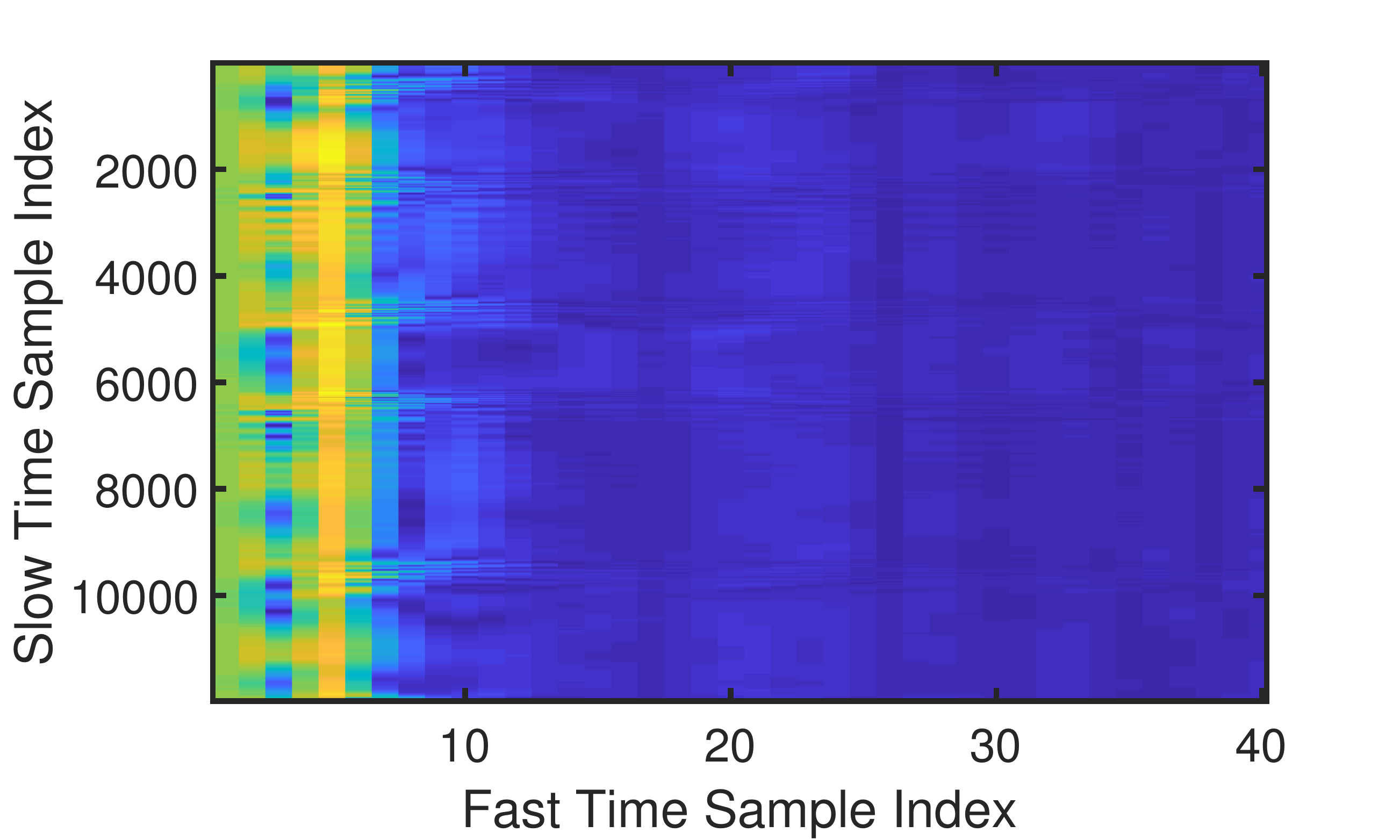}
    	\caption{Signal before background subtraction.}
    	\label{fig:bgrmvd}
	\end{minipage}
	\begin{minipage}[ht]{0.44\linewidth}
		\centering
		\includegraphics[width=\textwidth]{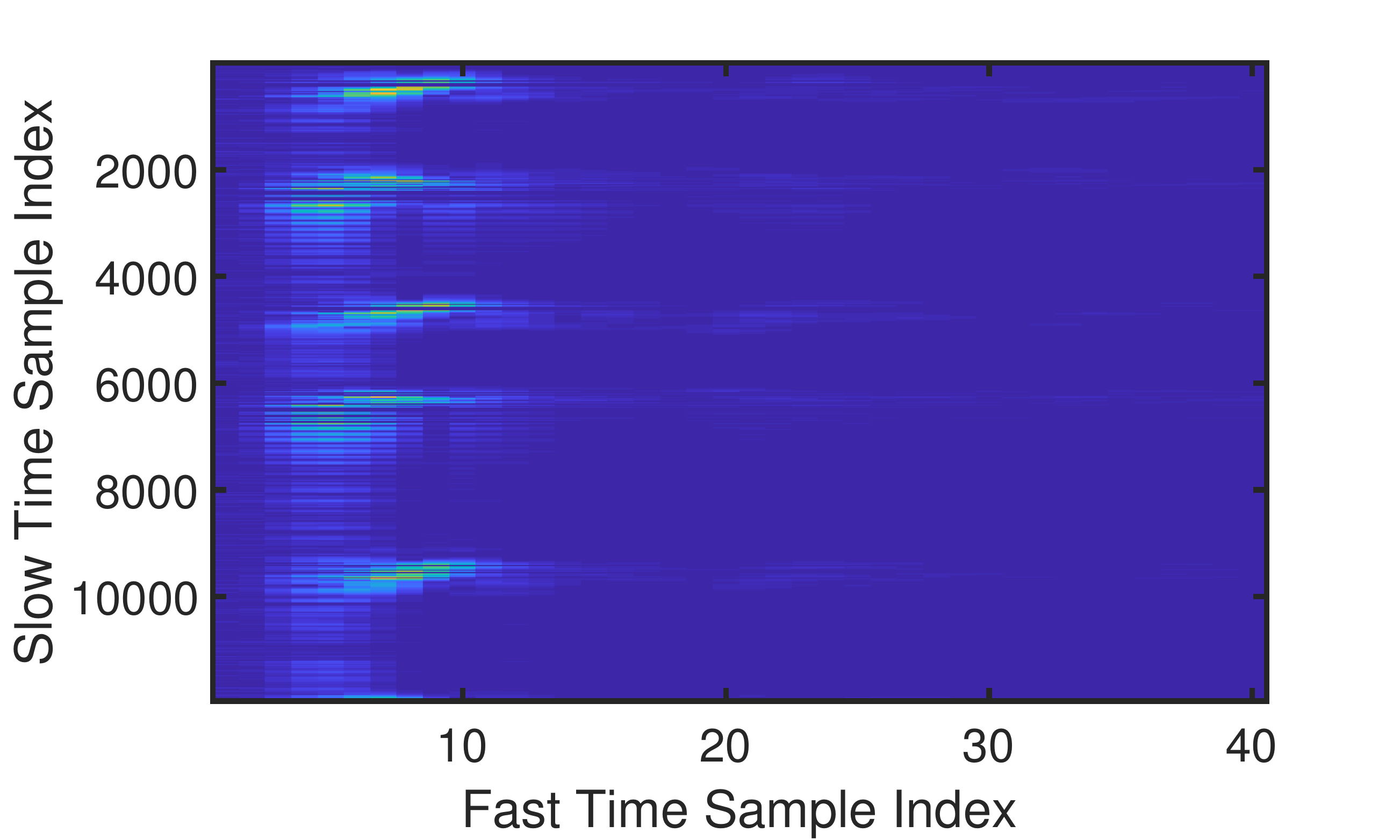}
		\caption{Signal after background subtraction. }
		\label{fig:without_bg}
	\end{minipage}
\end{figure}

Background subtraction removes all static reflectors in the background. Reflectors  includes static and moving objects. Therefore, besides reflected signal from the human body, there are many sources of unwanted static signals reflected in the vehicle shown in Fig.~\ref{fig:bgrmvd}. Unwanted signals in a radar are generally described as noise and clutter. Loopback filters can be used to remove clutter from the original signal~\cite{3Dtracking-NSDI}. The clutter of the system can be described as: $c_{k}(t)=\beta c_{k-1}(t)+(1-\beta) r_{k}(t)$ and the background subtracted signal can be represented as $y_{k}(t)=r_{k}(t)-c_{k}(t)$. In which $\beta$ is a constant used for weighting, $r_k(t)$ is the value at $k$-th frame. We can see that the signal after background subtraction is illustrated in Fig.~\ref{fig:without_bg}, where the clutters have been removed. A low $\beta$ value enables the filter to remove background fast, but the filter is not robust to noise. On the other hand, a high $\beta$ value makes background removal slow, but the process is robust to noise. In our work, $\beta$ is set to $\mathrm{0.97}$. 

\subsection{Signal Separation} \label{ssec:vmd}

\subsubsection{User Identification}\label{sssec:ui}

\begin{figure} [ht]
	\centering
	\includegraphics[width=0.9\linewidth]{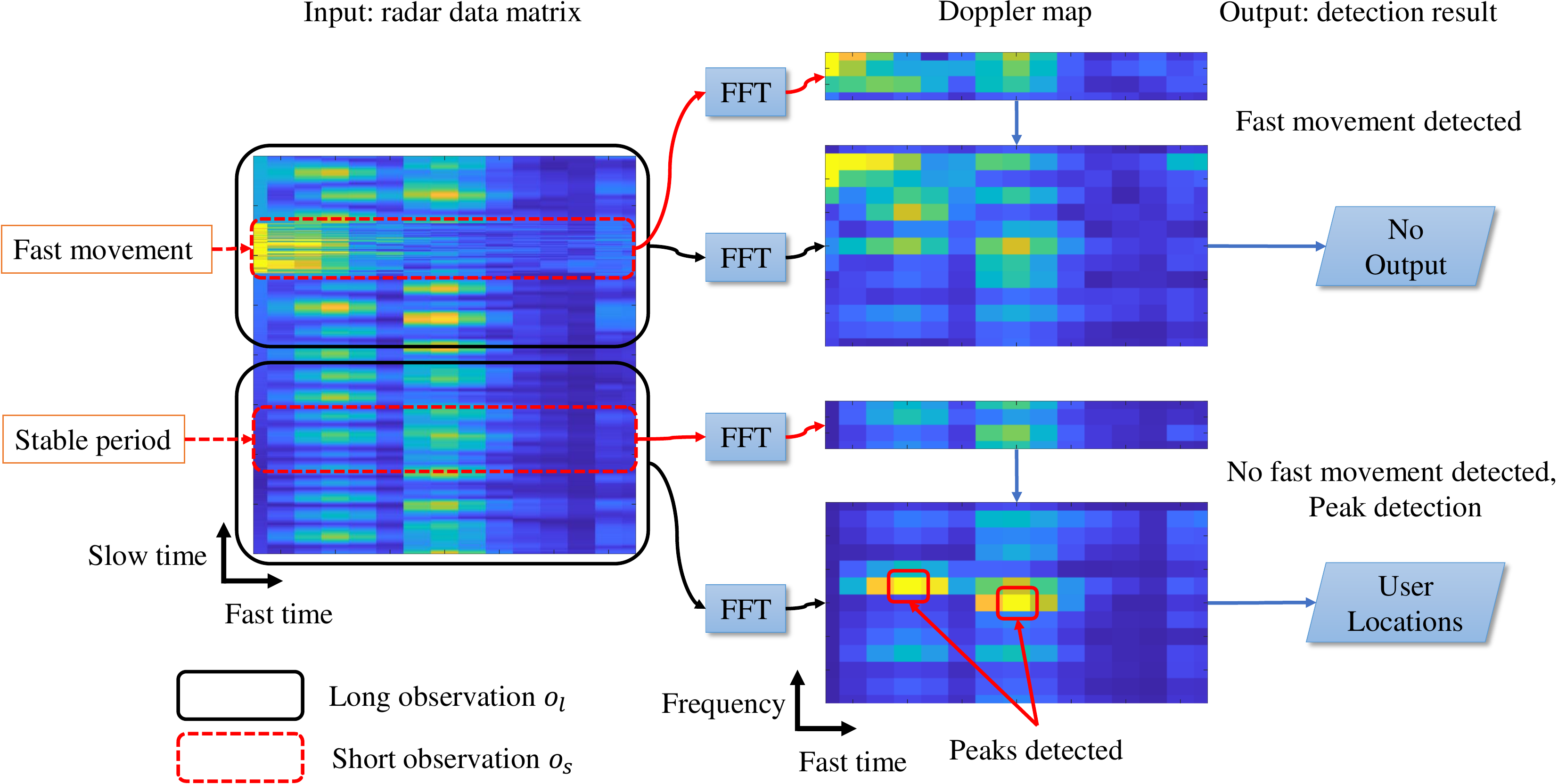}
	\caption{The diagram of the user identification process. In the diagram, FFT means column-wise FFT, i.e., FFT along the slow-time axis.}
	\label{fig:UI}
    \end{figure}

User identification is the process of identifying the driver's vital sign signal hidden in the radar data frames. It enables us to focus on the driver's signal for further vital sign extraction. The diagram of the user identification process is shown in Fig.~\ref{fig:UI}.

The first step of user identification is to filter out the radar data frames that are unsuitable for user identification and further processing. In the diagram, we can see that fast movements in the driving environments degrade vital sign signals greatly. To combat this, we take out 5~\!s data and 20~\!s matrix from the input radar data frames, and these two matrices are called the short observation $o_s$ and long observation $o_l$. We perform FFT along the slow-time axes of the two matrices to get the Doppler map of the long observation and short observation respectively. The Doppler map of the long observation is used for vital sign extraction, and the Doppler map of the short observation is used to detect fast movement.
	
By our definition, $b = \frac{\alpha_{\mathrm{peak}}}{\sum_{i}\sum_{j}{\alpha_{ij}}/(I\times J)}$ is the measure of relative signal strength, where $i$ denotes the $i$-th fast-time index, $j$ denotes the $j$-th frequency index, $\alpha_{\mathrm{peak}}$ is the peak amplitude on the Doppler map, and $\alpha_{ij}$ is the amplitude at fast-time index $i$ and frequency index $j$, $I$ is the total number of fast-time indices, and $J$ is the total number of frequency indices, hence $b_s$ (i.e., $b$ of the short observation) is a measure of the relative strength of fast movement signal, and $b_l$ (i.e., $b$ of the long observation) is a measure of the relative strength of vital sign signal. $b_s > c\times b_l$, where $c$ is empirically set to 1.2, is an empirical formula based on our experiments, it indicates that fast movement signal dominates vital sign signal. When this happens, the received signal is not stable and no valid vital signs can be extracted, hence we remove invalid frames. Otherwise, if there is no fast movement, peak detection algorithm is further used to locate the driver on the Doppler map. 

The peak detection algorithm uses an adaptive threshold to detect real peaks where humans exist, it determines the best threshold above which any peak can be considered to be from the desired target. A sliding window is used to scan all values and estimate the noise threshold near the fast-time sample indices under test. The window size is set to $60~\!\mathrm{cm} \times 0.16~\!\mathrm{Hz}$, where $60~\!\mathrm{cm}$ is the width of the human body and $0.16~\!\mathrm{Hz}$ is the maximum difference of a specific person's respiratory rate. For example, if two peak indices are separated by a distance less than 60$~\!\mathrm{cm}$, we only pick the peak index with higher amplitude. Last, we output the range value of the detected peak, which defines the location of the driver.

\begin{figure}[ht]
 	\centering
 	\includegraphics[width=0.6\linewidth]{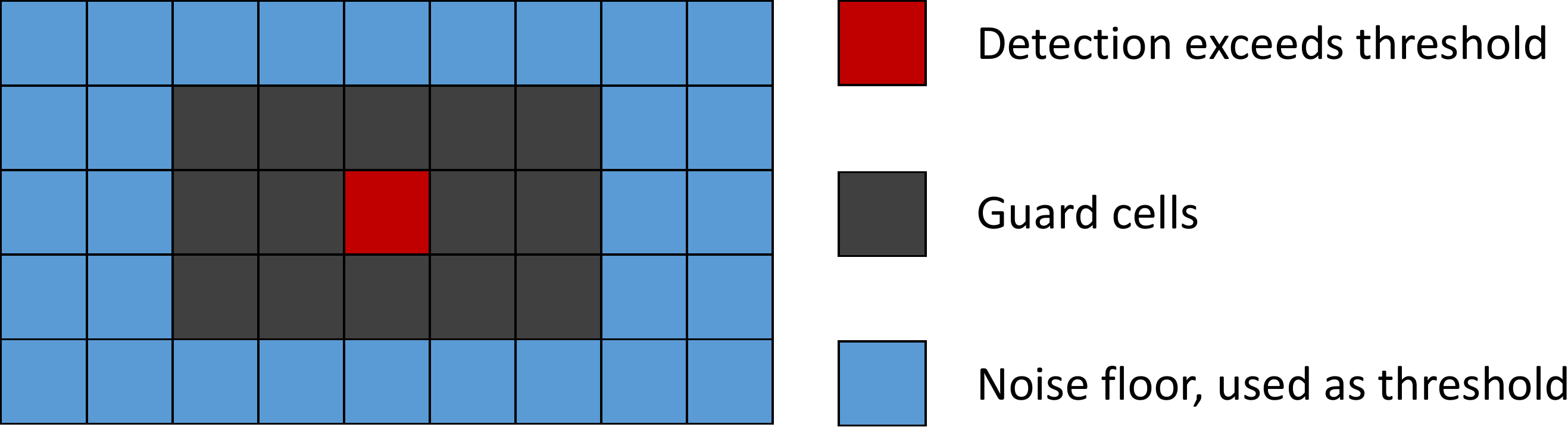}
 	\caption{The procedure of peak-average detection algorithm.}
 	\label{fig:cfar3}
 \end{figure}

That algorithm is illustrated in Fig.~\ref{fig:cfar3}. It can be seen that the noise floor threshold $ th_{\mathrm{motion}}$ can be estimated by averaging the values around the fast-time index under test, and the value $val$ of the fast-time bin under test is compared with $\mathit{coef} \cdot \mathit{th}_{\mathrm{motion}}$ where $\mathit{coef}$ is a constant coefficient. Empirically, $\mathit{coef}$ is set to 1.5 in \systemname\ . In Fig.~\ref{fig:UI}, we can see that the peaks circled in red represent users occupying a fast-time index (range) and a frequency, hence the peaks can be used to identify different users.

\subsubsection{Vibration Decomposition via Variational Mode Decomposition} \label{ssec:VMD}
\systemname\ separates breathing and heartbeat signals from noise in driving environments by utilizing the novel MS-VMD algorithm proposed by us, and further estimates respiratory rate, heart rate and fine-grained IBIs.
Although we separate each people and identify the driver, vibrations of the vehicle always influences the vital sign estimation. The received signal contain mainly four components~\cite{BreathListener-MobiSys19}. 

\begin{itemize}
 \item \textit{Respiration of the driver.} The respiration frequency of an adult ranges from 0.16~$\!$Hz to 0.6~$\!$Hz.
 \item \textit{Heartbeat of the driver.} The heartbeat frequency of an adult ranges from 1~$\!$Hz to 2~$\!$Hz.
 \item  \textit{The vehicle's vibrations.} Frequency of vibrations of a running vehicle, caused by engine, transmission, wheel and bumps on the road ranges from 0.5~$\!$Hz to 10~$\!$Hz.
 \item  \textit{Vibrations of driver's motions.} Frequency of human motions such as turning steering wheel may bring interference with ranges from 0.5~$\!$Hz to 2~$\!$Hz.
\end{itemize}
We can see that the four components of the received signal are overlapping in frequency, hence a simple band-pass filter with specified frequency range cannot filter out undesired noise caused by vibrations.  Although the vibrations of driver's motions may be separated via distances, there are some cases where the motions and vital signs occur at the same distance.

To separate the vital sign signals from other interference,  VMD algorithm is used to decompose the signal into various modes by calculus of variation. Each mode of the signal is assumed to be around a central frequency and have a narrow bandwidth. VMD calculates these central frequencies and the mode functions concurrently, by using an optimization technique called Alternating Direction Method Method of Multipliers (ADMM). However, the VMD only supports single  ``sequence''  decomposition (terminology defined in Sec.~\ref{ssec:rf_channel}). Due to the complicated in-vehicle environment, 
a single sequence of signal may not be enough to extract vital signs, so we need to combine multiple sequences to enable accurate vital sign extractions. Therefore,  to better refine the vital signs of the driver, we hereby design a Multi-Sequence VMD (MS-VMD) algorithm, an enhanced version of the conventional VMD algorithm.

It is noticed
that in the radar data matrix a person does not just occupy a fast-time index, but multiple fast-time indices. Generally speaking, indices occupied by the same person carry similar information of the person's vital signs. This phenomenon is called time diversity in wireless communication theory. Our MS-VMD algorithm employs time diversity, and jointly optimizes the problem of minimum bandwidth with multiple sequences taken into account, hence better than other algorithms (e.g. Hilbert-Huang transform~\cite{WiFind-IEEEToBD}, signal fitting~\cite{leem2017vital} and traditional VMD algorithm~\cite{b6:vmd}),  where only one data sequence is considered.

We leverage this time diversity to reformulate the objective function of VMD according to \cite{b6:vmd}. Our modified VMD algorithm decomposes the data $z(t)$ into multiple Intrinsic Mode Functions~$\!$(IMFs). We have
$	z(t) = \sum_{n = 1}^{ N} u_n(t)$
where $u_n(t)$ is the $n$-th IMF.  However, for data of multiple lags (time delays in \systemname)  $\mathbf{z} (t) = [z_1(t), z_2(t), \cdots, z_M(t)]^T$ where $M$ represents the number of lags, we can obtain $ \mathbf{z} (t) = \sum_{n = 1}^{ N} \mathbf{u}_n(t) $
where $ \mathbf{u}_n(t) = [ u_1(t) = u_n(t), u_2(t)= u_n(t), \cdots, u_M(t)= u_n(t) ]^T $. Same as the conventional VMD, to estimate the bandwidth of IMFs, we need to follow the three  steps: 1) each $\mathbf{u}_n(t)$ uses the Hilbert transform to compute the associated analytic signal, 2) $\mathbf{u}_n(t)$ is mixed with  a complex exponential of center frequency $\omega_n$ to baseband, and 3) The bandwidth of $\mathbf{u}_n(t)$ can be estimated via the squared $l^2$-norm of the gradient. In this way, our constrained problem can be formulated as\footnote{$j$ is the imaginary number unit, not the $j$ as we used as frequency index in Section~\ref{sssec:ui}}:
 \begin{align} \label{eq:cvmd_obj}
\min_{u_n, \omega_n}	&  \sum_{n = 1}^{N} \sum_{m = 1}^M  \left \Arrowvert \frac{\partial  \left[ \left( \delta(t)  + \frac{j}{\pi t} \right) * u_n(t)  \right] e^{-j \omega_n t} }{\partial t}  \right  \Arrowvert_2^2 \cr
\text{s.t.}\ \ &  {z}_m (t) = \sum_{n = 1}^{ N} u_n(t), \ m = 1, 2, \cdots, M.
\end{align}
Different from \cite{b6:vmd}, the constraints become multiple linear equations corresponding to the number of lags. Therefore, the augmented Lagrangian function $\mathcal{L}	$ is 
\begin{align} \label{eq:cvmd_lag}
\mathcal{L}	&=   \sum_{n = 1}^{N} \sum_{m = 1}^M  \left \Arrowvert \frac{\partial  \left[ \left( \delta(t)  + \frac{j}{\pi t} \right) * u_n(t)  \right] e^{-j \omega_n t} }{\partial t}  \right  \Arrowvert_2^2  +  \sum_{m = 1}^{M} \left \Arrowvert  z_m(t) - \sum_{n = 1}^{ N} u_n(t) \right  \Arrowvert_2^2 + \sum_{m = 1}^{M} \left<  \lambda_m (t), z_m(t)  - \sum_{n = 1}^{ N} u_n(t) \right >.
\end{align}
Similar to \cite{b6:vmd}, the E.q.~\eqref{eq:cvmd_lag} is also can be transformed to multiple simpler sub-optimization problems and solved by ADMM algorithm. Next, we divide the problem into two sub-problems: IMF update and center frequency update. 

\textit{IMF update.} We fix the center frequency $\omega_i $ firstly, and solve the minimization problem with respect to IMF $u_i(t)$. Then, we can obtain 
\begin{align} \label{eq:imf_update}
	u_i^{q+1}(t) = 	\arg \min_{u_i}    \left \Arrowvert \frac{\partial  \left[ \left( \delta(t)  + \frac{j}{\pi t} \right) * u_i(t)  \right] e^{-j \omega_i t} }{\partial t}  \right  \Arrowvert_2^2  +  \sum_{m = 1}^{M}  \left \Arrowvert  z_m(t) - \sum_{n = 1}^{ N} u_n(t)  + \frac{\lambda_m(t) }{2}  \right  \Arrowvert_2^2,
\end{align}
where $q$ is the current iteration.

According to Parseval/Plancherel Fourier isometry under the $l^2$-norm, the problem E.q.~\eqref{eq:imf_update} is equal to solving the following problem in frequency domain:
\begin{align} \label{eq:imf_update_fre}
	u_i^{q+1}(\omega) 
&= 	  \arg \min_{u_i}    \left \Arrowvert { j \omega \left[ \left( 1 +  \text{sgn}(\omega +  \omega_i ) \right)  u_i( \omega +  \omega_i  )  \right]  }  
\right  \Arrowvert_2^2  +  \sum_{m = 1}^{M}  \left \Arrowvert  z_m(\omega) - \sum_{n = 1}^{ N} u_n(\omega)  + \frac{\lambda_m(\omega) }{2}  \right  \Arrowvert_2^2,
\end{align}
where $\text{sgn}(\cdot)$ is the sign function. We can use $ \omega = \omega - \omega_i $ in the first term:
\begin{align} \label{eq:imf_update_fre1}
 	u_i^{q+1}(\omega)
&= \arg \min_{u_i}  \left \{ \int_{-\infty}^{+\infty} \left [  4 (\omega - \omega_i )^2 | u_i(\omega) |^2 + \sum_{m = 1}^{M} \left |   z_m(\omega) - \sum_{n = 1}^{ N} u_n(\omega)  + \frac{\lambda_m(\omega) }{2}   \right |^2  \right ] d\omega  \right  \}
.	
\end{align}
We can rewrite both terms in E.q.~\eqref{eq:imf_update_fre1} using half-space integrals over the non-negative frequencies, due to the Hermitian symmetry of real signals in the reconstruction term:
\begin{align} \label{eq:imf_update_fre_int}
		u_i^{q+1}(\omega) &= \arg \min_{u_i}  \left \{  \int_{0}^{+\infty} \left [ 4 (\omega - \omega_i )^2 | u_i(\omega) |^2 + 2 \sum_{m = 1}^{M} \left |   z_m(\omega) - \sum_{n = 1}^{ N} u_n(\omega)  + \frac{\lambda_m(\omega) }{2}   \right |^2  \right ] d\omega  \right  \}.
\end{align}
Considering the problem E.q.~\eqref{eq:imf_update_fre_int} is to find the minimal function $u_n$, and then, according to the Euler-Lagrange equation~\cite{troutman2012variational} of the calculus of variations $\frac{\partial L }{ \partial f} - \frac{d}{d \omega} \frac{\partial L}{ \partial f^{'}} = 0$, we can define 
\begin{align} \label{eq:variation}
  L &=  4 (\omega - \omega_i )^2 | u_i(\omega) |^2 + 2 \sum_{m = 1}^{M} \left |   z_m(\omega) - \sum_{n = 1}^{ N} u_n(\omega)  + \frac{\lambda_m(\omega) }{2}   \right|^2   \ \text{and} 
  \ f = u_i(\omega).
\end{align}
Since $f^{'}$ is not explicit in $L$ , the second term in the Euler-Lagrange equation vanishes. Thus, we let $\frac{\partial L }{ \partial f}  = 0$, and  the first sub-problem can be solved

\begin{align} \label{eq:imf_eq}
	\hat{u}_i^{q+1}(\omega) = \frac{ \sum_{m=1}^{M} \left [ \hat{z}_m(\omega) - \sum_{n \neq i} \hat{u}_n^{q+1}(\omega) + \frac{\hat{\lambda}_m(\omega)}{2} \right ] }{ M + 2 (\omega - \omega_m)^2  }.
\end{align}	

\textit{Center frequency update.}  The center frequency $\omega_i$ is only related to the first term in E.q. \eqref{eq:cvmd_lag}, hence the second sub-problem is 
\begin{align} \label{eq:fc_obj}
	\omega_i^{q+1} = \arg \min_{\omega_i}  \left \Arrowvert \frac{\partial  \left[ \left( \delta(t)  + \frac{j}{\pi t} \right) * u_i(t)  \right] e^{-j \omega_i t} }{\partial t}  \right  \Arrowvert_2^2. 
\end{align}	
The optimization problem can also solved in frequency domain efficiently using the Plancherel theorem: 
\begin{align} \label{eq:fc_obj_fre}
	\omega_i^{q+1} = \arg \min_{\omega_i}  \int_{0}^{+\infty} (\omega - \omega_i )^2 | \hat{u}_i(\omega) |^2 d \omega . 
\end{align}	
We also give the update expression at $(q+1)$-th iteration result as 
\begin{align} \label{eq:fc_update}
	\hat{\omega}_i^{q+1} = \frac{ \int_{0}^{ \infty} \omega  \mathopen| \hat{u}_i(\omega) \mathclose|^2 d \omega }{ \int_{0}^{ \infty   } \mathopen| \hat{ u}_i(\omega) \mathclose|^2d \omega}. 
\end{align}	

\begin{figure}[ht]
	\centering
	\includegraphics[width=0.8\linewidth]{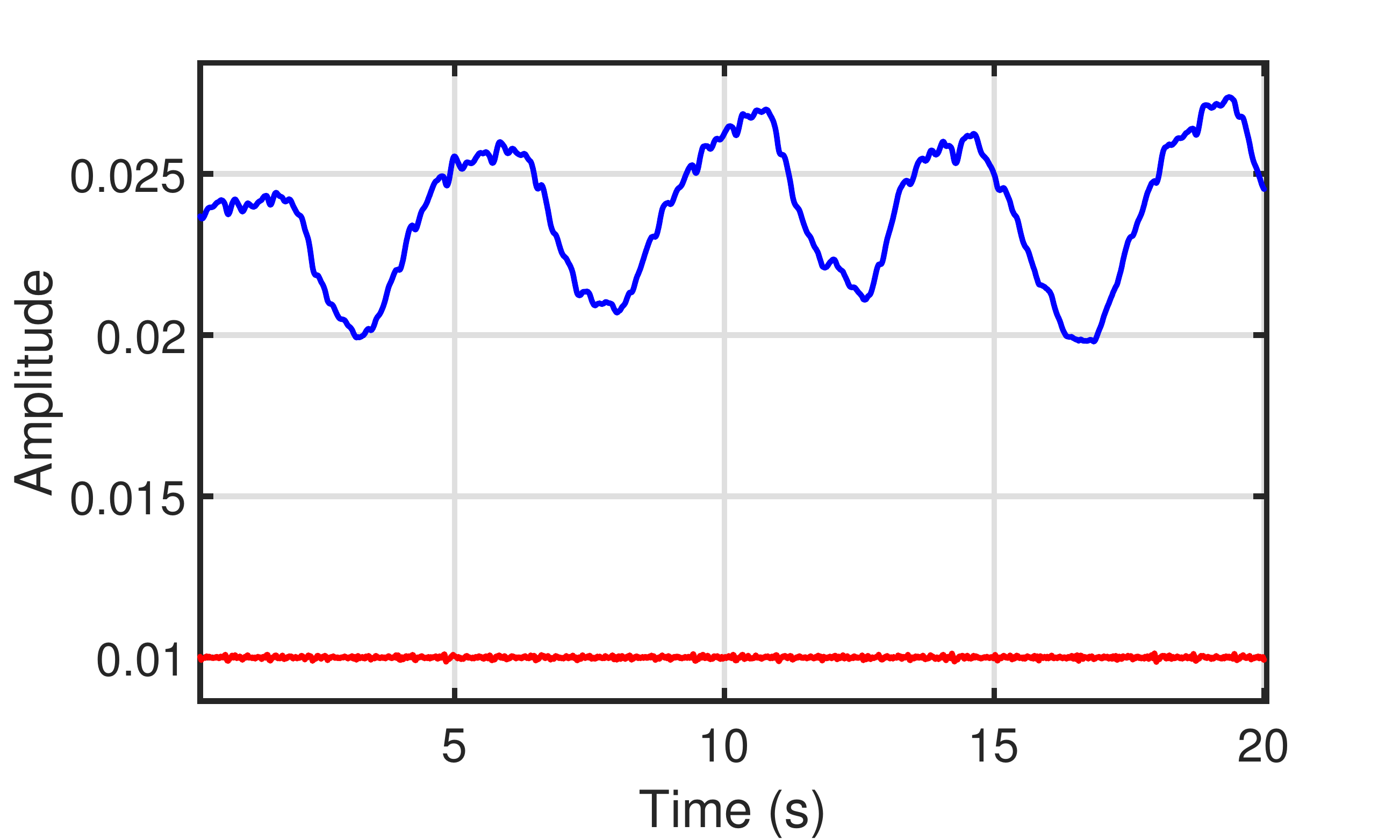}
	\caption{Reflected signal decomposed by MS-VMD.}
	\label{fig:VMD}
\end{figure}

After that, we can plug the above two IMFs and center frequency updates into AMDD \cite{boyd2011distributed} to solve the MS-VMD problem. The complete MS-VMD algorithm is illustrated in Algorithm~\ref{alg:ms-vmd}.  Line~\ref{alg:update_u1} and line~\ref{alg:update_w} are used to update $\hat{u}_i^{q+1}(\omega)$ and $\hat{\omega}_i^{q+1}$ according to E.q.~\eqref{eq:imf_eq} and E.q.~\eqref{eq:fc_update}, respectively in $(q+1)$-th iteration. In line~\ref{alg:eta}, $\eta$ is the update parameter. Line~\ref{alg:convergence} is a condition to check convergence of this algorithm.  Fig.~\ref{fig:VMD} illustrates that our MS-VMD decompose the noisy signals into four modes, including respiration, heartbeat and other vehicle vibrations treated as noise components.

\begin{algorithm}[ht]
	\SetKwInOut{Input}{input}\SetKwInOut{Output}{output}
	\Input{ \{$\hat{u}_i^0$\} ,  \{$\hat{\omega}_i^0$\}, $\{ \lambda^0_m \}$, $q \gets 0$}
	\Output{ \{$\hat{u}_i^q$\} ,  \{$\hat{\omega}_i^q$\}}
	\BlankLine
	\While{$q\leftarrow q + 1 $ }{
		\For{$i\leftarrow 1$ \KwTo $N$}{
			\emph{Update $\hat{u}_i$ for all $\omega \geq 0$:}\label{alg:update_u}\\
				 $\hat{u}_i^{q+1}(\omega) \gets \frac{ \sum_{m=1}^{M} \left [ \hat{z}_m(\omega) - \sum_{  n < i} \hat{u}_n^{q+1}(\omega) - \sum_{  n > i} \hat{u}_n^{q}(\omega) \frac{{\lambda}^{q}_m(\omega)}{2} \right ] }{ M + 2 (\omega - \omega_m)^2  }$\; \label{alg:update_u1}
				 
			\emph{Update $\hat{\omega}_i$:}\\
				$	\hat{\omega}_i^{q+1} \gets \frac{ \int_{0}^{ \infty} \omega  \mathopen| \hat{u}_i(\omega) \mathclose|^2 d \omega }{ \int_{0}^{ \infty   } \mathopen| \hat{ u}_i(\omega) \mathclose|^2d \omega}$ \label{alg:update_w}

			}

		\For{$m \leftarrow $ 1 \KwTo $M$}{
			\emph{Dual ascent for all $\omega \ge 0$:}\\
			$\lambda_m^{q+1} (\omega) \leftarrow \lambda_m^{q} (\omega) + \eta \left(   z_m(\omega) - \sum_{n}^{N} \hat{u}^{q+1}_n (\omega) \right)$\;}\label{alg:eta}
	\emph{Convergence:}\\
	\If{$ \frac{\sum_{n}^{N} \|  \hat{u}^{q+1}_n - \hat{u}^{q}_n    \|^2_2}{\| \hat{u} \|^2_2 }  < \epsilon $ \label{alg:convergence} } {break\;} 
	}
	\caption{Complete MS-VMD }\label{alg:ms-vmd}
\end{algorithm}

\subsection{Vital Signs Extraction} \label{ssec:vs}
So far we have decomposed the RF signal into its components, including respiration signal, heartbeat signal and noise caused by vehicle vibration. In this section, we introduce how we identify the respiratory and heartbeat component and extract vital signs.%

\subsubsection{Respiratory Rate Estimation}\label{ssec:rre}
Respiration can be modeled as follows: when the driver or passengers inhales, his diaphragm and other related muscle contracts to create space in the lungs and the lungs expands and let the air flow in. Looking from outside, chest and abdomen of the driver move forward. If the impulse radio is placed at the front of the vehicle, chest and abdomen get closer to the impulse radio. Conversely, if the driver or passengers exhale, diaphragm and related muscles relax, therefore the space in the chest gets smaller and the air inside is squeezed out, his chest and abdomen move backward and further away from the device. Since the distance between human and \systemname\ and the signal is linearly related, \systemname\ can track the respiration of the driver or the passenger by detecting the amplitude. To determine the respiratory rate of the driver, we apply Fast Fourier Transform (FFT) to the IMFs obtained from MS-VMD. Since the frequency of respiration ranges from 0.16~$\!$Hz to 0.6~$\!$Hz, the component whose frequency matches the range is the respiratory component. And the peak frequency of this IMF is respiratory rate. Respiration is the blue line shown both in time domain and frequency domain in Fig.~\ref{fig:breath_t} and Fig.~\ref{fig:breath_f}.

\subsubsection{Heart Rate Estimation}\label{hb_sec}\label{ssec:hbe}
\begin{figure} 
	\begin{minipage}[ht]{0.33\linewidth}
		\centering
		\includegraphics[width=\textwidth]{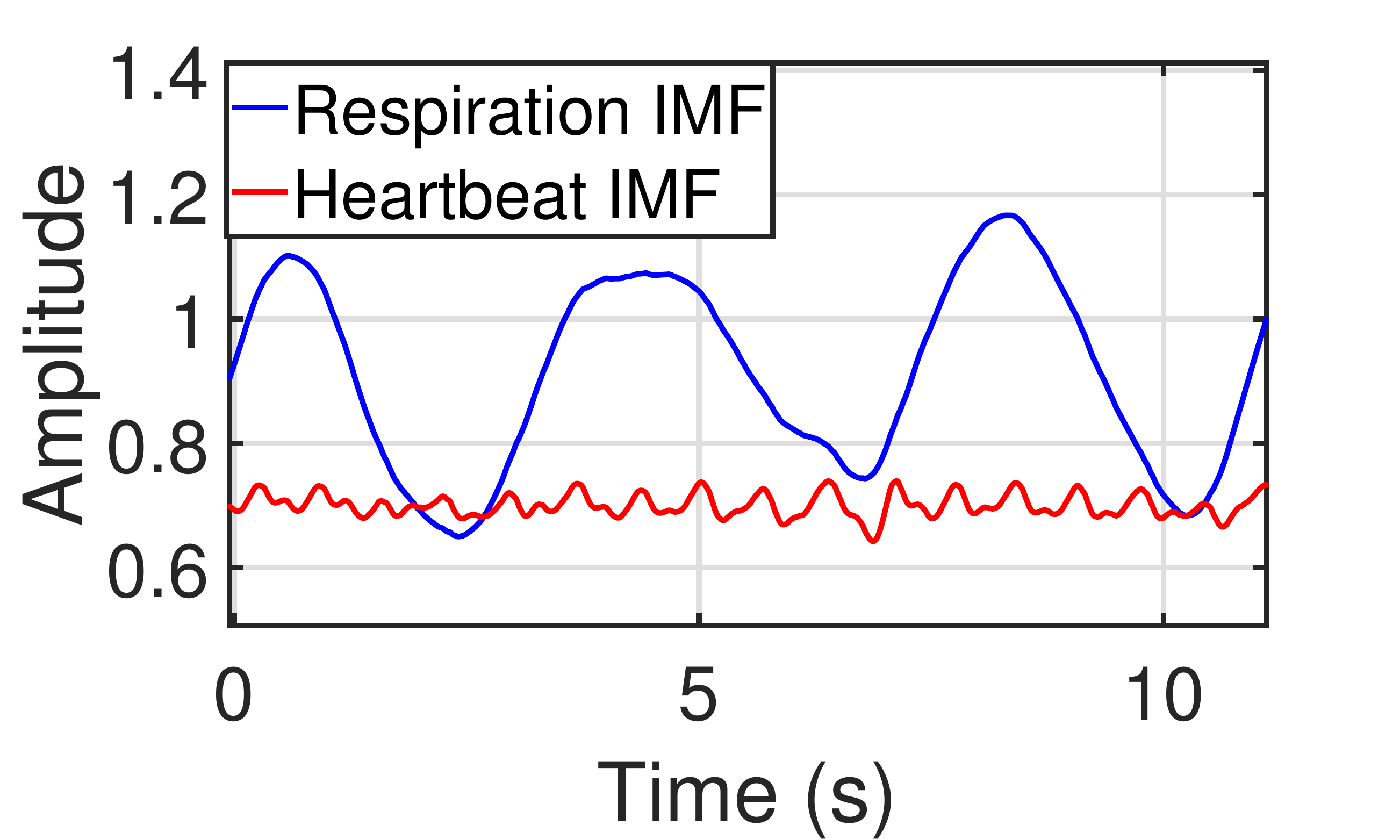}
		\caption{ Respiration and heartbeat IMFs in time domain.}
		\label{fig:breath_t}
	\end{minipage}
	\begin{minipage}[ht]{0.33\linewidth}
		\centering
		\includegraphics[width=\textwidth]{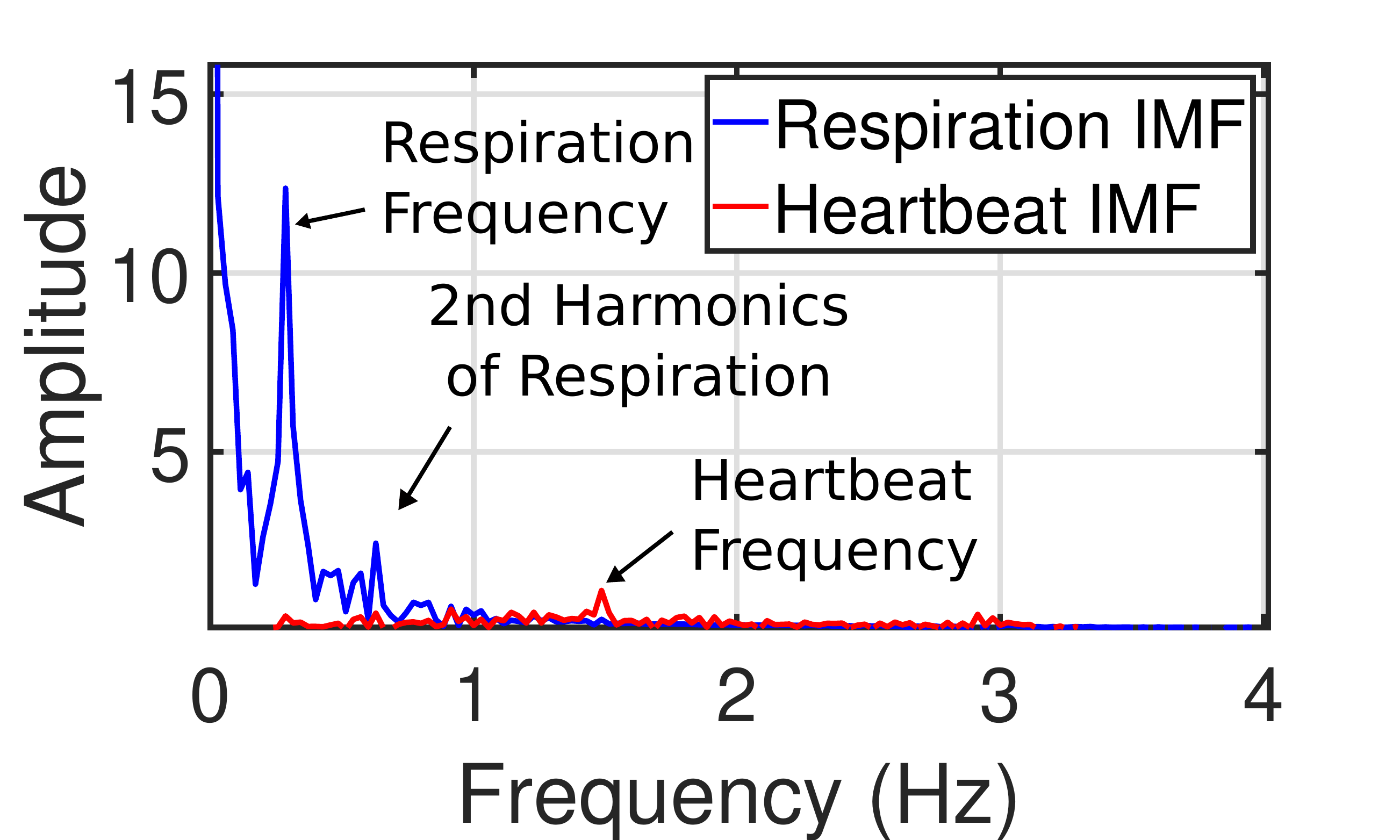}
		\caption{ Respiration and heartbeat IMFs in frequency domain.}
		\label{fig:breath_f}
	\end{minipage}
	\begin{minipage}[ht]{0.33\linewidth}
		\centering
		\includegraphics[width=\textwidth]{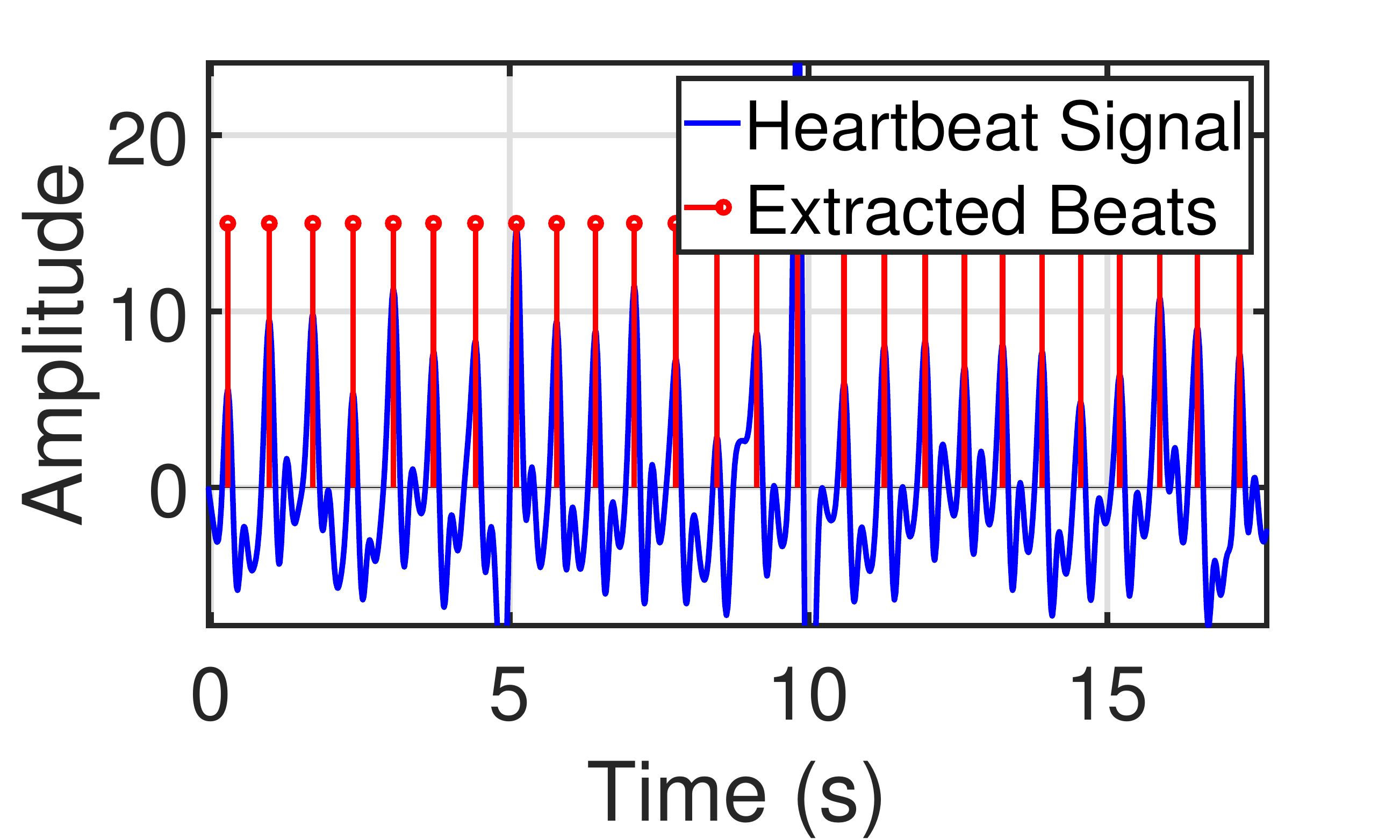}
		\caption{ Heartbeat signal and extracted timings of each beat.}
		\label{fig:ibi}
	\end{minipage}
\end{figure}

Similar to respiration, the contraction and relaxation of human heart cause small displacements on the surface of different parts of the person's body. This phenomenon is called ventricular pump activity~\cite{pinheiro2010theory}. Ventricular pump activities are found in the head, torso, leg, buttock and etc. The minute movement caused by heartbeat also makes the signal change periodically, so \systemname\ can also track heartbeat. To determine the heart rate of the driver, we again apply FFT to the IMFs obtained from MS-VMD (except the breathing-related IMF). Since the frequency of heartbeat ranges from 1~$\!$Hz to 2~$\!$Hz, the component whose frequency matches the range is the heartbeat component. And the peak frequency of this IMF is heart rate. Heartbeat is the red line shown both in time domain and frequency domain in Fig.~\ref{fig:breath_t} and Fig.~\ref{fig:breath_f}.

\subsubsection{Interbeat Interval Segmentation}\label{ssec:ibis}
To estimate interbeat intervals, we look into the heartbeat IMF, as shown in \ref{fig:breath_t}. IBIs are determined by the timings of each heartbeat. To obtain the exact timings, we detect peaks in the decomposed heartbeat IMF. However, the heartbeat waveform signal is weak and easily affected by noise and artifacts. Hence there are challenges in detecting peaks: i) heartbeat peak detection should be adaptive, since the strength of the signal is not stable and ii) fake peaks should be removed. To solve these challenges, we design a peak detection algorithm in which peaks are separated by at least $d_\mathrm{min}$ samples, that is to say, we find peaks that are the local maxima in a region of $2d_\mathrm{min}+1$. $d_\mathrm{min}$ is determined by the heart rate estimated in Sec.~\ref{hb_sec}. If the heart rate of the driver is $HR$, the heartbeat frequency is $HR/60$. Since the sampling rate of \systemname\ is $400~\!\mathrm{Hz}$, the average number of samples in one interval should be $N_\mathrm{avg}=HR \times 400/60$. The relation between the minimum samples in one IBI and the average samples is given by $N_\mathrm{min}^\mathrm{IBI} = c N_\mathrm{avg}^\mathrm{IBI}$, where $c$ is a constant, and set to 0.7 empirically. To summarize, we detect local peaks separated by at least $0.7 HR \times 400/60$ samples. The result of the peak detection algorithm is shown in Fig.~\ref{fig:ibi}. It can be seen that the timings of the beats are recovered accurately.

\section{Evaluation} \label{sec:eval}
In this section, we evaluate \systemname's performance by conducting road test. We tested on different road conditions, participants and procedures. As shown in Table \ref{tab:eval_sum}, \systemname\ is able to measure respiratory rate, heart rate and heart rate variability for the driver, but it can only measure respiratory rate for all the other 3 people in the vehicle.

\subsection{Experiment Setup} \label{ssec:expset}
\begin{figure}[ht]
	\centering
	\includegraphics[width=0.6\linewidth]{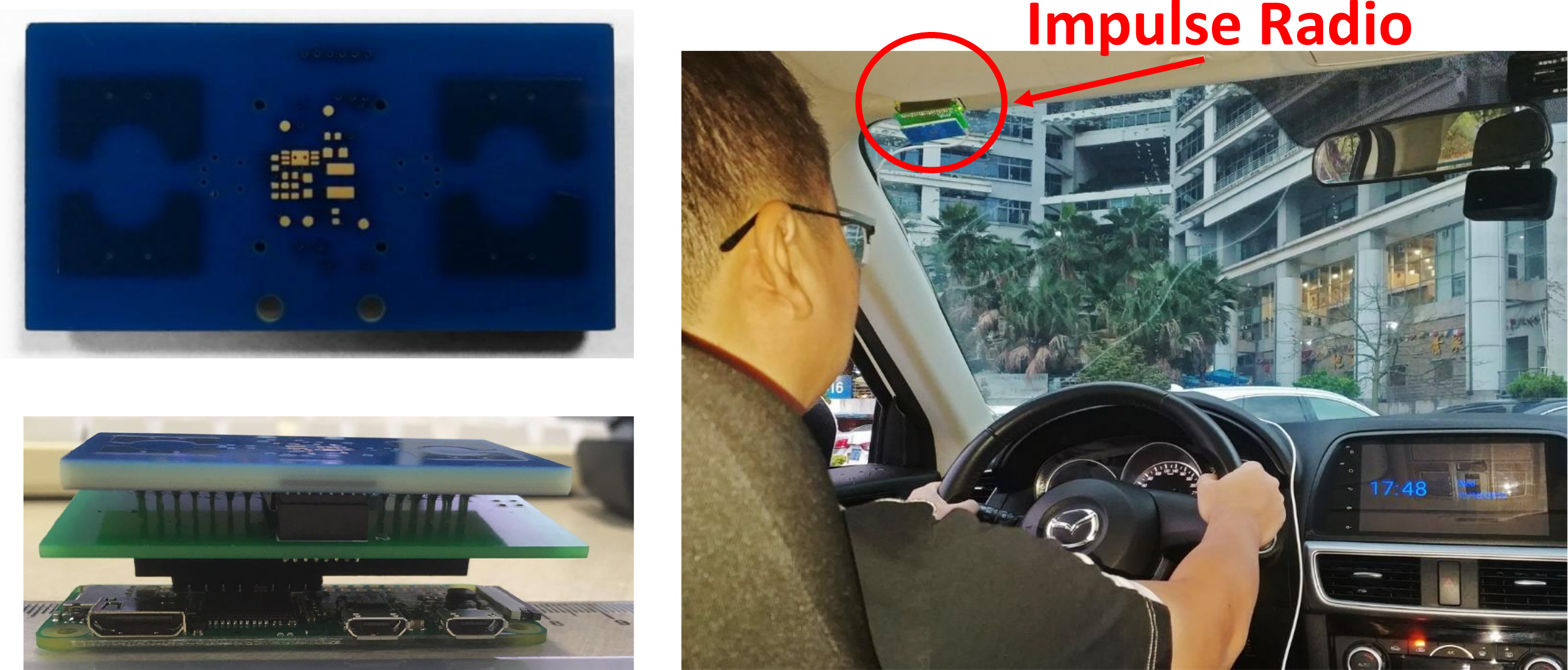}
	\caption{COTS impulse radio used in \systemname\ and radio setup inside the vehicle. }
	\label{fig:setup}
\end{figure}

\systemname\ adopts a COTS impulse radio XETHRU~\cite{xethru} model X4M05 as its front-end. The radio transmits pulses with a width of $\mathrm{0.4~\!ns}$. The center frequency of the radio is $\mathrm{7.3~\!GHz}$, the bandwidth is $\mathrm{1.4~\!GHz}$, the sampling frequency is $\mathrm{23.328~\!GHz}$, and we set frame per second as 400. The radio is connected to a Raspberry Pi via Serial Peripheral Interface (SPI). The hardware PCB is rather small with a size of $\mathrm{6.5 \times 3~\!cm^2}$. The whole hardware including power supply, $\mathrm{5~\!V}$ fan, Raspberry Pi, and impulse radio is shown in Fig.~\ref{fig:setup}.  
We place the whole \systemname\ on the windshield of a vehicle. Since the impulse radio is facing the driver (also shown in Fig.~\ref{fig:setup}), it can capture maximum amount of weak signals from multiple body parts (e.g., chest, abdomen, forehead and neck). We implement the software system presented in Sec.~\ref{sec:systemdesign} using C\texttt{++}; it operates in real-time. 

Extensive road tests have been performed while developing \systemname. The driving route is designed to cover different road conditions and different traffic flows. Different conditions covered in the experiment include: parking, driving on smooth highway, driving on traffic-heavy road, driving in school area, driving at zebra and pedestrian crossing, entering and leaving intersections, driving uphill, driving downhill, driving in roundabouts, changing lanes, turning left, turning right and making U-turns. We have 4 drivers in total; each of them drive 15 miles and 30 minutes, so the total road test takes 60 miles and 2 hours. The vehicle used in the evaluation is a Mazda Axela. While the driver is driving, a copilot is responsible for collecting data. Two more passengers are in the back seats. We also tested effects of different clothing (lightweight T-shirt, heavyweight T-shirt, sweatshirt + lightweight T-shirt and sweatshirt + heavyweight T-shirt) on road tests.

We use a NEULOG respiration monitor belt logger sensor NUL-236 to obtain the ground truth of breathing. For heartbeat ground truth, we use the Heal Force PC-80B portable ECG monitor and extend the electrodes by exterior electrodes with lead wires to record the ECG data. The data is further processed to obtain heart rate and heart rate variability, which are then used as ground truth.

\subsection{Overall Performance}
\subsubsection{Respiratory Rate Estimation Evaluation}
Respiratory rate estimation error is used for the evaluation. The error is defined as the absolute value of the difference between the estimated respiratory rate $R_E$ and the actual respiratory rate $R_A$, i.e., $\left|R_{E}-R_{A}\right|$. We evaluate the respiratory rate estimation performance of \systemname\ against WiFind* and Leem et al.~\cite{leem2017vital}. In WiFind*, we port the existing algorithm of WiFind~\cite{WiFind-IEEEToBD} to UWB radio, so as to compare algorithms rather than radios. Leem et al. is one of the very few state-of-the-art research proposals leveraging UWB for vital sign detection~\cite{leem2017vital}. Fig.~\ref{fig:cdf_r} plots the CDF of respiratory rate measurement errors over all experiments. We can see that \systemname\ achieves a median error of $0.06~\!\mathrm{rpm}$ for the driver. As a comparison, the median error of respiratory rate of WiFind* and Leem et al. are $0.08~\!\mathrm{rpm}$ and $0.06~\!\mathrm{rpm}$ respectively, and they all have long tails indicating their performance is unstable in extreme conditions.

\subsubsection{Heart Rate Estimation Evaluation}
\begin{figure}[ht]
	\centering
	\subfigure[CDF of respiratory rate estimation error.]
	{
		\centering
		\includegraphics[width=0.31\textwidth]{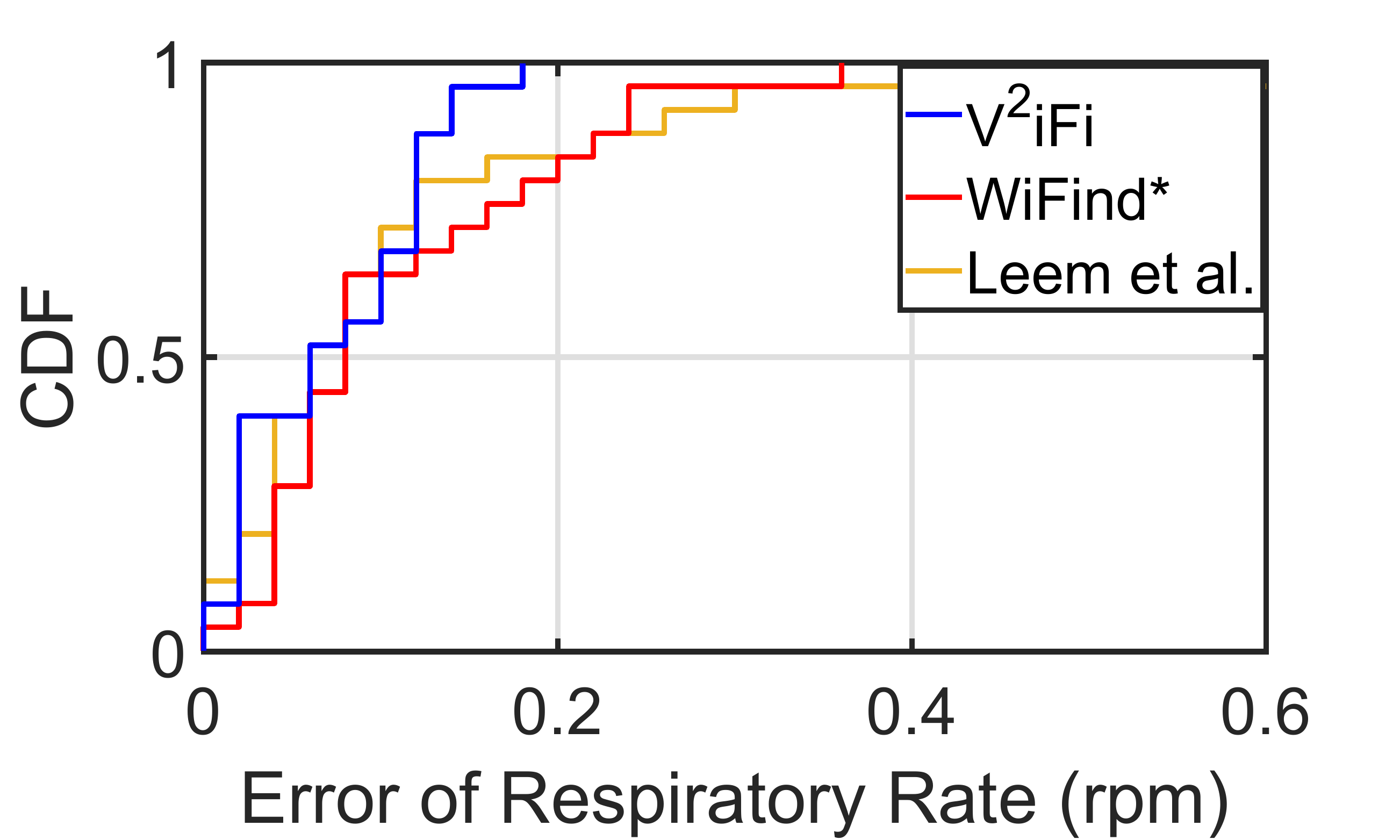}
		\label{fig:cdf_r}	
	}
	\subfigure[CDF of heart rate estimation error.]
	{
		\centering
		\includegraphics[width=0.31\textwidth]{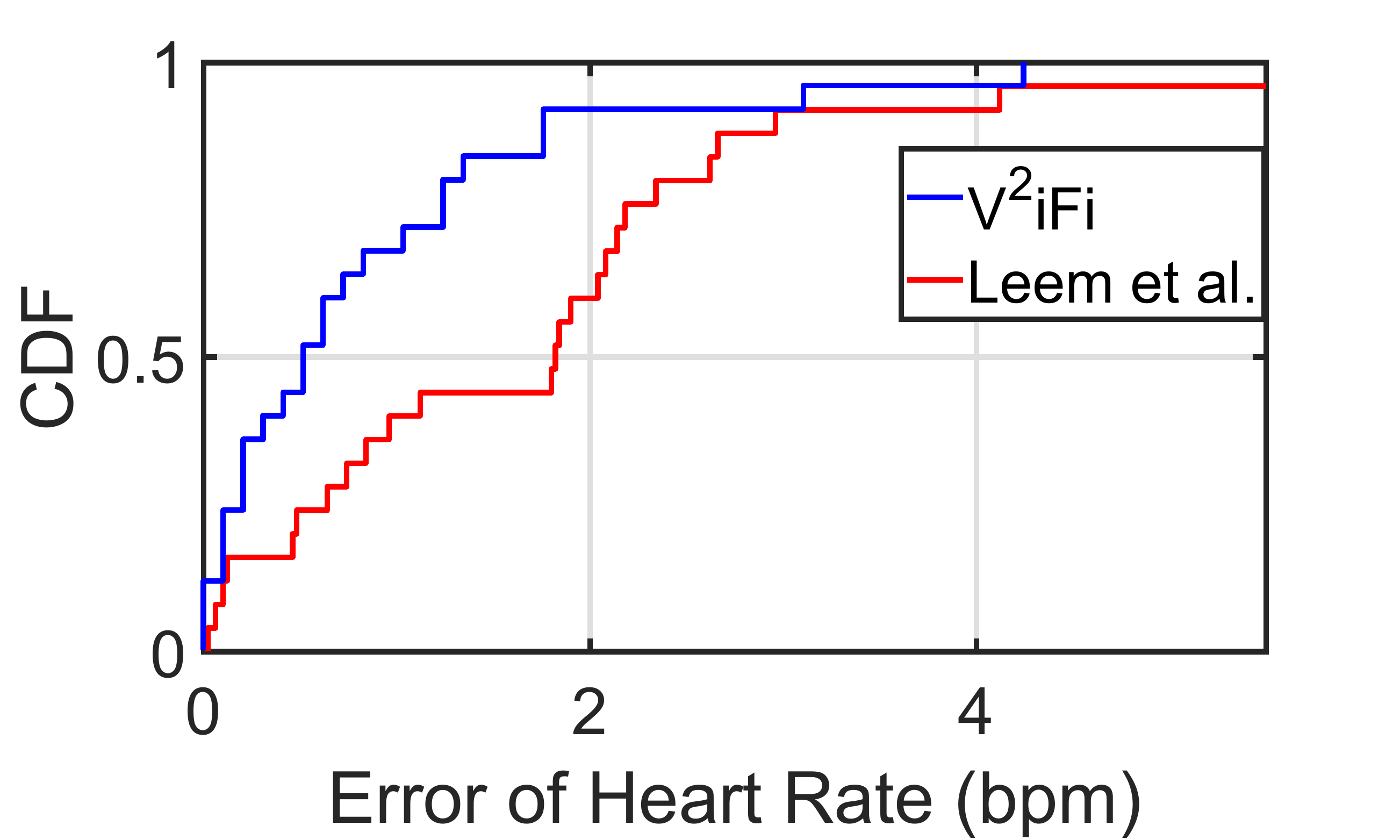}
		\label{fig:cdf_h}	
	}
	\subfigure[ CDF of IBI estimation error.]
	{
		\centering
		\includegraphics[width=0.31\textwidth]{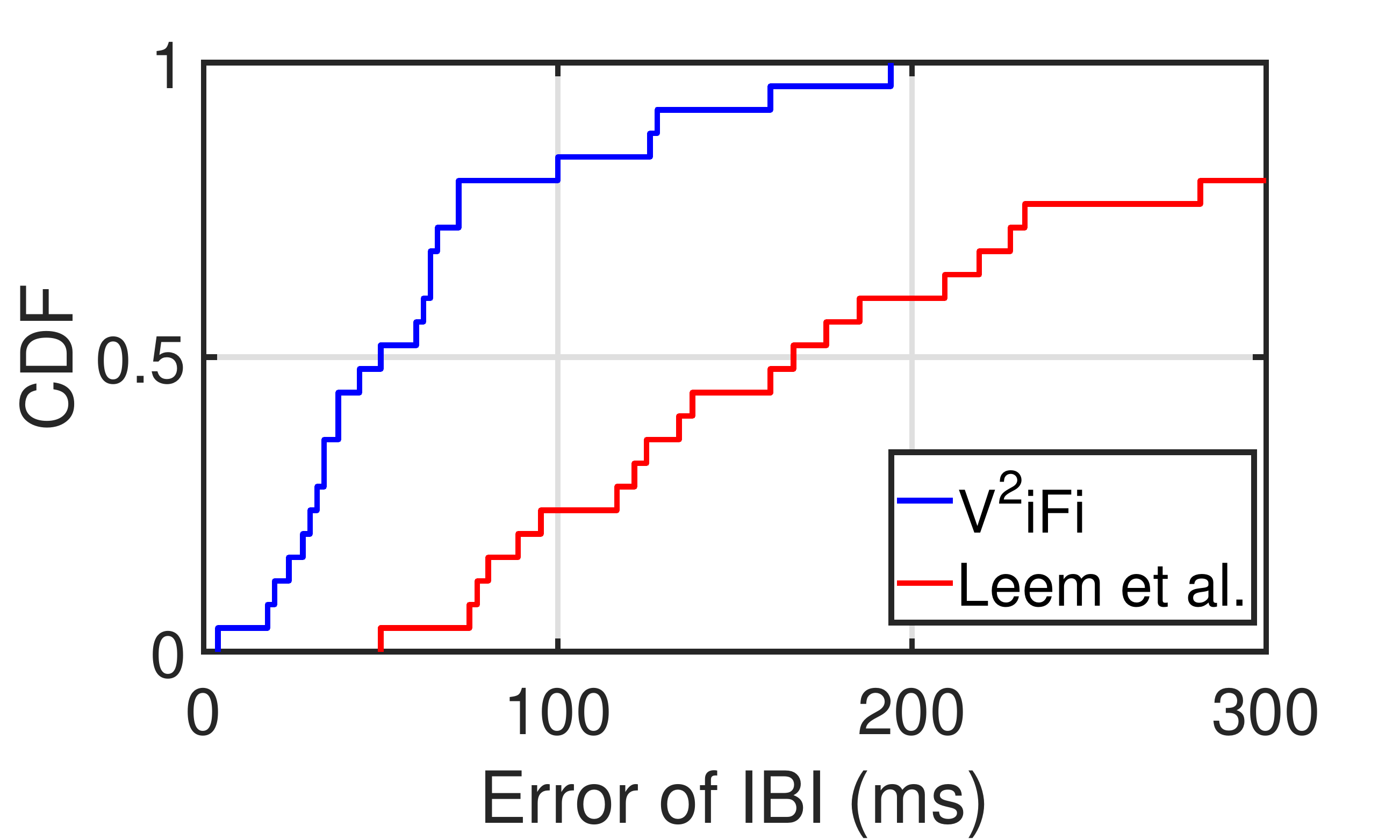}
		\label{fig:cdf_IBI}	
	}
	\caption{Comparison of vital sign estimation performance among \systemname, WiFind* and Leem et al.}
\end{figure}

We evaluate the heart rate estimation performance of \systemname\ against Leem et al. The heart rate estimation error is defined as the absolute value of the difference between the estimated heart rate $R_E$ and the actual heart rate $R_A$, i.e., $\left|R_{E}-R_{A}\right|$. Fig.~\ref{fig:cdf_h} plots the CDF of respiratory rate measurement errors over all experiments. We can see that \systemname\ achieves a median error of $0.6\!~\mathrm{bpm}$ for the driver. As a comparison, the median error of heart rate of Leem et al.~\cite{leem2017vital} is $1.8~\!\mathrm{bpm}$, and has a relatively long tail indicating the performance is unstable in extreme conditions.

\subsubsection{Heart Rate Variability Estimation Evaluation}
Next, we evaluate \systemname's heart rate variability estimation performance against Leem et al.. Since all other indicators, such as SDNN, SDRR, SDANN are all derived from IBI, we evaluate the accuracy of IBI measurement accuracy. Fig.~\ref{fig:cdf_IBI} plots the CDF of IBI measurement errors over all experiments. We can see that \systemname\ achieves a median error of 50\!~ms for the IBI of the driver. As a comparison, the median error of IBI of Leem et al.~\cite{leem2017vital} is $166~\!\mathrm{ms}$, which is almost useless for vital sign monitoring.

\subsubsection{Breathing and Heartbeat Transitions Evaluation} 
\begin{figure}[ht]
	\centering
	\subfigure[Driver controls breathing, from 10\!~rpm to 15\!~rpm.]
	{
		\centering
		\includegraphics[width=0.44\textwidth]{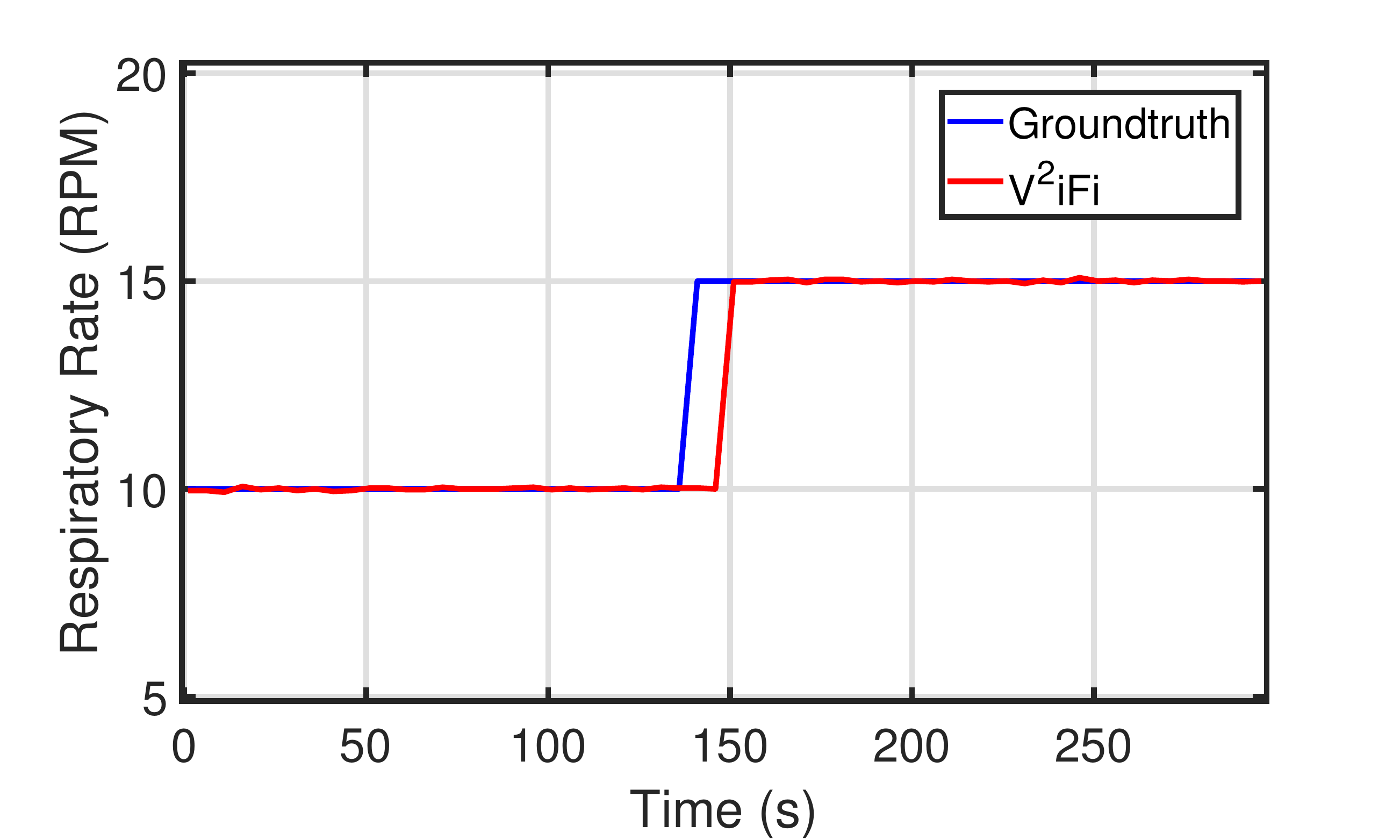}
		\label{fig:tracking_r}	
	}
	\subfigure[Heart rate decreases gradually after exercise.]
	{
		\centering
		\includegraphics[width=0.44\textwidth]{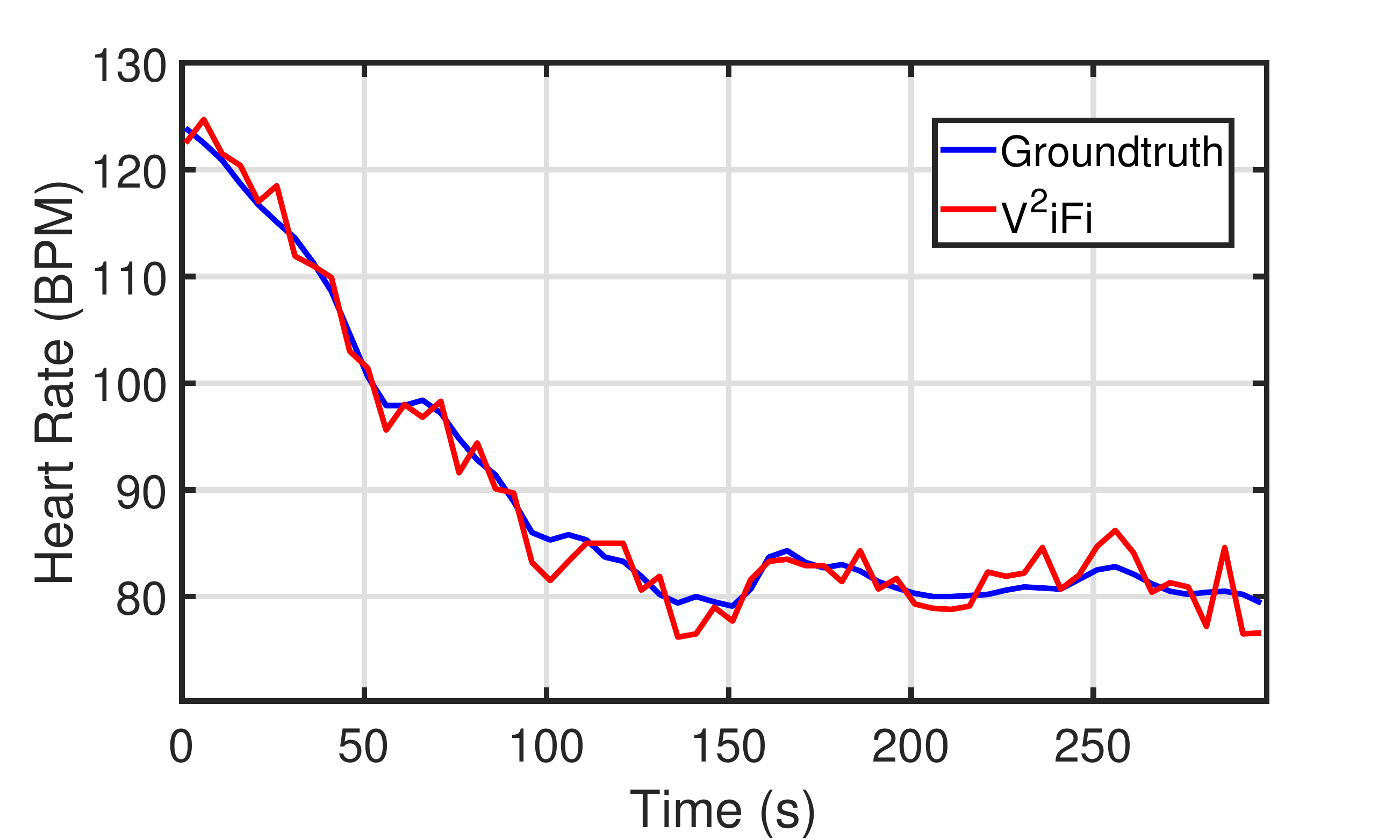}
		\label{fig:tracking_h}	
	}
	\caption{Vital sign estimation performance of \systemname\ when respiratory rate and heart rate change.}
	\label{fig:tracking}
\end{figure}

One thing to keep in mind is that \systemname\ is designed for detecting abnormal human vital signs and alert the user when suspicious signals exist. Therefore it should not only detect normal signals, but also abnormal signals of the driver and passengers. Next we evaluate the performance of \systemname\ for detecting abnormal vital signs. We ask the driver in the vehicle to intentionally change their respiratory rate (from $10\!~\mathrm{rpm}$ to $15\!~\mathrm{rpm}$ after 2 minutes) and respiration pattern. For heartbeat, we ask the people in the vehicle to do some vigorous exercising to increase heart rate before entering the vehicle, and we observe how heart rate decreases with time. The result is shown in Fig.~\ref{fig:tracking_r} and Fig.~\ref{fig:tracking_h}. It can be seen that the detected vital signs of \systemname\ follow the same trends as the actual abnormal vital signs for all these respiration heartbeat transitions. During the transitions, the average error of the estimated respiratory rate is $0.2\!~\mathrm{rpm}$, the average error of the estimated heart rate is $1.6\!~\mathrm{bpm}$.

\subsection{User and Environment Issue Study}
\subsubsection{Number of Users}Although \systemname\ focuses on monitoring vital signs of the driver, the system can potentially be extended to monitor copilot and passengers in theory. In this section, we study the multi-user vital sign monitoring performance of \systemname.
However, due to the Field of View (FOV) and noise in the vehicle,  for the copilot and passengers, \systemname\ can measure respiratory rate only. Table \ref{tab:eval_sum} summarize the vital signs \systemname\ can measure for different users. In Fig.~\ref{users}, it can be clearly seen that estimation error increases with distance, the respiratory rate estimation error of the driver is the smallest while passengers in the rear seats have the largest respiratory rate estimation error. 

\begin{minipage}{\textwidth}
\begin{minipage}[t]{0.46\textwidth}
\centering
\vspace{6ex}
\captionsetup{type=table}
        \captionof{table}{Measured vital signs for different users.}  \label{tab:eval_sum}
		\begin{tabular} {p{1.8cm} p{1.5cm} p{0.8cm} p{1.5cm}}
		\hline
		    & Respiratory Rate & Heart Rate & Heart Rate Variability \\
		\hline
		Driver &  Yes & Yes & Yes  \\
		\hline
		Copilot &    Yes  & No & No \\
		\hline
		Passenger 1 &  Yes  & No & No\\
		\hline
		Passenger 2 &  Yes  & No & No  \\
		\hline
        \end{tabular}
\end{minipage}
\hfill
\begin{minipage}[t]{0.5\textwidth}
\centering
\captionsetup{type=figure}
	\includegraphics[width=\textwidth]{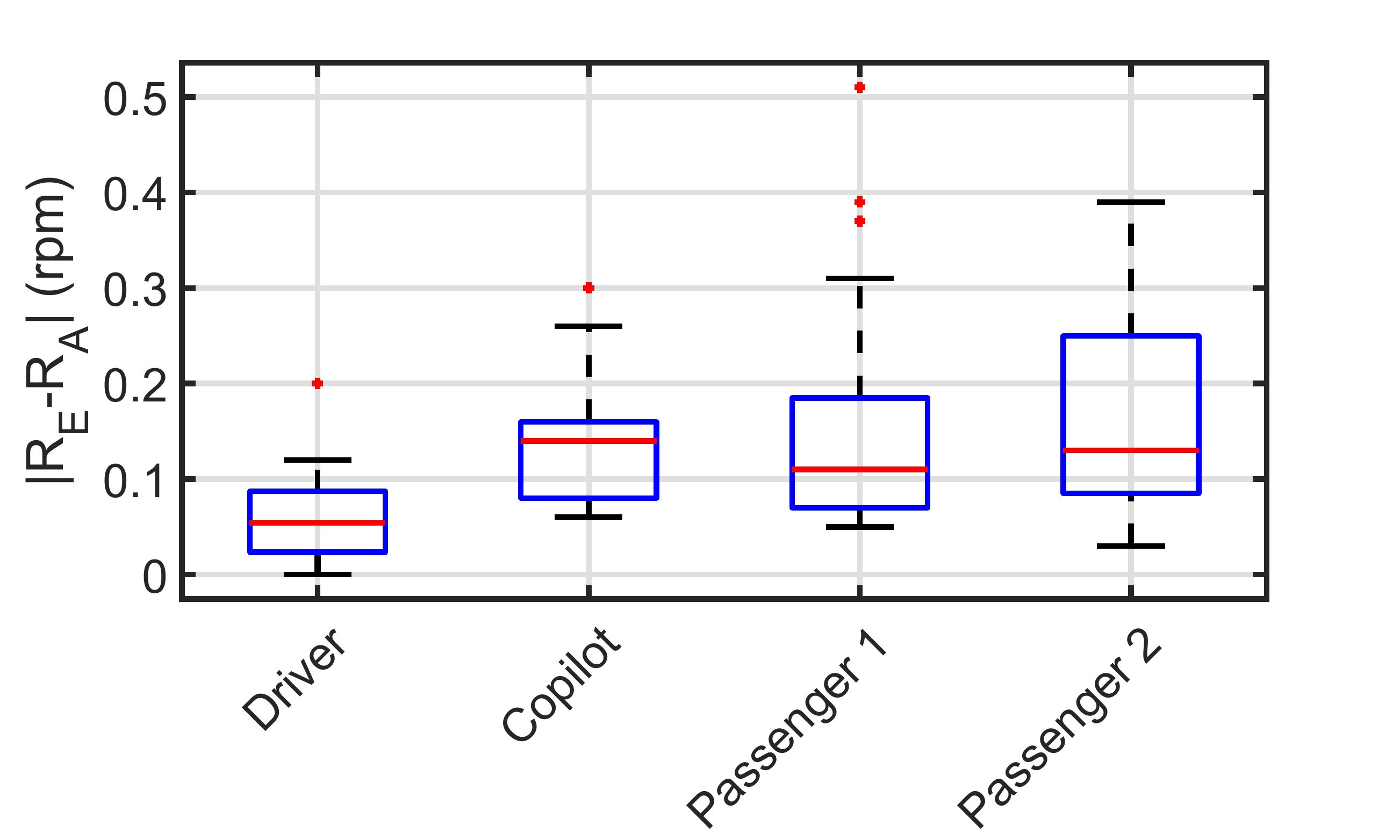}
	\captionof{figure}{Respiratory rate estimation error of different users.}
	\label{users}
\end{minipage}
\end{minipage}

\subsubsection{Impact of Radio Placement}
\begin{figure}[ht]
	\centering
	\subfigure[Respiratory rate estimation error.]
	{
		\centering
		\includegraphics[width=0.31\textwidth]{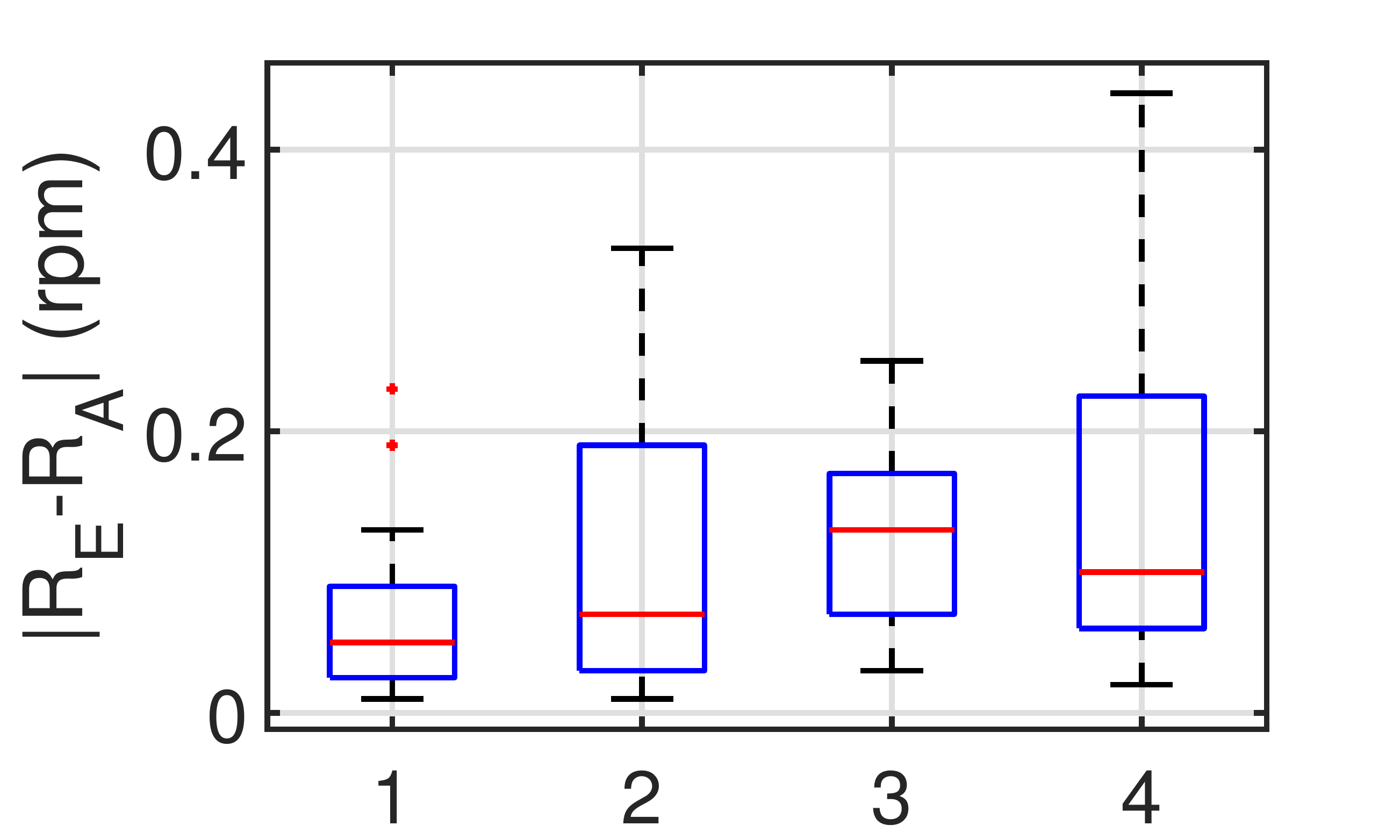}
		\label{fig:placement_r}	
	}
	\subfigure[Heart rate estimation error.]
	{
		\centering
		\includegraphics[width=0.31\textwidth]{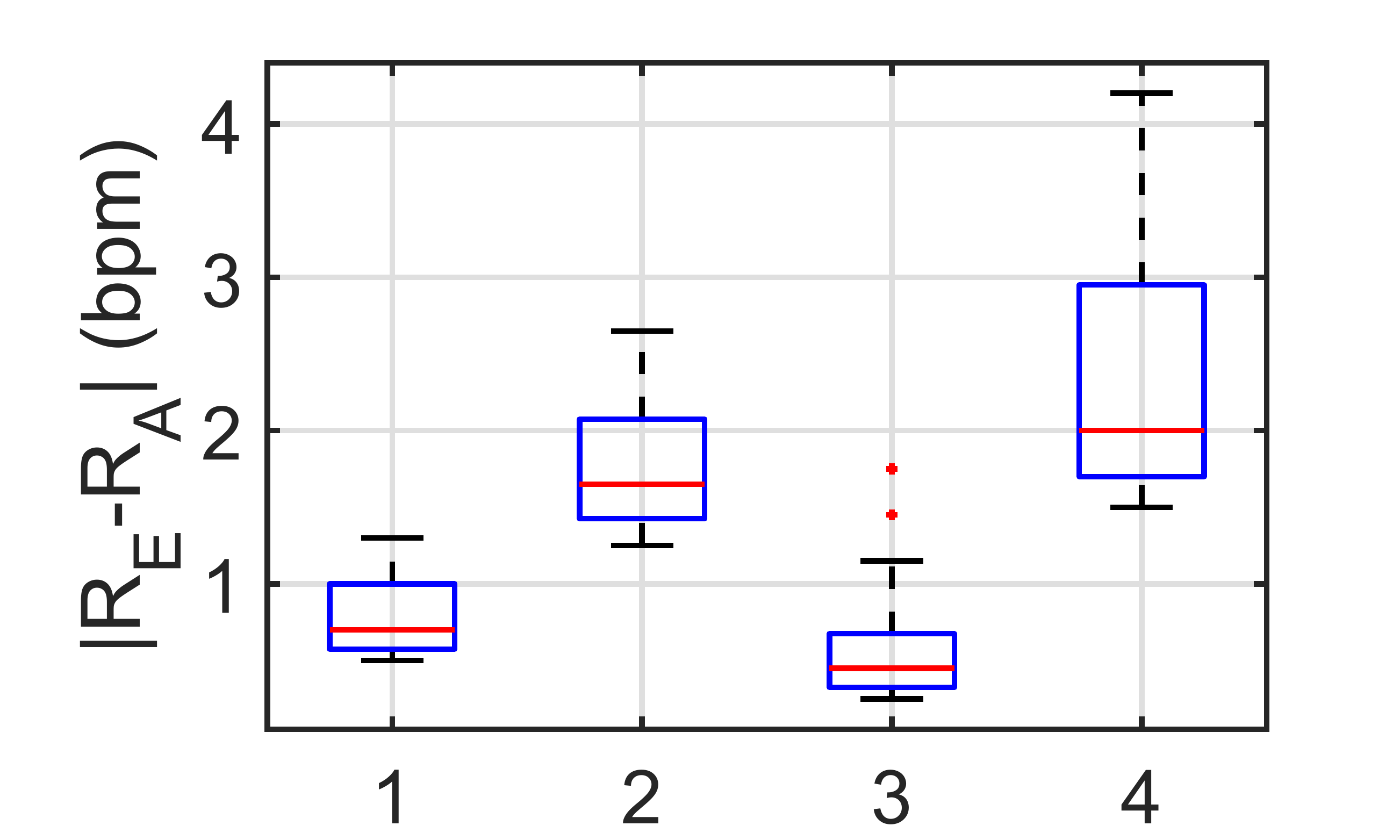}
		\label{fig:placement_h}	
	}
	\subfigure[IBI estimation error.]
	{
		\centering
		\includegraphics[width=0.31\textwidth]{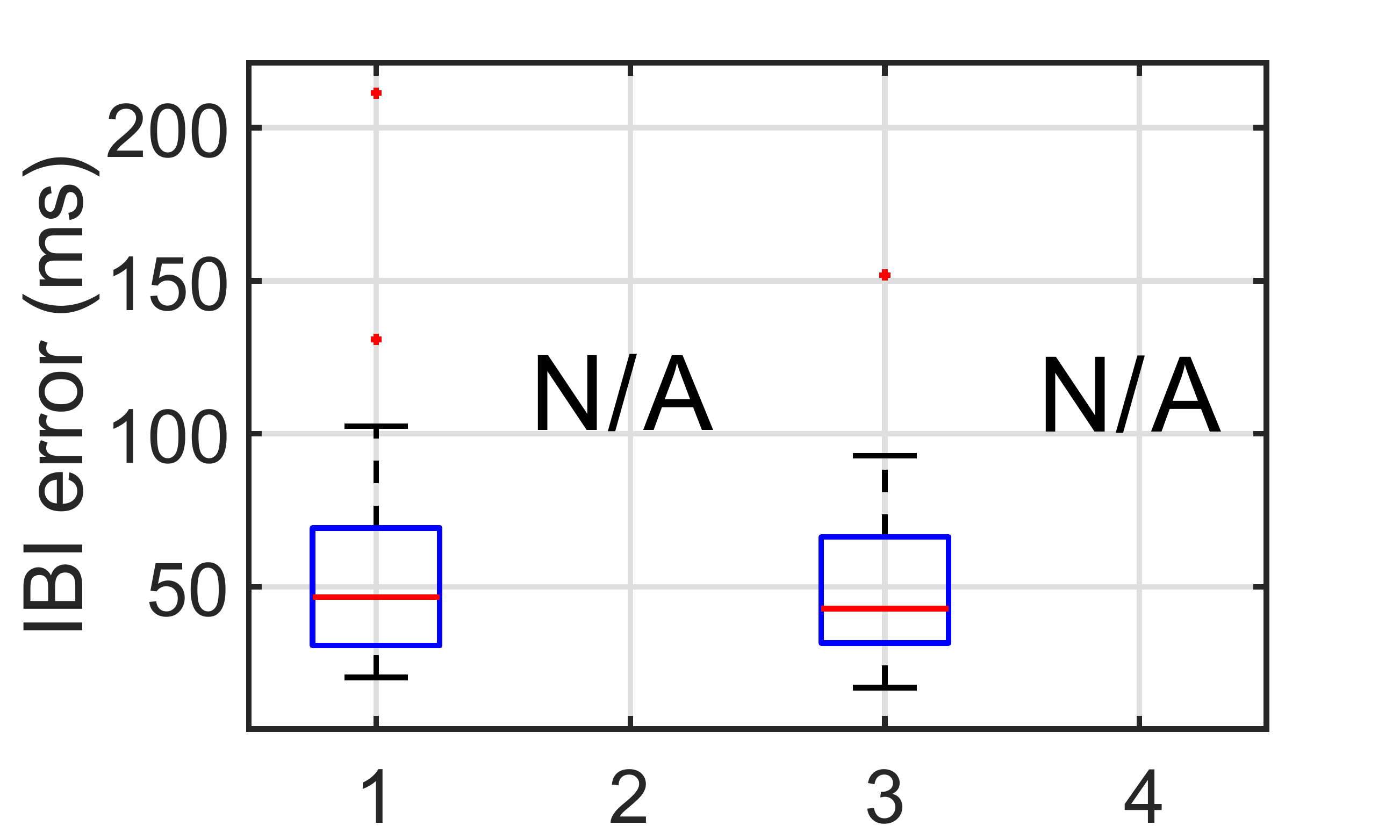}
		\label{fig:placement_IBI}	
	}
	\caption{Respiratory rate, heart rate and IBI estimation error of the driver. In the figure, class 1 is the case the radio is mounted at the top of the windshield; class 2 is the case when the radio is mounted at the bottom of the windshield; class 3 is the case when the radar is mounted at the top of the chair; class 4 is the case when the radio is mounted on the door of the vehicle.}
	\label{placement}
\end{figure}

To get the best performance of \systemname, we study the impact of different radio placements for \systemname\ by placing the radio at three different locations in the vehicle, i.e., at the top of the windshield, at the bottom of the windshield, at the top of the chair and on the door. The estimation errors of the vital signs are shown in Fig.~\ref{placement}. It can be seen that the performance of \systemname\ is the best when the radio is placed at the top of the windshield and faces the driver. In addition, when the radio is placed at the top of chair, the estimation errors of heart rate and IBI are relatively low. This is because when the impulse radio is placed at the top of the chair, it faces the driver's neck, where the carotid artery is located and movements caused by heartbeat is most obvious.

\subsubsection{Impact of Clothing}

\begin{figure}[ht]
	\centering
	\subfigure[Respiratory rate estimation error.]
	{
		\centering
		\includegraphics[width=0.31\textwidth]{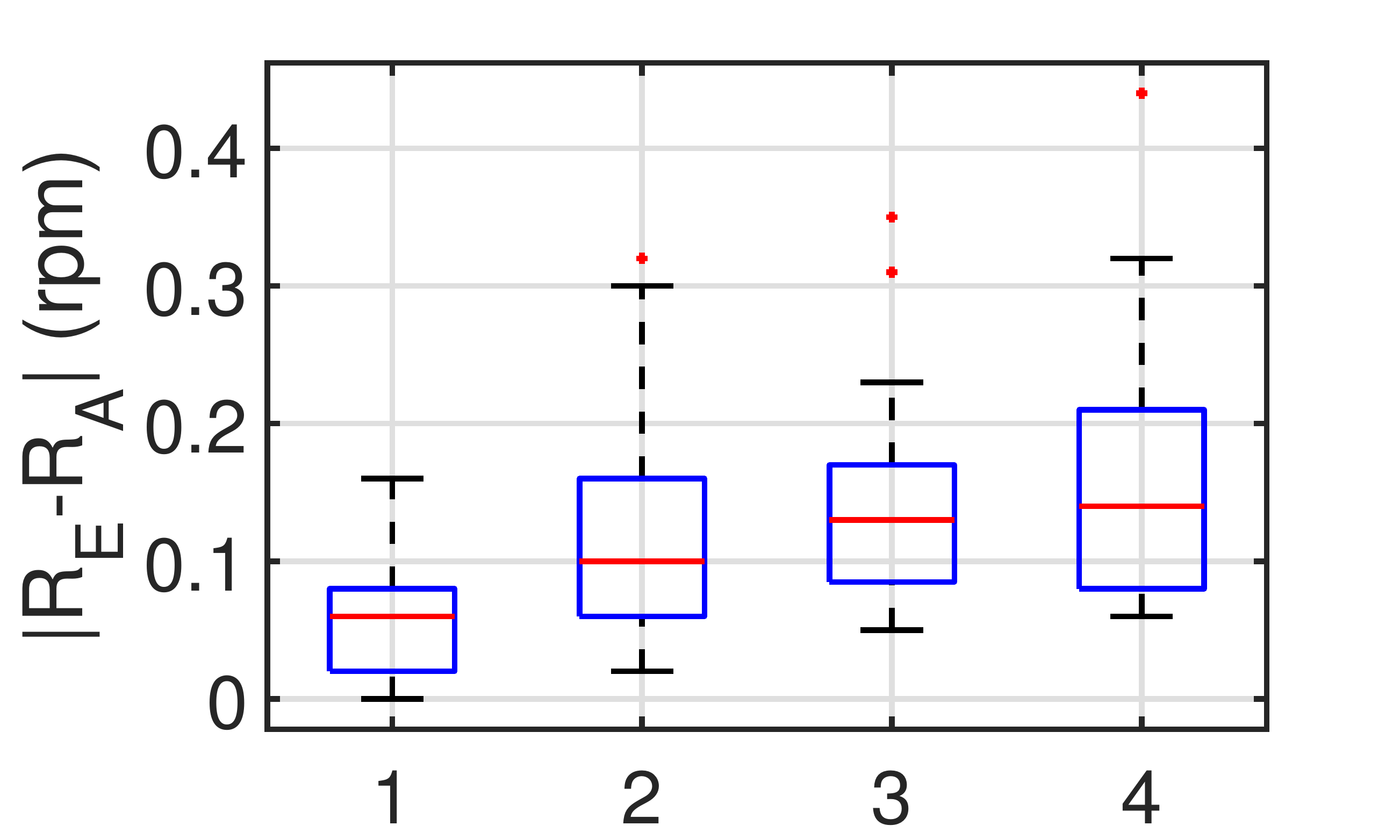}
		\label{fig:clothing_r1}	
	}
	\subfigure[Heart rate estimation error.]
	{
		\centering
		\includegraphics[width=0.31\textwidth]{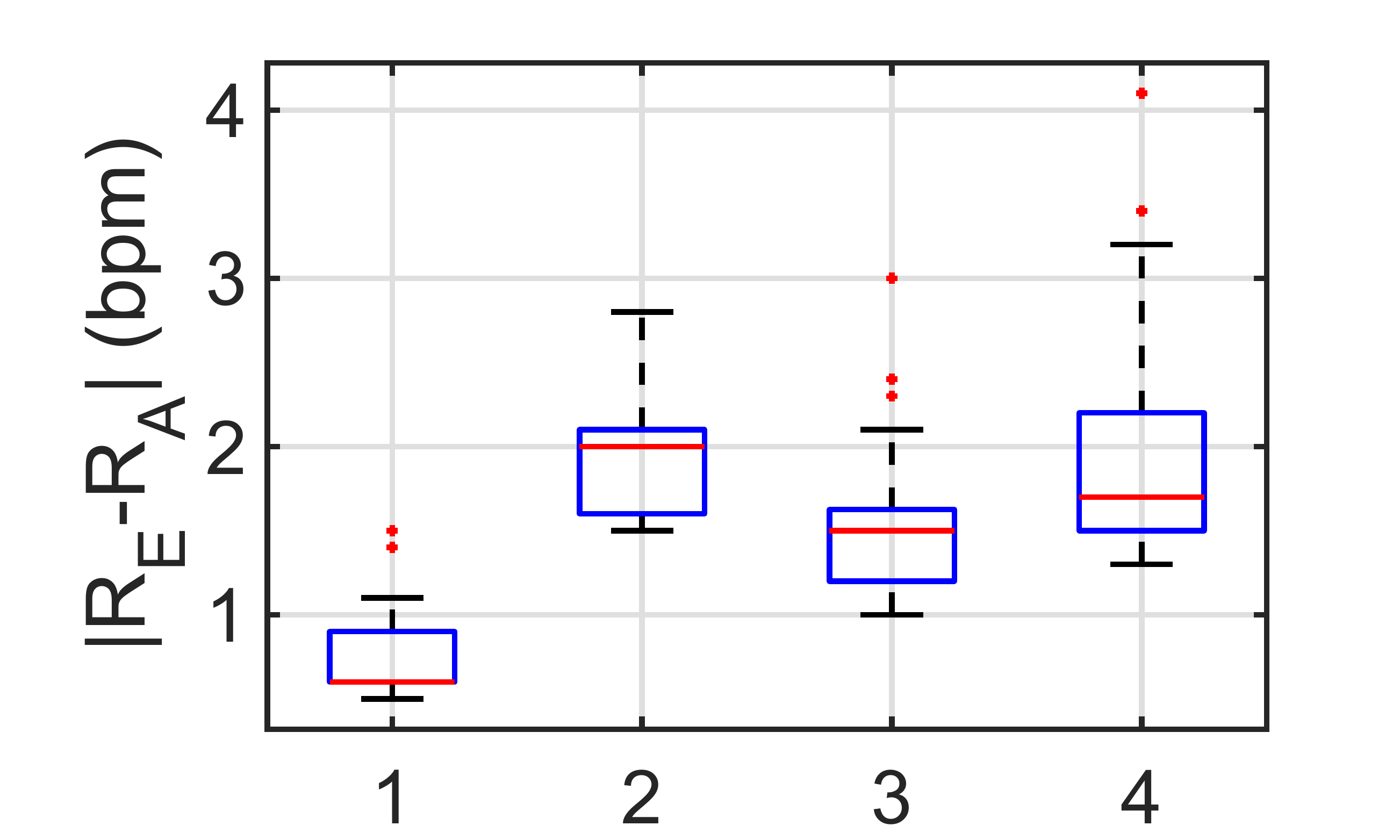}
		\label{fig:clothing_h}	
	}
	\subfigure[IBI estimation error.]
	{
		\centering
		\includegraphics[width=0.31\textwidth]{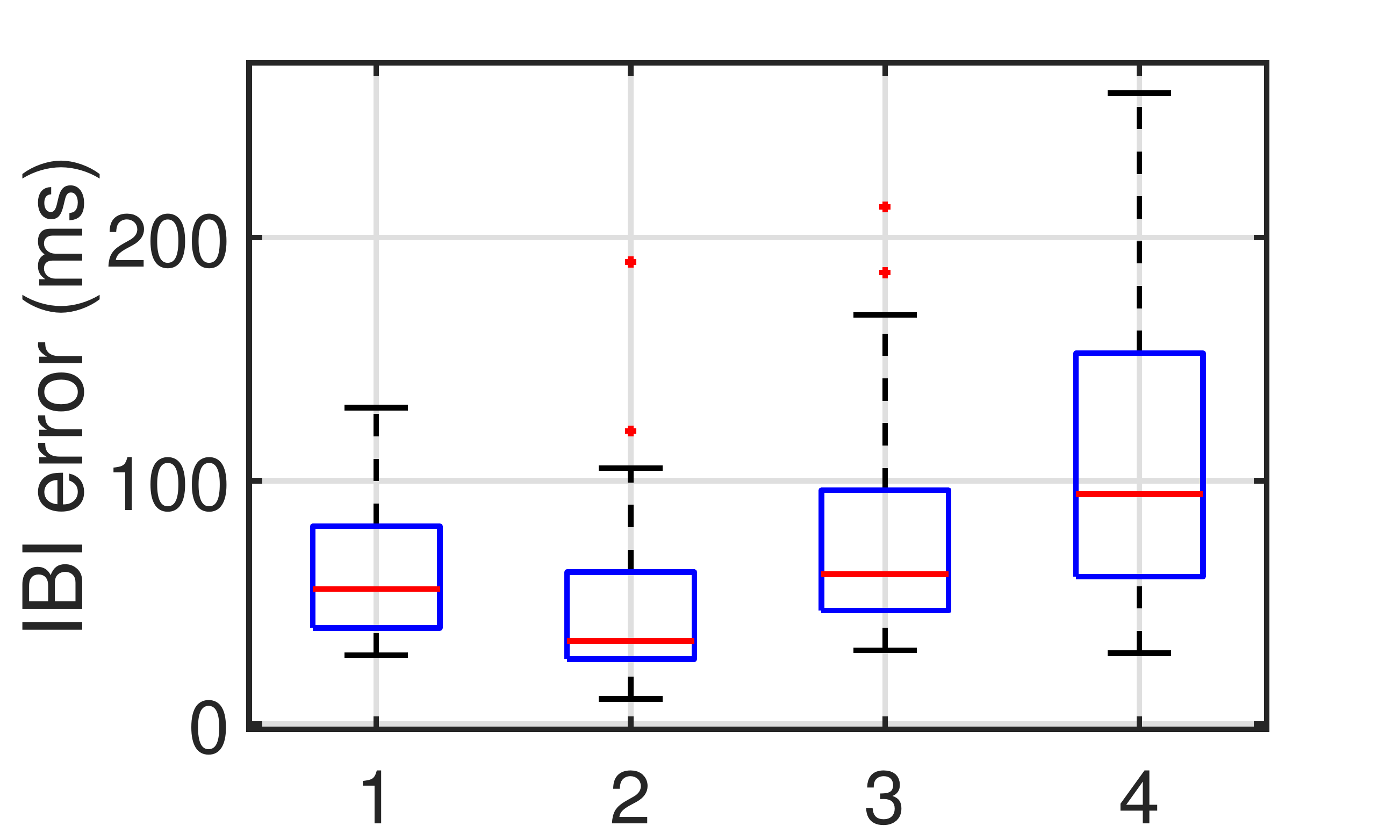}
		\label{fig:clothing_IBI}	
	}
	\caption{Respiratory rate, heart rate and IBI estimation error of the driver. In the figure, class 1 is the case when the driver wear lightweight T-shirt; class 2 is the case when the driver wear heavyweight T-shirt; class 3 is the case when the driver wear sweatshirt + lightweight T-shirt; class 4 is the case when the driver wear sweatshirt + heavyweight T-shirt.}
	\label{clothing}
\end{figure}

In our experiment, we ask our driver to wear different clothes and observe their respective results. The performance of \systemname\ is studied extensively under different clothing conditions. The clothing in our experiment are: (1) lightweight T-shirt (2) heavyweight T-shirt (3) sweatshirt +lightweight T-shirt (4) sweatshirt + heavyweight T-shirt. The result is shown in Fig.~\ref{clothing}.

It can be seen for all the different clothing, \systemname\ reaches a median estimation error $0.2~\!\mathrm{rpm}$ for respiratory rate, a median estimation error $0.4\!~\mathrm{bpm}$ for heart rate and $100\!~\mathrm{ms}$ estimation error for IBI. Moreover, it can be seen that \systemname\ performs better when the driver wear less, e.g., the result when the driver wear lightweight T-shirt is better than that of sweatshirt + heavyweight T-shirt. This is because some signals unable to penetrate heavyweight clothing and reach people's chest and abdomen and still able to detect minute movements like heartbeat. Moreover, even in the worst case, where the driver or passenger wears sweatshirt + heavy weight T-shirt, the estimation error for respiratory rate and heart rate is still acceptable. 

\subsubsection{Impact of Road Types and Traffic Conditions}

\begin{figure}[ht]
	\centering
	\subfigure[Respiratory rate estimation error.]
	{
		\centering
		\includegraphics[width=0.31\textwidth]{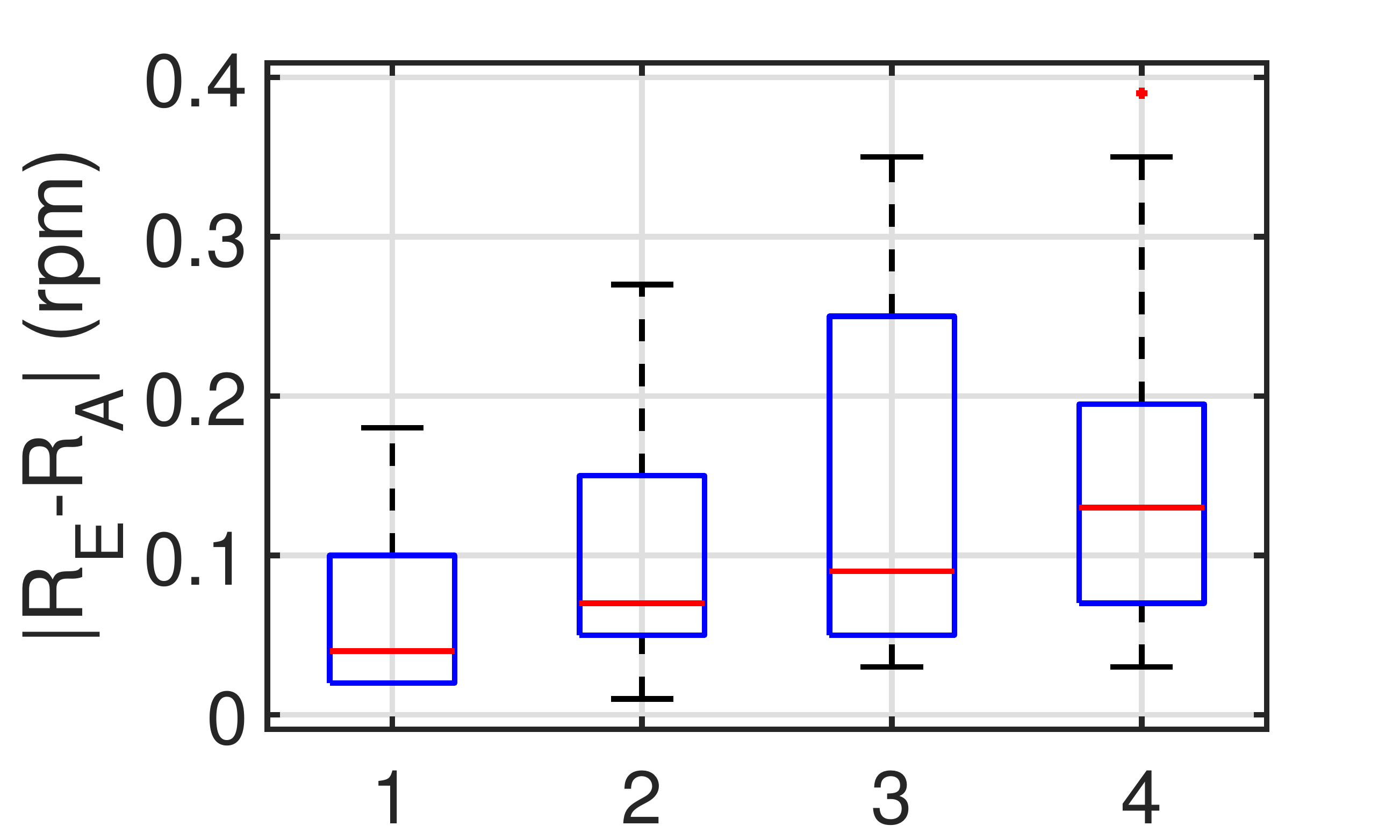}
		\label{fig:road_r}	
	}
	\subfigure[Heart rate estimation error.]
	{
		\centering
		\includegraphics[width=0.31\textwidth]{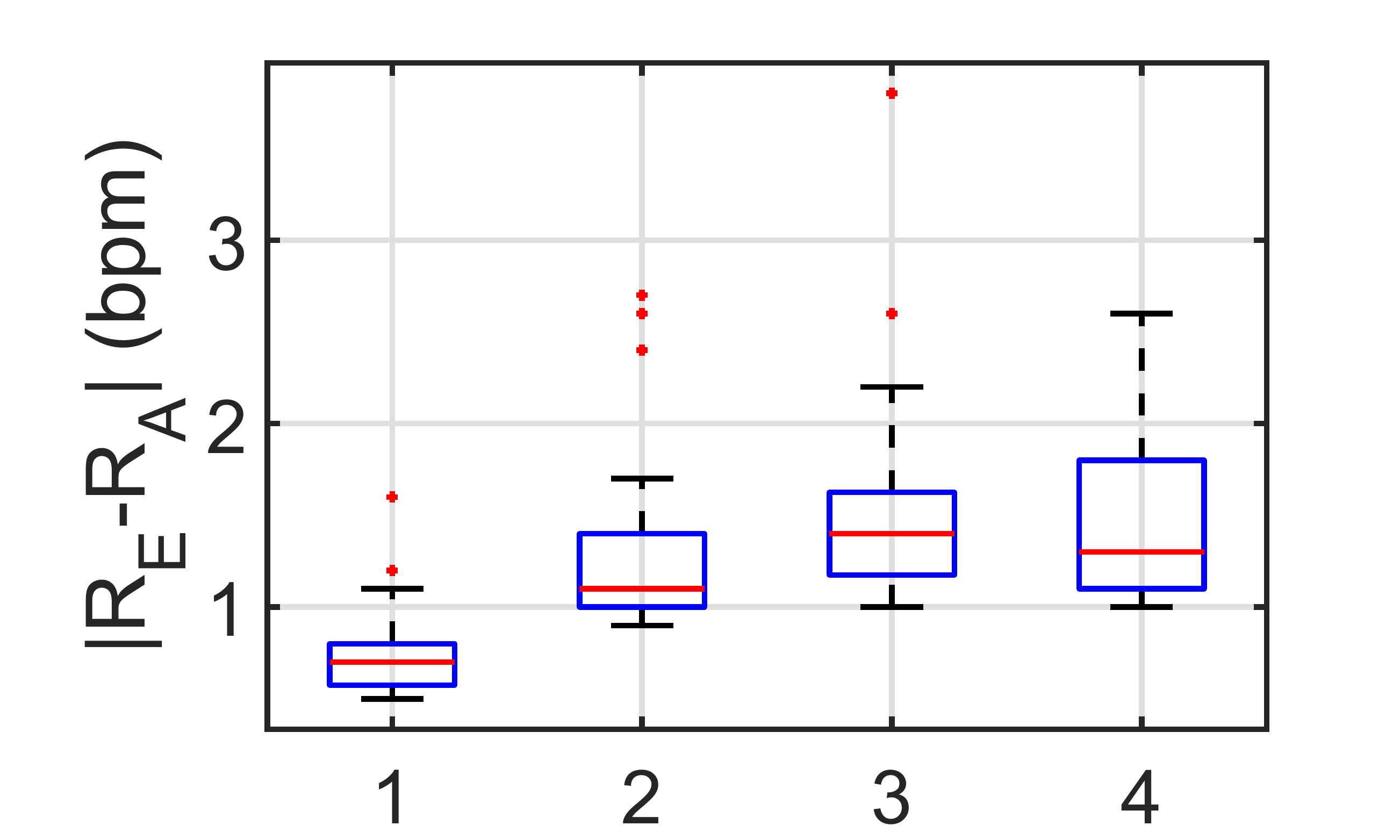}
		\label{fig:road_h}	
	}
	\subfigure[IBI estimation error.]
	{
		\centering
		\includegraphics[width=0.31\textwidth]{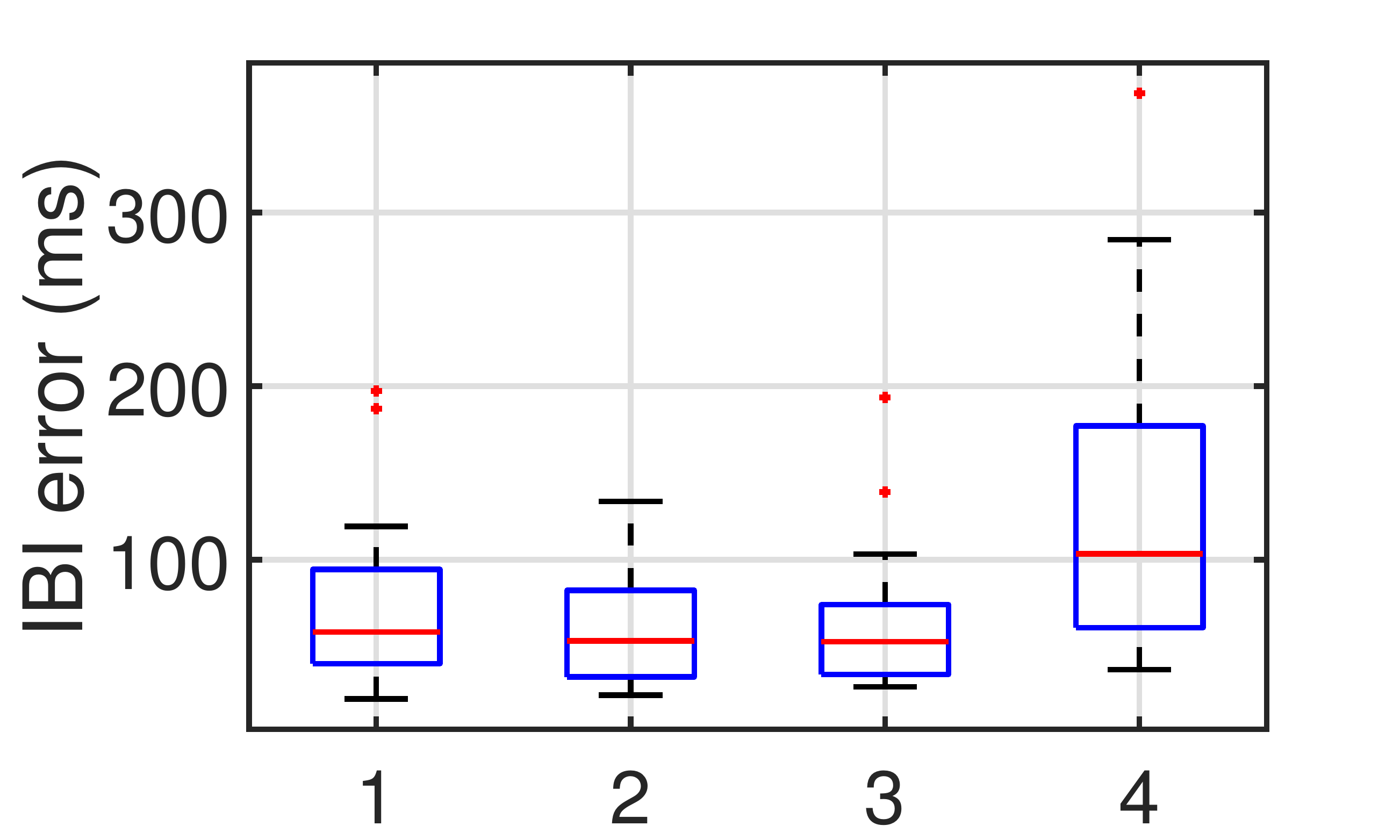}
		\label{fig:road_IBI}	
	}
	\caption{Respiratory rate, heart rate and IBI estimation error of the driver. In the figure, class 1 is the case when driving on smooth highway; class 2 is the case when driving uphill/downhill; class 3 is the case when taking turns (e.g., left turns, right turns, U-turns and roundabouts); class 4 is the case when driving on bumpy road.}
	\label{road}
\end{figure}

The signal quality of \systemname\ can be affect by different road types and traffic conditions, and thus the performance of \systemname\ can be impacted. To make sure \systemname\ works on different road types and under different traffic conditions. We collect data of different road types (e.g., smooth highway, bumpy road, uphill road, downhill road, intersections, left turns, right turns, roundabouts, U-turns) and analyzed the results respectively. The results are shown in Fig.~\ref{road}. It can be seen that if the surface of the road is smooth and not many manoeuvres are made, the estimation error of vital signs are low, while bumpy roads and driving manoeuvres increase the estimation error.

\subsection{Key Processes Study}

\subsubsection{Baseband Signal Processing}
According to the analysis in Sec.~\ref{sec:systemdesign}, both amplitude and phase of the reflected signal after IQ downconversion contain vital sign information. \systemname\ chooses amplitude to estimate breathing and heartbeat. Contrary to common belief that amplitude is more vulnerable to random noises, our evaluation show that estimating vital signs by amplitude gives a better performance. Our explanation for this is as follows: if phase signal is to be used, we have to select a static fast-time bin as reference and do phase correction first. However, in driving environments, it is nearly impossible to find fast-time bin which represents a static reflector. Therefore, in \systemname, amplitude of the baseband signal is selected for vital sign extraction. 

\begin{figure}[ht]
	\centering
	\subfigure[Respiratory rate estimation error.]
	{
		\centering
		\includegraphics[width=0.31\textwidth]{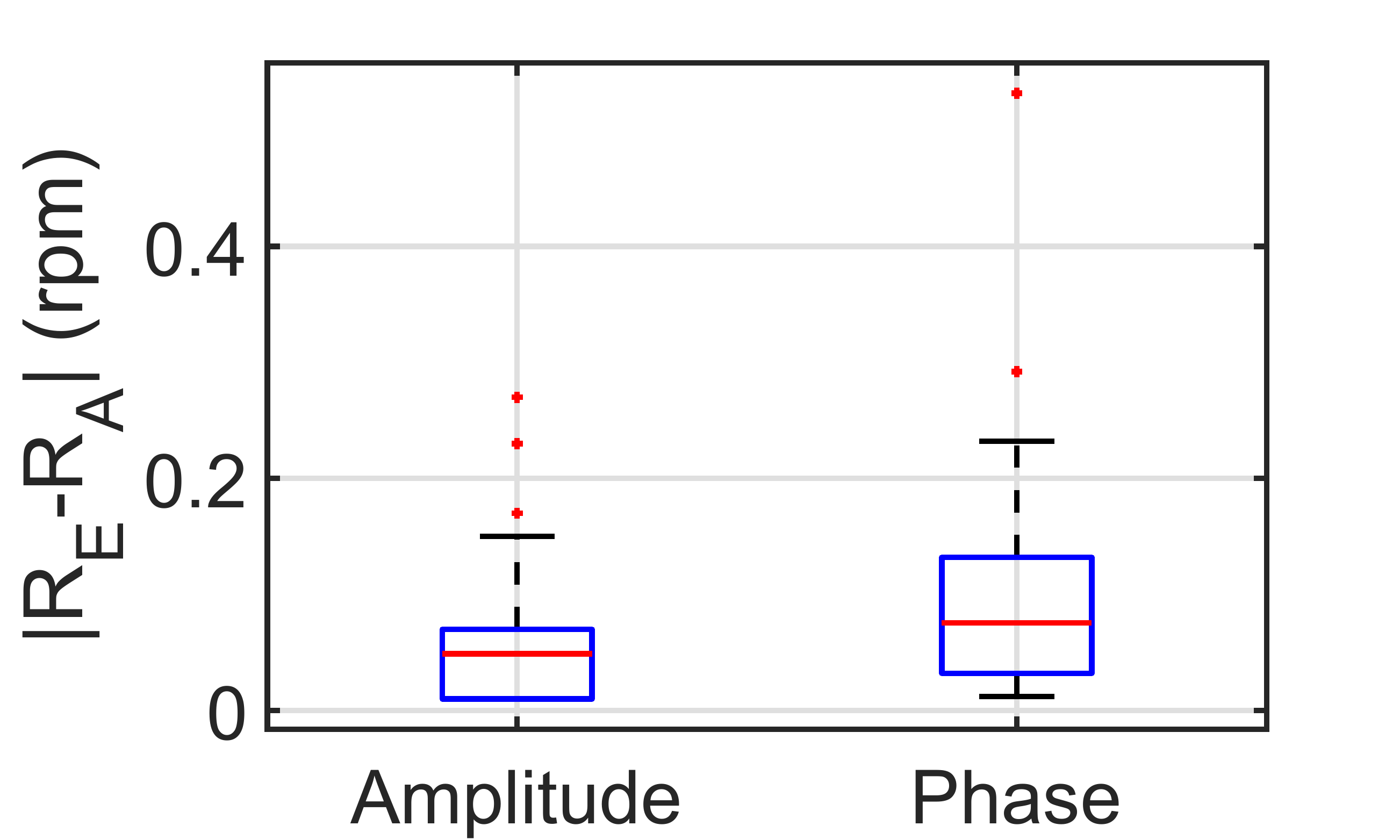}
		\label{fig:baseband_r}	
	}
	\subfigure[Heart rate estimation error.]
	{
		\centering
		\includegraphics[width=0.31\textwidth]{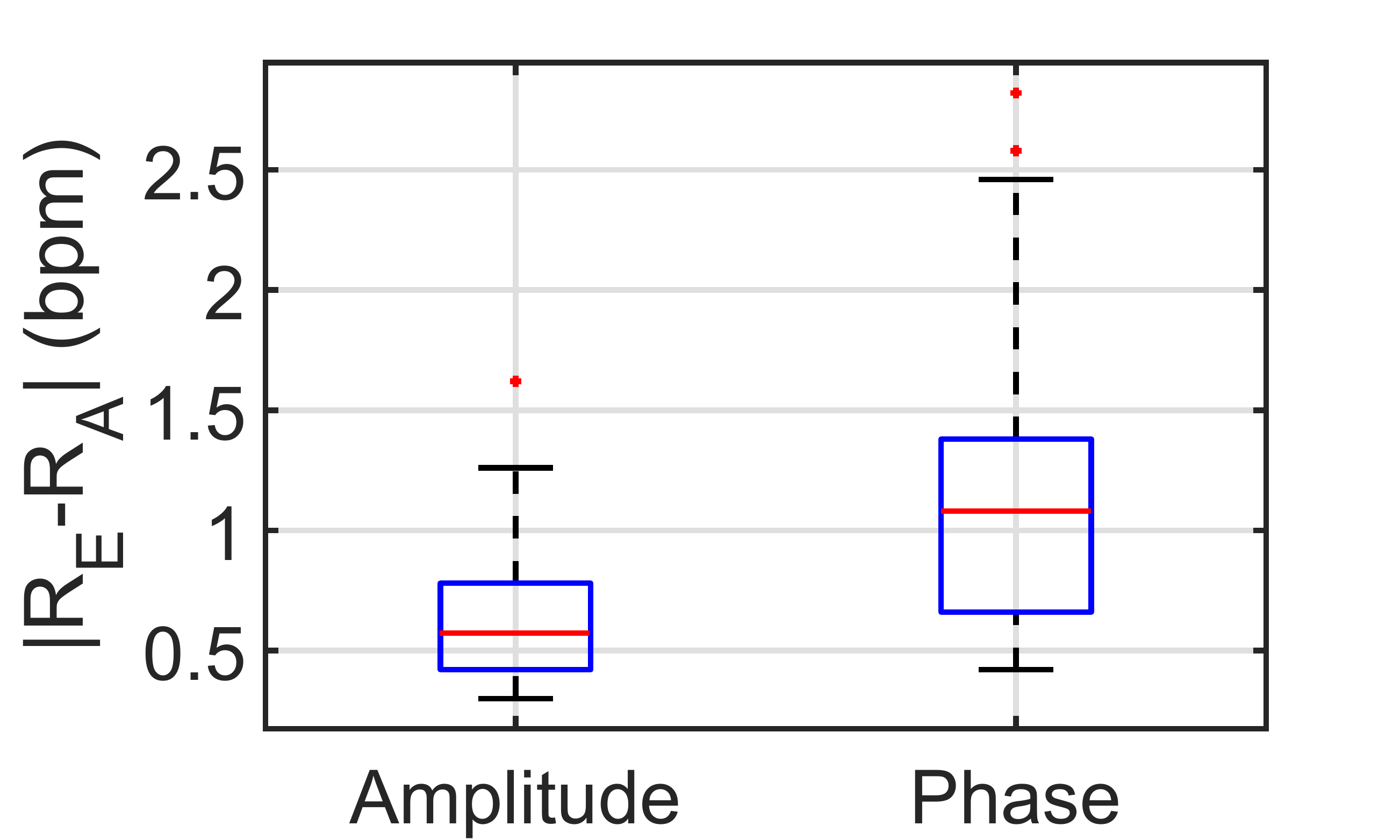}
		\label{fig:baseband_h}	
	}
	\subfigure[IBI estimation error.]
	{
		\centering
		\includegraphics[width=0.31\textwidth]{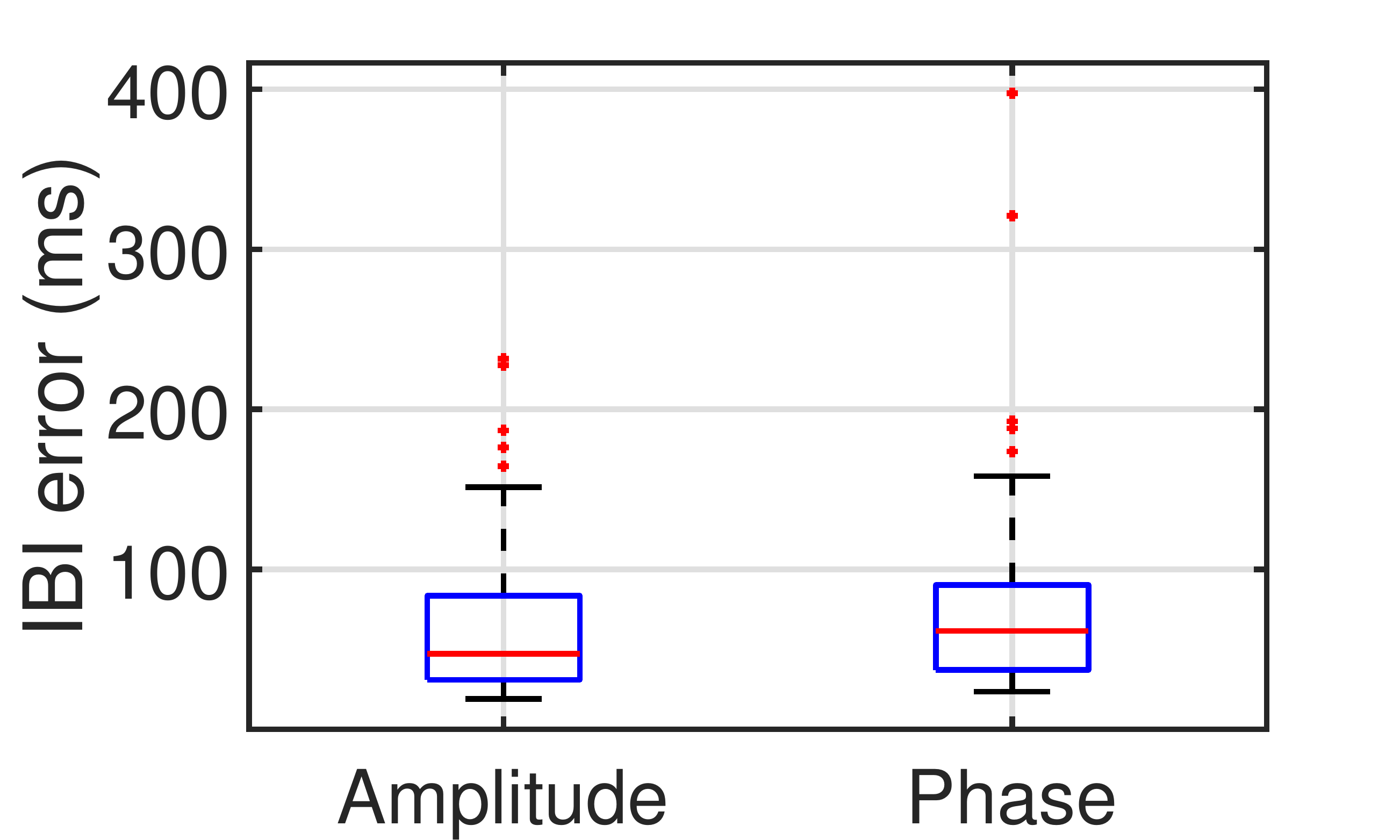}
		\label{fig:baseband_IBI}	
	}
	\caption{Respiratory rate, heart rate and IBI estimation error of using the amplitude and phase of the baseband signal.}
	\label{fig:baseband}
\end{figure}

The experiment result is shown in Fig.~\ref{fig:baseband}. It can be seen that system using amplitude signal achieves less overall errors than systems using phase signal. For respiratory rate estimation error, the median with amplitude baseband processing is $0.06\!~\mathrm{rpm}$, and the median with phase baseband processing is $0.08\!~\mathrm{rpm}$. For heart rate estimation error, the median with amplitude baseband processing is $0.6\!~\mathrm{bpm}$, and the median with phase baseband processing is $1.1\!~\mathrm{bpm}$. For IBI estimation error, the median with amplitude baseband processing is $50\!~\mathrm{ms}$, and median with phase baseband processing is $60\!~\mathrm{ms}$.

\subsubsection{Signal Decomposition}
\begin{figure}[ht]
	\centering
	\subfigure[Decomposition result of VMD.]
	{
		\centering
		\includegraphics[width=0.44\textwidth]{pic/VMD.eps}
		\label{fig:VMD_performance}	
	}
	\subfigure[Decomposition result of MS-VMD.]
	{
		\centering
		\includegraphics[width=0.44\textwidth]{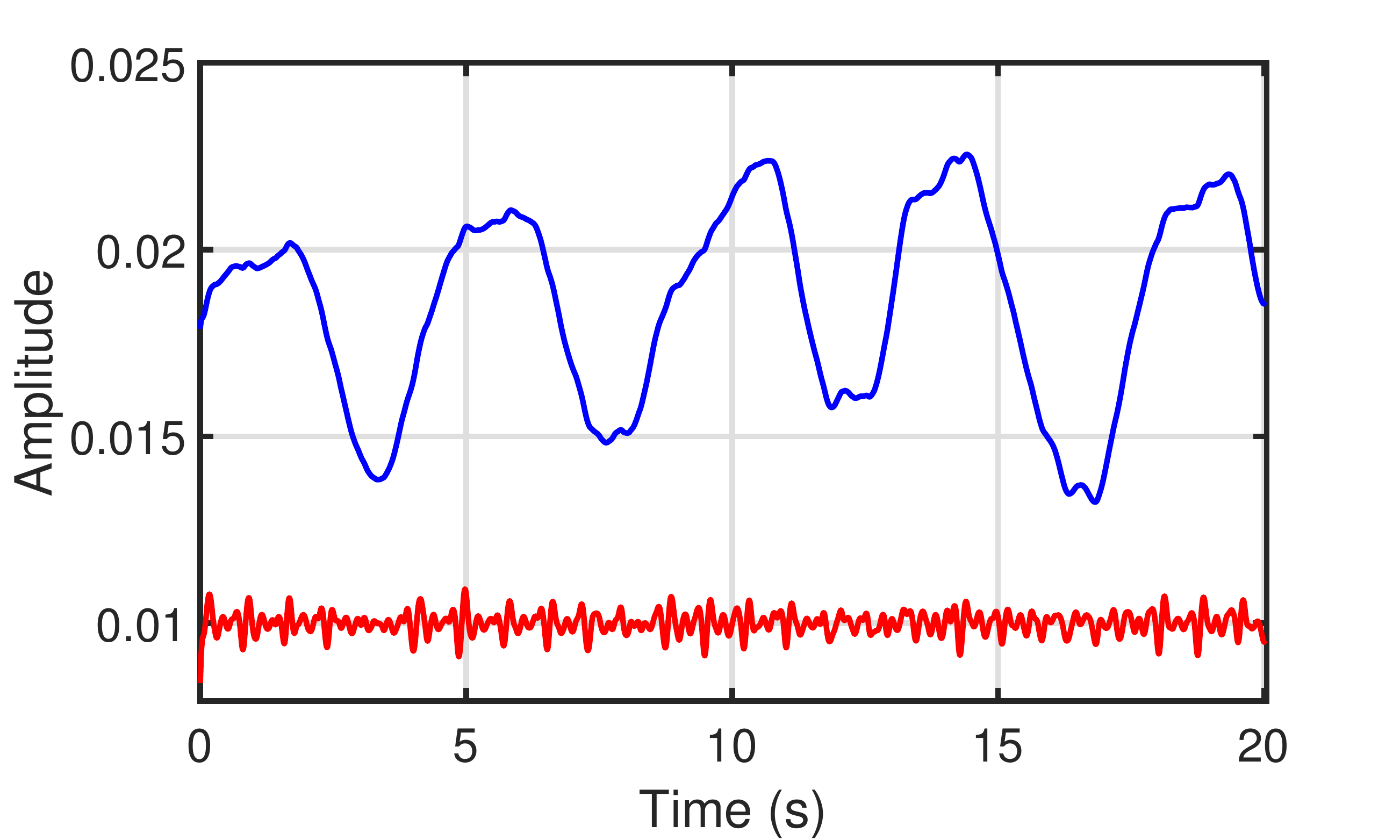}
		\label{fig:MS-VMD_performance}	
	}
	\caption{Example of VMD and MS-VMD on decomposing signals to respiration and heartbeat component.}
	\label{VMD_eval}
\end{figure}

In \systemname, our novel MS-VMD algorithm is used to decompose the filtered signal into its vital sign components. According to the analysis in Sec.~\ref{ssec:vmd}, our novel MS-VMD algorithm employs time diversity to enhance vital signals. It jointly optimizes the problem of minimum bandwidth with multiple sequences taken into account, hence better than VMD, where only one data sequence is considered. Fig.~\ref{VMD_eval} gives the resulting waveforms of a case where conventional VMD algorithm only obtains a mix of respiration and heartbeat, while MS-VMD separates the two signals successfully.

\begin{figure}[ht]
	\centering
	\subfigure[Respiratory rate estimation error.]
	{
		\centering
		\includegraphics[width=0.31\textwidth]{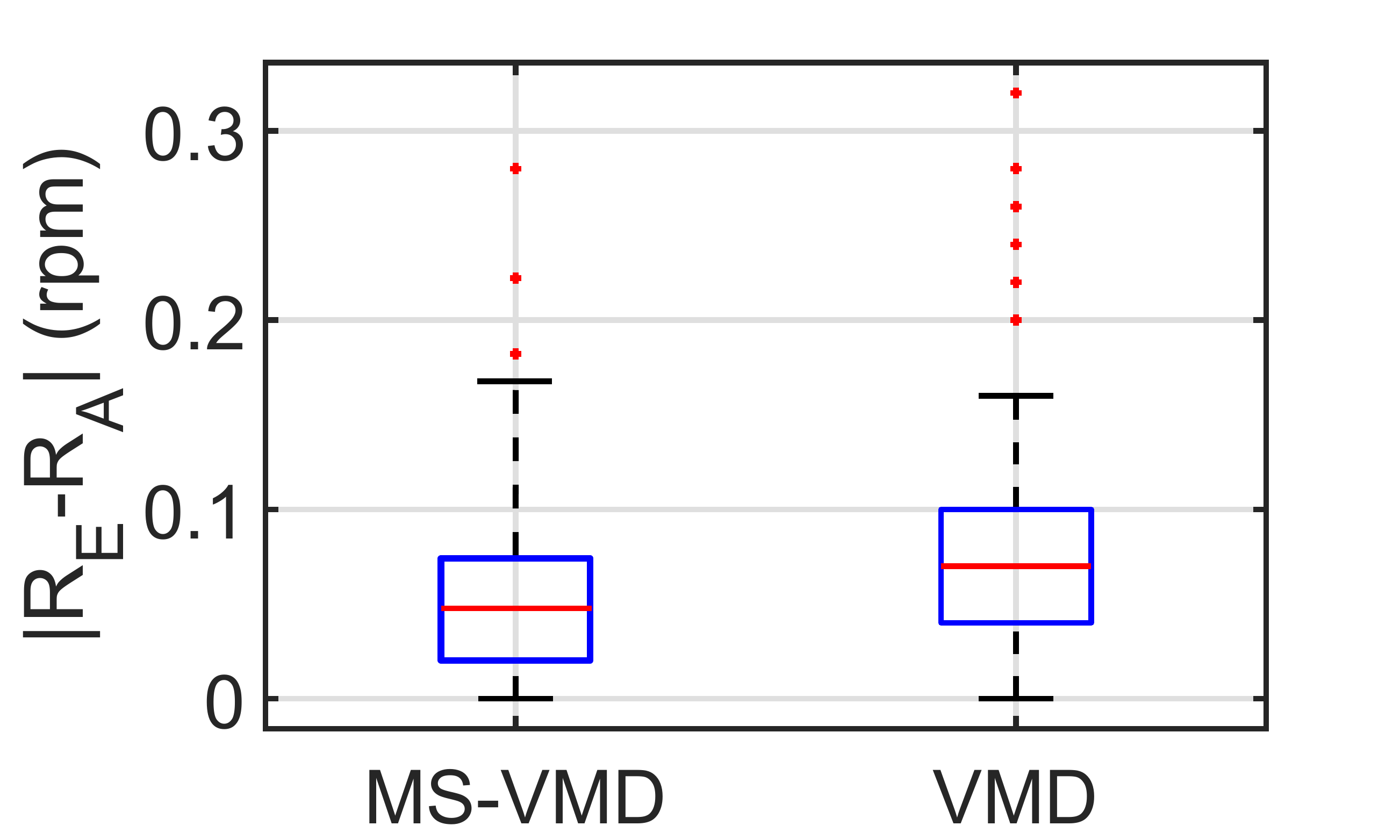}
		\label{fig:MSVMDVSVMD_r}	
	}
	\subfigure[Heart rate estimation error.]
	{
		\centering
		\includegraphics[width=0.31\textwidth]{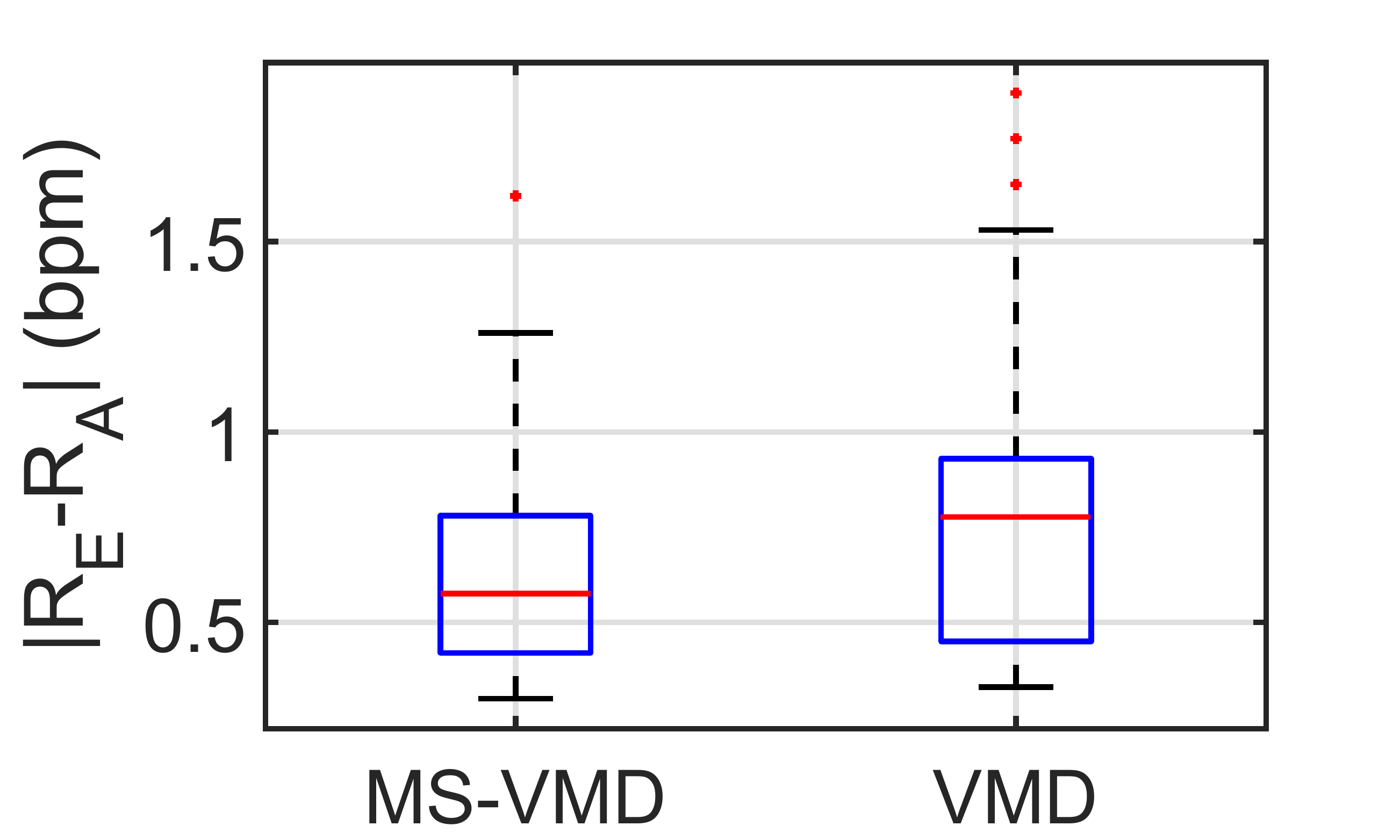}
		\label{fig:MSVMDVSVMD_h}	
	}
	\subfigure[IBI estimation error.]
	{
		\centering
		\includegraphics[width=0.31\textwidth]{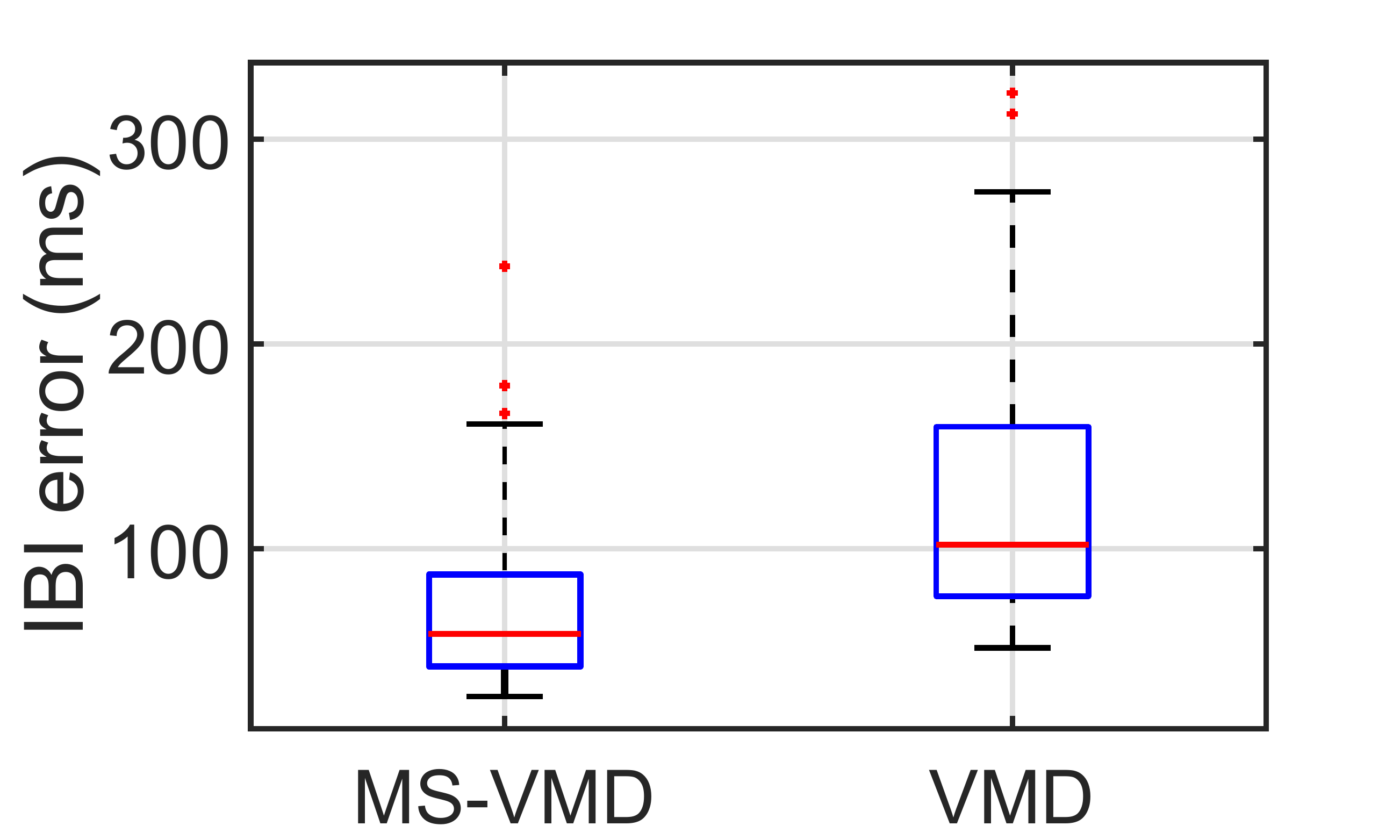}
		\label{fig:MSVMDVSVMD_IBI}	
	}
	\caption{Respiratory rate, heart rate and IBI estimation error of MS-VMD algorithm and VMD algorithm. }
	\label{fig:MSVMDVSVMD}
\end{figure}

To evaluate their performance, we test these two algorithms on the same data, and the result is shown in Fig.~\ref{fig:MSVMDVSVMD}. In the figure, it can be seen that system using MS-VMD algorithm achieves less overall errors than systems using VMD algorithm. For respiratory rate estimation error, the median with MS-VMD algorithm is $0.06\!~\mathrm{rpm}$, and the median with VMD algorithm is $0.07\!~\mathrm{rpm}$. For heart rate estimation error, the median with MS-VMD algorithm is $0.6\!~\mathrm{bpm}$, and the median with VMD algorithm is $0.8\!~\mathrm{bpm}$. For IBI estimation error, the median with MS-VMD algorithm is $50\!~\mathrm{ms}$, and the median with VMD algorithm is $100\!~\mathrm{ms}$.

\section{Limitations and Potentials} \label{sec:limit_fw}
\subsection{Limitation of \texorpdfstring{\systemname}{V2iFi}}
We designed \systemname\ in the hope of bringing well-being to the driving environments. \systemname\ is successful in terms of monitoring vital signs. However, monitoring is not the final goal: the vital sign data collected should be used by the vehicle to infer user states and take actions correspondingly.  To achieve this, vital signs should be mapped to physiological and psychological states. In previous studies it is shown that the vital signs can indeed reflect physical health (e.g., cadiac health and drowsiness level)~\cite{Smartwatch-IEEESens,fieselmann1993respiratory,cysarz2007regular} and mental health (e.g., emotion and mental workload)~\cite{zhao2016emotion, monitoring-Accid}. A commonly used approach for such mapping is supervised learning, e.g., decision tree and SVM~\cite{WiFind-IEEEToBD}. However, learning algorithms do not work without properly labelled data. Awais et~al.~\cite{awais2017hybrid} use subjective measures (i.e., what the participant says) to label both physical and mental states, but subjective measures are unreliable and inaccurate, potentially leading to inconsistent inference results. To measure the person-specific states objectively and accurately, we face many challenges such as infringement of privacy and lack of gauging device. Moreover, medical organizations do not easily provide related data because of potential misuse of information and violation of law. Nevertheless, once we have properly labelled data at hand, we envision that \systemname\ can become the foundation of more interesting applications, as we present in the following.

\subsection{Application 1: Driver Drowsiness Detection}
\begin{figure}[ht]
	\centering
	\includegraphics[width=0.6\linewidth]{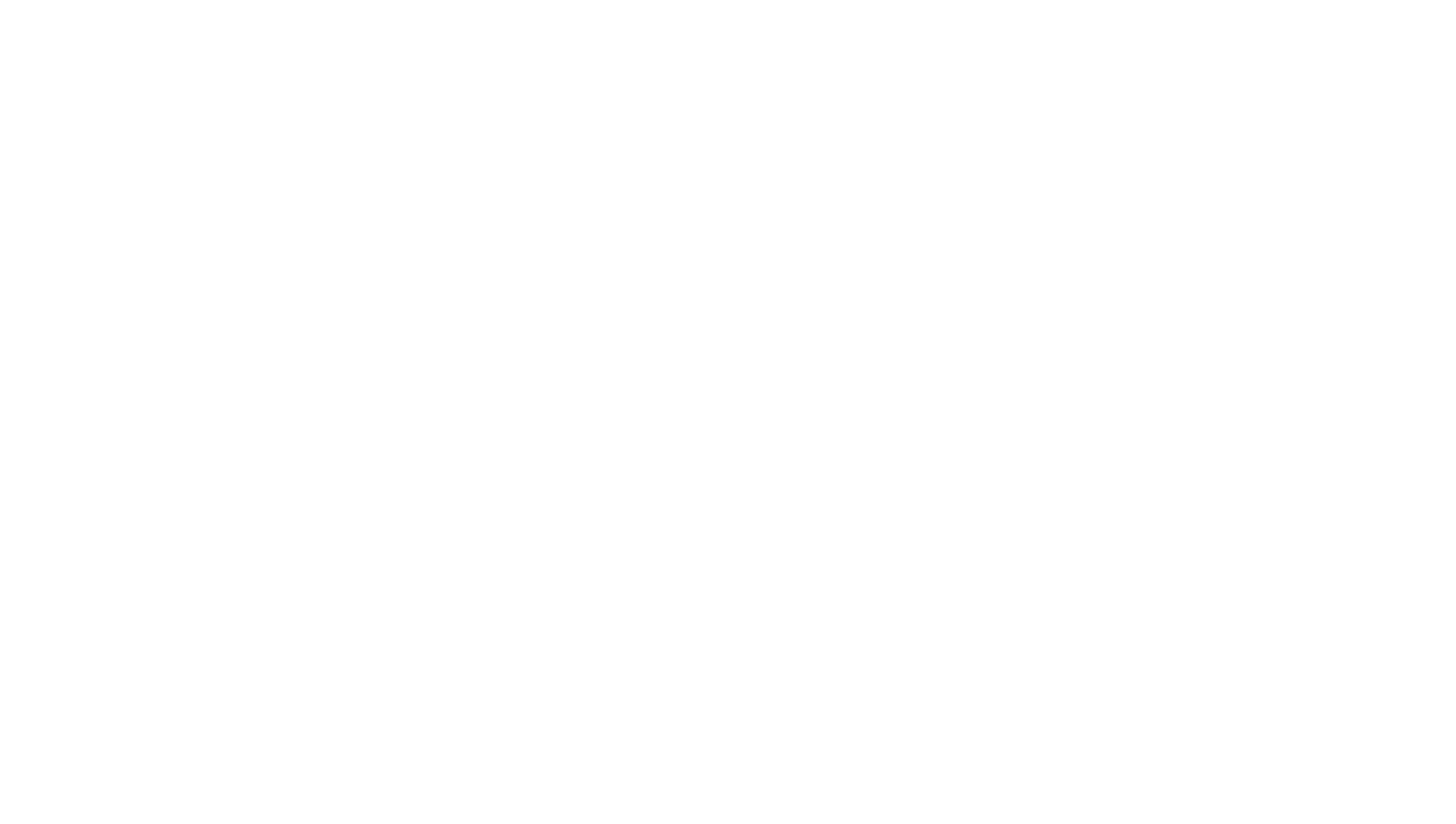}
	\caption{Driver drowsiness decision tree: an example.}
	\label{fig:tree}
\end{figure}

Driver drowsiness is one of the biggest culprit in car accidents, research shows that driver drowsiness is a contributing factor in $\mathrm{20 \%}$ of all the crashes~\cite{jackson2011fatigue}, and about $\mathrm{25 \%}$ of all fatalities on road~\cite{flatley2004sleep}. A series of vital sign changes indicate a drowsy driver, e.g., decreased respiratory rate, decreased heart rate and decreased HRV. \systemname\ can detect all these changes, and a high-level learning algorithm such as decision tree and SVM can be used to combine the aforementioned change in vital signs to make a binary decision(i.e., drowsy or awake) in a specific time span. An example decision tree is shown in Fig.~\ref{fig:tree}.

\subsection{Application 2: Road Rage Prevention}

Road rage is violent behaviors perpetrated by drivers and it is extremely dangerous. Being slowed down and blocked by others on overly crowded streets make drivers on the road irritated and road rage arises thereafter. DePasquale et al.~\cite{depasquale2001measuring} show road rage incidents is on the rise in recent years. To prevent road rage, legal measures, better vehicle design, education in everyday life and infrastructure improvement have been tested. However, because road rage events are sporadic, all these preventive measures cannot handle a real road rage event in time. Since reflected RF signals of \systemname\ carry information of the emotion of the driver~\cite{zhao2016emotion}, it is possible to build a road rage detection system on top of \systemname\ and the vehicle can respond, take precautions and combat road rage. Similar to driver drowsiness detection, we can use a learning algorithm, such as a decision tree similar to that in Fig.~\ref{fig:tree} to infer the driver's road rage level. The vehicle can then sends an alert to the driver in time or brakes if necessary.

\subsection{Application 3: Health Monitoring for Children and Seniors in the Vehicle}
It has long been known that environments inside vehicle cause significant health harm and hazards. Particularly, seniors and children are more vulnerable and tend to become victims. There are two reasons: i) seniors and children are more physically inactive inside a vehicle and ii) children and seniors are more sensitive to the car exhaust in vehicle.  With \systemname, vital signs of children and seniors inside can be monitored, and common diseases such as childhood asthma and heart failure among the senior people can be infered from abnormal vital signs, hence be prevented. \systemname\ is a potential solution to solve the health issues and build a healthy and secure environments for the children and seniors.

\section{Related Work}  \label{sec:rw}

Existing work in health monitoring in driving environments can be roughly categorized into three classes: (1) vision-based method, (2) sensor-based method and (3) RF-based method.%

The accuracy of the detection systems depends largely on the invasiveness allowed. The detection systems with direct contact with the driver are the most accurate ones, but they hinder the body movements of the driver and thus cause inconvenience while driving, so they are not accepted very well in the industry. On the other side of the invasiveness spectrum are those "aerial" sensors, they are not in direct contact with the driver. These systems get indirect signals of the driver, such as eye movement to infer other physiological variables, so they cannot be very accurate. Although these systems are more easily accepted in the industry, they need more data and calibration to characterize health states of the people effectively. In the following, we elaborate the subcategories of the systems:  

\textbf{Vision-based methods}: Vision-based methods deploy one camera, usually at the front of the vehicle, to capture the activities, various signals and performance of the driver and passengers. Eriksson et al.~\cite{eriksson1997eye} proposed a system to track movements of eyes and eyelids of the driver. Following preprocessing, they locate the driver's face in the grayscale image, and then eyes are further located and tracked. The states of the eyes, whether open or closed are determined by template matching. However, taking photo of the driver's face raises privacy concerns,  which makes it not good for private room such as driving environments. Meanwhile, vision-based systems are also sensible to external conditions, lighting and weather conditions, different dress and fashion accessories like glasses, winter clothes will decrease the detection accuracy.  

\textbf{Sensor-based methods}:  Sensor-based methods can be further categorized to two groups: EEG-based and ECG-based method. EEG-based methods use brain physiological changes to detect vital signs. Multiple thin metal wires (electrodes) are placed on the scalp. The waveform obtained by EEG reflects the mental activities of the brain. In~\cite{mardi2011eeg}, nonlinear and chaotic vital signs of drivers are extracted. Another portable low-cost EEG method is proposed in~\cite{ogino2018portable}, and SVM is used to classify EEG data as different health states. Similar to EEG-based methods, ECG-based methods provide accurate physiological signals of the driver under test. Heart rate variability is extracted from the ECG signal in~\cite{elsenbruch1999heart,tsunoda2001effects}. In another work, a smart cushion containing an integrated micro-bending fiber sensor is used to imitate ECG sensors~\cite{cushion-biocas}. However, EEG and ECG deploy sensors on human skin to detect biological signals. They require conductive paste to reduce skin impedance to build electrical contact, making the methods intrusive, time-consuming and inconsistent when long recording is needed~\cite{sullivan2007low}. These methods always make the users aware of its existence and degrade user experience.

\textbf{Wi-Fi-based methods}: Wi-Fi sensing based vital sign estimation has been investigated in recent works such as ~\cite{WiCare-MobiQuitous, Fullbreathe-UbiComp, RFECG-UbiComp,wu2017device}. These works  focus on static environment but not in-vehicle environment. In~\cite{WiFind-IEEEToBD}, Jia et al. propose a fatigue detection approach Wi-Find based on Wi-Fi CSI. Channel State Information (CSI) describes the channel properties of a communication link from the transmitter to the receiver. For a Wi-Fi router, CSI information can be obtained at different carrier frequencies and along different paths. The CSI measured by Wi-Fi devices is widely used for localization, occupation sensing, activity recognition and other sensing purposes. Compared to the aforementioned sensor-based systems, the system proposed in~\cite{WiFind-IEEEToBD} only requires a wireless STA and AP, thus is device-free. 

Despite the existence of several vital signs monitoring systems, few of them are practical in the driving environments. Our work answers challenges such as passenger interference, in-vehicle vibration, human motion, etc., and monitors vital signs in driving environments successfully.

\section{Conclusion} \label{sec:con}
In this paper, we present the design and implementation of \systemname, which a vital sign monitoring system in driving environments leveraging the power of impulse radio. We address several challenges including extracting fine-grained driver's vital sign from RF signal polluted with vehicle vibration noise and mixed with vital signals from other passengers. \systemname\ consists of multiple algorithms for removing noise, including the novel MS-VMD algorithm. The system is evaluated in real road test and demonstrated low respiratory rate, heart rate and heart rate variability estimation error. In the future, we will extend \systemname\ to a fully-grown in-vehicle health monitoring system. We believe \systemname\ takes a big step toward enabling health monitoring in driving environments, it allows vehicles to ``listen'' and understand the driver and passengers and thus become more intelligent.

\begin{acks}
This paper is supported by the Ministry of Education, Singapore, under its AcRF Tier 2 Grant MOE2016-T2-2-022 and AcRF Tier 1 Grant RG17/19. We would also like to extend our thanks to the technicians of WiRUSH.ai  (\url{http://www.wirush.ai}) for their help in offering us the hardware and technical support.
\end{acks}

\bibliographystyle{ACM-Reference-Format}
\bibliography{citations}


\begin{thebibliography}{65}


\ifx \showCODEN    \undefined \def \showCODEN     #1{\unskip}     \fi
\ifx \showDOI      \undefined \def \showDOI       #1{#1}\fi
\ifx \showISBNx    \undefined \def \showISBNx     #1{\unskip}     \fi
\ifx \showISBNxiii \undefined \def \showISBNxiii  #1{\unskip}     \fi
\ifx \showISSN     \undefined \def \showISSN      #1{\unskip}     \fi
\ifx \showLCCN     \undefined \def \showLCCN      #1{\unskip}     \fi
\ifx \shownote     \undefined \def \shownote      #1{#1}          \fi
\ifx \showarticletitle \undefined \def \showarticletitle #1{#1}   \fi
\ifx \showURL      \undefined \def \showURL       {\relax}        \fi
\providecommand\bibfield[2]{#2}
\providecommand\bibinfo[2]{#2}
\providecommand\natexlab[1]{#1}
\providecommand\showeprint[2][]{arXiv:#2}

\bibitem[\protect\citeauthoryear{Adib, Kabelac, Katabi, and Miller}{Adib
  et~al\mbox{.}}{2014}]%
        {3Dtracking-NSDI}
\bibfield{author}{\bibinfo{person}{Fadel Adib}, \bibinfo{person}{Zach Kabelac},
  \bibinfo{person}{Dina Katabi}, {and} \bibinfo{person}{Robert~C Miller}.}
  \bibinfo{year}{2014}\natexlab{}.
\newblock \showarticletitle{{3D Tracking via Body Radio Reflections}}. In
  \bibinfo{booktitle}{\emph{Proc. of the 10th USENIX NSDI}}.
  \bibinfo{pages}{317--329}.
\newblock


\bibitem[\protect\citeauthoryear{Adib, Mao, Kabelac, Katabi, and Miller}{Adib
  et~al\mbox{.}}{2015}]%
        {adib2015smart}
\bibfield{author}{\bibinfo{person}{Fadel Adib}, \bibinfo{person}{Hongzi Mao},
  \bibinfo{person}{Zachary Kabelac}, \bibinfo{person}{Dina Katabi}, {and}
  \bibinfo{person}{Robert~C Miller}.} \bibinfo{year}{2015}\natexlab{}.
\newblock \showarticletitle{{Smart Homes That Monitor Breathing and Heart
  Rate}}. In \bibinfo{booktitle}{\emph{Proc. of the 33rd ACM CHI}}.
  \bibinfo{pages}{837--846}.
\newblock


\bibitem[\protect\citeauthoryear{Awais, Badruddin, and Drieberg}{Awais
  et~al\mbox{.}}{2017}]%
        {awais2017hybrid}
\bibfield{author}{\bibinfo{person}{Muhammad Awais}, \bibinfo{person}{Nasreen
  Badruddin}, {and} \bibinfo{person}{Micheal Drieberg}.}
  \bibinfo{year}{2017}\natexlab{}.
\newblock \showarticletitle{{A Hybrid Approach to Detect Driver Drowsiness
  Utilizing Physiological Signals to Improve System Performance and
  Wearability}}.
\newblock \bibinfo{journal}{\emph{Sensors}} \bibinfo{volume}{17},
  \bibinfo{number}{9} (\bibinfo{year}{2017}), \bibinfo{pages}{1991}.
\newblock


\bibitem[\protect\citeauthoryear{Banning and Ng}{Banning and Ng}{2012}]%
        {banning2012driving}
\bibfield{author}{\bibinfo{person}{Amerjeet~S Banning} {and}
  \bibinfo{person}{G~Andre Ng}.} \bibinfo{year}{2012}\natexlab{}.
\newblock \showarticletitle{{Driving and Arrhythmia: A Review of Scientific
  Basis for International Guidelines}}.
\newblock \bibinfo{journal}{\emph{European Heart Journal}}
  \bibinfo{volume}{34}, \bibinfo{number}{3} (\bibinfo{year}{2012}),
  \bibinfo{pages}{236--244}.
\newblock


\bibitem[\protect\citeauthoryear{Benichou, Pereira, Mermillod, Tauveron,
  Pfabigan, Maqdasy, and Dutheil}{Benichou et~al\mbox{.}}{2018}]%
        {benichou2018heart}
\bibfield{author}{\bibinfo{person}{Thomas Benichou}, \bibinfo{person}{Bruno
  Pereira}, \bibinfo{person}{Martial Mermillod}, \bibinfo{person}{Igor
  Tauveron}, \bibinfo{person}{Daniela Pfabigan}, \bibinfo{person}{Salwan
  Maqdasy}, {and} \bibinfo{person}{Frederic Dutheil}.}
  \bibinfo{year}{2018}\natexlab{}.
\newblock \showarticletitle{{Heart Rate Variability in Type 2 Diabetes
  Mellitus: A Systematic Review and Meta-Analysis}}.
\newblock \bibinfo{journal}{\emph{PLOS One}} \bibinfo{volume}{13},
  \bibinfo{number}{4} (\bibinfo{year}{2018}), \bibinfo{pages}{e0195166}.
\newblock


\bibitem[\protect\citeauthoryear{Bj{\"o}rklund, Nelander, and
  Pettersson}{Bj{\"o}rklund et~al\mbox{.}}{2015}]%
        {fast-Radarcon}
\bibfield{author}{\bibinfo{person}{Svante Bj{\"o}rklund},
  \bibinfo{person}{Anders Nelander}, {and} \bibinfo{person}{Mats~I
  Pettersson}.} \bibinfo{year}{2015}\natexlab{}.
\newblock \showarticletitle{{Fast-Time and Slow-Time Space-Time Adaptive
  Processing for Bistatic Radar Interference Suppression}}. In
  \bibinfo{booktitle}{\emph{2015 IEEE Radar Conference (Radarcon)}}.
  \bibinfo{pages}{0674--0678}.
\newblock


\bibitem[\protect\citeauthoryear{{BMW AG}}{{BMW AG}}{2019}]%
        {bmw}
\bibfield{author}{\bibinfo{person}{{BMW AG}}.} \bibinfo{year}{2019}\natexlab{}.
\newblock \bibinfo{title}{{Smartsenior: Intelligente Dienste Und
  Dienstleistungen Für Senioren}}.
\newblock
  \bibinfo{howpublished}{\url{http://www1.smart-senior.de/pdf/presse/SmartSenior_CeBIT_Partnerflyer_BMW_DE_EN_2012_final.pdf}}.
\newblock
\newblock
\shownote{Accessed: 2019-10-09.}


\bibitem[\protect\citeauthoryear{Boyd, Parikh, Chu, Peleato, Eckstein,
  et~al\mbox{.}}{Boyd et~al\mbox{.}}{2011}]%
        {boyd2011distributed}
\bibfield{author}{\bibinfo{person}{Stephen Boyd}, \bibinfo{person}{Neal
  Parikh}, \bibinfo{person}{Eric Chu}, \bibinfo{person}{Borja Peleato},
  \bibinfo{person}{Jonathan Eckstein}, {et~al\mbox{.}}}
  \bibinfo{year}{2011}\natexlab{}.
\newblock \showarticletitle{{Distributed Optimization and Statistical Learning
  via the Alternating Direction Method of Multipliers}}.
\newblock \bibinfo{journal}{\emph{Foundations and Trends{\textregistered} in
  Machine learning}} \bibinfo{volume}{3}, \bibinfo{number}{1}
  (\bibinfo{year}{2011}), \bibinfo{pages}{1--122}.
\newblock


\bibitem[\protect\citeauthoryear{Brookhuis and de~Waard}{Brookhuis and
  de~Waard}{2010}]%
        {monitoring-Accid}
\bibfield{author}{\bibinfo{person}{Karel~A Brookhuis} {and}
  \bibinfo{person}{Dick de Waard}.} \bibinfo{year}{2010}\natexlab{}.
\newblock \showarticletitle{{Monitoring Drivers' Mental Workload in Driving
  Simulators using Physiological Measures}}.
\newblock \bibinfo{journal}{\emph{Accident Analysis \& Prevention}}
  \bibinfo{volume}{42}, \bibinfo{number}{3} (\bibinfo{year}{2010}),
  \bibinfo{pages}{898--903}.
\newblock


\bibitem[\protect\citeauthoryear{Catterall, Calverley, Brezinova, Douglas,
  Brash, Shapiro, and Flenley}{Catterall et~al\mbox{.}}{1982}]%
        {catterall1982irregular}
\bibfield{author}{\bibinfo{person}{JR Catterall}, \bibinfo{person}{PMA
  Calverley}, \bibinfo{person}{V Brezinova}, \bibinfo{person}{NJ Douglas},
  \bibinfo{person}{HM Brash}, \bibinfo{person}{CM Shapiro}, {and}
  \bibinfo{person}{DC Flenley}.} \bibinfo{year}{1982}\natexlab{}.
\newblock \showarticletitle{{Irregular Breathing and Hypoxaemia During Sleep in
  Chronic Stable Asthma}}.
\newblock \bibinfo{journal}{\emph{The Lancet}} \bibinfo{volume}{319},
  \bibinfo{number}{8267} (\bibinfo{year}{1982}), \bibinfo{pages}{301--304}.
\newblock


\bibitem[\protect\citeauthoryear{Chalmers, Quintana, Abbott, Kemp,
  et~al\mbox{.}}{Chalmers et~al\mbox{.}}{2014}]%
        {chalmers2014anxiety}
\bibfield{author}{\bibinfo{person}{John~A Chalmers}, \bibinfo{person}{Daniel~S
  Quintana}, \bibinfo{person}{Maree~J Abbott}, \bibinfo{person}{Andrew~H Kemp},
  {et~al\mbox{.}}} \bibinfo{year}{2014}\natexlab{}.
\newblock \showarticletitle{{Anxiety Disorders Are Associated with Reduced
  Heart Rate Variability: A Meta-Analysis}}.
\newblock \bibinfo{journal}{\emph{Frontiers in Psychiatry}}
  \bibinfo{volume}{5} (\bibinfo{year}{2014}), \bibinfo{pages}{80}.
\newblock


\bibitem[\protect\citeauthoryear{Cho, Seo, Kumar, and Rajkumar}{Cho
  et~al\mbox{.}}{2014}]%
        {multisensor-ICRA}
\bibfield{author}{\bibinfo{person}{Hyunggi Cho}, \bibinfo{person}{Young-Woo
  Seo}, \bibinfo{person}{BVK~Vijaya Kumar}, {and}
  \bibinfo{person}{Ragunathan~Raj Rajkumar}.} \bibinfo{year}{2014}\natexlab{}.
\newblock \showarticletitle{{A Multi-Sensor Fusion System for Moving Object
  Detection and Tracking in Urban Driving Environments}}. In
  \bibinfo{booktitle}{\emph{IEEE International Conference on Robotics and
  Automation (ICRA)}}. \bibinfo{pages}{1836--1843}.
\newblock


\bibitem[\protect\citeauthoryear{Chua, Tan, Yeo, Lau, Lee, Mien, Puvanendran,
  and Gooley}{Chua et~al\mbox{.}}{2012}]%
        {b3:hrv_drowsy}
\bibfield{author}{\bibinfo{person}{Eric Chern-Pin Chua},
  \bibinfo{person}{Wen-Qi Tan}, \bibinfo{person}{Sing-Chen Yeo},
  \bibinfo{person}{Pauline Lau}, \bibinfo{person}{Ivan Lee},
  \bibinfo{person}{Ivan~Ho Mien}, \bibinfo{person}{Kathiravelu Puvanendran},
  {and} \bibinfo{person}{Joshua~J. Gooley}.} \bibinfo{year}{2012}\natexlab{}.
\newblock \showarticletitle{{Heart Rate Variability Can Be Used to Estimate
  Sleepiness-related Decrements in Psychomotor Vigilance during Total Sleep
  Deprivation}}.
\newblock \bibinfo{journal}{\emph{Sleep}} \bibinfo{volume}{35},
  \bibinfo{number}{3} (\bibinfo{date}{03} \bibinfo{year}{2012}),
  \bibinfo{pages}{325--334}.
\newblock


\bibitem[\protect\citeauthoryear{Clark and Kruse}{Clark and Kruse}{1990}]%
        {clark1990clinical}
\bibfield{author}{\bibinfo{person}{Vivian~L Clark} {and}
  \bibinfo{person}{James~A Kruse}.} \bibinfo{year}{1990}\natexlab{}.
\newblock \showarticletitle{{Clinical Methods: The History, Physical, and
  Laboratory Examinations}}.
\newblock \bibinfo{journal}{\emph{Jama}} \bibinfo{volume}{264},
  \bibinfo{number}{21} (\bibinfo{year}{1990}), \bibinfo{pages}{2808--2809}.
\newblock


\bibitem[\protect\citeauthoryear{Cretikos, Bellomo, Hillman, Chen, Finfer, and
  Flabouris}{Cretikos et~al\mbox{.}}{2008}]%
        {cretikos2008respiratory}
\bibfield{author}{\bibinfo{person}{Michelle~A Cretikos},
  \bibinfo{person}{Rinaldo Bellomo}, \bibinfo{person}{Ken Hillman},
  \bibinfo{person}{Jack Chen}, \bibinfo{person}{Simon Finfer}, {and}
  \bibinfo{person}{Arthas Flabouris}.} \bibinfo{year}{2008}\natexlab{}.
\newblock \showarticletitle{{Respiratory Rate: The Neglected Vital Sign}}.
\newblock \bibinfo{journal}{\emph{Medical Journal of Australia}}
  \bibinfo{volume}{188}, \bibinfo{number}{11} (\bibinfo{year}{2008}),
  \bibinfo{pages}{657--659}.
\newblock


\bibitem[\protect\citeauthoryear{Cysarz, Lange, Matthiessen, and
  Leeuwen}{Cysarz et~al\mbox{.}}{2007}]%
        {cysarz2007regular}
\bibfield{author}{\bibinfo{person}{Dirk Cysarz}, \bibinfo{person}{Silke Lange},
  \bibinfo{person}{Peter~F Matthiessen}, {and} \bibinfo{person}{Peter~van
  Leeuwen}.} \bibinfo{year}{2007}\natexlab{}.
\newblock \showarticletitle{{Regular Heartbeat Dynamics Are Associated with
  Cardiac Health}}.
\newblock \bibinfo{journal}{\emph{Am. J. Physiol}} \bibinfo{volume}{292},
  \bibinfo{number}{1} (\bibinfo{year}{2007}), \bibinfo{pages}{R368--R372}.
\newblock


\bibitem[\protect\citeauthoryear{{Daimler AG}}{{Daimler AG}}{2019}]%
        {benz}
\bibfield{author}{\bibinfo{person}{{Daimler AG}}.}
  \bibinfo{year}{2019}\natexlab{}.
\newblock \bibinfo{title}{{Attention Assist: Drowsiness-Detection System Warns
  Drivers to Prevent Them Falling Asleep Momentarily - Daimler Global Media
  Site}}.
\newblock
  \bibinfo{howpublished}{\url{https://media.daimler.com/marsMediaSite/en/instance/ko/ATTENTION-ASSIST-Drowsiness-detection-system-warns-drivers-to-prevent-them-falling-asleep-momentarily.xhtml?oid=9361586}}.
\newblock
\newblock
\shownote{Accessed: 2019-10-09.}


\bibitem[\protect\citeauthoryear{Deepu, Chen, Teo, Ng, Yang, and Lian}{Deepu
  et~al\mbox{.}}{2012}]%
        {cushion-biocas}
\bibfield{author}{\bibinfo{person}{Chacko~John Deepu}, \bibinfo{person}{Zhihao
  Chen}, \bibinfo{person}{Ju~Teng Teo}, \bibinfo{person}{Soon~Huat Ng},
  \bibinfo{person}{Xiefeng Yang}, {and} \bibinfo{person}{Yong Lian}.}
  \bibinfo{year}{2012}\natexlab{}.
\newblock \showarticletitle{{A Smart Cushion for Real-Time Heart Rate
  Monitoring}}. In \bibinfo{booktitle}{\emph{2012 IEEE Biomedical Circuits and
  Systems Conference}}. \bibinfo{pages}{53--56}.
\newblock


\bibitem[\protect\citeauthoryear{DePasquale, Geller, Clarke, and
  Littleton}{DePasquale et~al\mbox{.}}{2001}]%
        {depasquale2001measuring}
\bibfield{author}{\bibinfo{person}{Jason~P DePasquale},
  \bibinfo{person}{E~Scott Geller}, \bibinfo{person}{Steven~W Clarke}, {and}
  \bibinfo{person}{Lawrence~C Littleton}.} \bibinfo{year}{2001}\natexlab{}.
\newblock \showarticletitle{{Measuring Road Rage: Development of The Propensity
  for Angry Driving Scale}}.
\newblock \bibinfo{journal}{\emph{J. Saf. Res.}} \bibinfo{volume}{32},
  \bibinfo{number}{1} (\bibinfo{year}{2001}), \bibinfo{pages}{1--16}.
\newblock


\bibitem[\protect\citeauthoryear{Dragomiretskiy and Zosso}{Dragomiretskiy and
  Zosso}{2013}]%
        {b6:vmd}
\bibfield{author}{\bibinfo{person}{Konstantin Dragomiretskiy} {and}
  \bibinfo{person}{Dominique Zosso}.} \bibinfo{year}{2013}\natexlab{}.
\newblock \showarticletitle{{Variational Mode Decomposition}}.
\newblock \bibinfo{journal}{\emph{IEEE Transactions on Signal Processing}}
  \bibinfo{volume}{62}, \bibinfo{number}{3} (\bibinfo{year}{2013}),
  \bibinfo{pages}{531--544}.
\newblock


\bibitem[\protect\citeauthoryear{Elliott and Coventry}{Elliott and
  Coventry}{2012}]%
        {b1:critical}
\bibfield{author}{\bibinfo{person}{Malcolm Elliott} {and}
  \bibinfo{person}{Alysia Coventry}.} \bibinfo{year}{2012}\natexlab{}.
\newblock \showarticletitle{{Critical Care: the Eight Vital Signs of Patient
  Monitoring}}.
\newblock \bibinfo{journal}{\emph{British Journal of Nursing}}
  \bibinfo{volume}{21}, \bibinfo{number}{10} (\bibinfo{year}{2012}),
  \bibinfo{pages}{621--625}.
\newblock


\bibitem[\protect\citeauthoryear{Elsenbruch, Harnish, and Orr}{Elsenbruch
  et~al\mbox{.}}{1999}]%
        {elsenbruch1999heart}
\bibfield{author}{\bibinfo{person}{Sigrid Elsenbruch},
  \bibinfo{person}{Michael~J Harnish}, {and} \bibinfo{person}{William~C Orr}.}
  \bibinfo{year}{1999}\natexlab{}.
\newblock \showarticletitle{{Heart Rate Variability During Waking and Sleep in
  Healthy Males and Females}}.
\newblock \bibinfo{journal}{\emph{Sleep}} \bibinfo{volume}{22},
  \bibinfo{number}{8} (\bibinfo{year}{1999}), \bibinfo{pages}{1067--1071}.
\newblock


\bibitem[\protect\citeauthoryear{Eriksson and Papanikotopoulos}{Eriksson and
  Papanikotopoulos}{1997}]%
        {eriksson1997eye}
\bibfield{author}{\bibinfo{person}{Martin Eriksson} {and}
  \bibinfo{person}{Nikolaos~P Papanikotopoulos}.}
  \bibinfo{year}{1997}\natexlab{}.
\newblock \showarticletitle{{Eye-Tracking for Detection of Driver Fatigue}}. In
  \bibinfo{booktitle}{\emph{Proc. of Conference on Intelligent Transportation
  Systems}}. \bibinfo{pages}{314--319}.
\newblock


\bibitem[\protect\citeauthoryear{Ernst}{Ernst}{2017}]%
        {ernst2017heart}
\bibfield{author}{\bibinfo{person}{Gernot Ernst}.}
  \bibinfo{year}{2017}\natexlab{}.
\newblock \showarticletitle{{Heart-Rate Variability-More than Heart Beats?}}
\newblock \bibinfo{journal}{\emph{Frontiers in Public Health}}
  \bibinfo{volume}{5} (\bibinfo{year}{2017}), \bibinfo{pages}{240}.
\newblock


\bibitem[\protect\citeauthoryear{Fieselmann, Hendryx, Helms, and
  Wakefield}{Fieselmann et~al\mbox{.}}{1993}]%
        {fieselmann1993respiratory}
\bibfield{author}{\bibinfo{person}{John~F Fieselmann},
  \bibinfo{person}{Michael~S Hendryx}, \bibinfo{person}{Charles~M Helms}, {and}
  \bibinfo{person}{Douglas~S Wakefield}.} \bibinfo{year}{1993}\natexlab{}.
\newblock \showarticletitle{{Respiratory Rate Predicts Cardiopulmonary Arrest
  for Internal Medicine Inpatients}}.
\newblock \bibinfo{journal}{\emph{Journal of General Internal Medicine}}
  \bibinfo{volume}{8}, \bibinfo{number}{7} (\bibinfo{year}{1993}),
  \bibinfo{pages}{354--360}.
\newblock


\bibitem[\protect\citeauthoryear{Flatley, Reyner, and Horne}{Flatley
  et~al\mbox{.}}{2004}]%
        {flatley2004sleep}
\bibfield{author}{\bibinfo{person}{Diane Flatley}, \bibinfo{person}{LA Reyner},
  {and} \bibinfo{person}{James~A Horne}.} \bibinfo{year}{2004}\natexlab{}.
\newblock \showarticletitle{{Sleep-Related Crashes on Sections of Different
  Road Types in the UK (1995-2001)}}.
\newblock \bibinfo{journal}{\emph{Road Safety United States Naval School of
  Aviation Medicine 52}} (\bibinfo{year}{2004}).
\newblock


\bibitem[\protect\citeauthoryear{Fujiwara, Abe, Kamata, Nakayama, Suzuki,
  Yamakawa, Hiraoka, Kano, Sumi, Masuda, et~al\mbox{.}}{Fujiwara
  et~al\mbox{.}}{2018}]%
        {fujiwara2018heart}
\bibfield{author}{\bibinfo{person}{Koichi Fujiwara}, \bibinfo{person}{Erika
  Abe}, \bibinfo{person}{Keisuke Kamata}, \bibinfo{person}{Chikao Nakayama},
  \bibinfo{person}{Yoko Suzuki}, \bibinfo{person}{Toshitaka Yamakawa},
  \bibinfo{person}{Toshihiro Hiraoka}, \bibinfo{person}{Manabu Kano},
  \bibinfo{person}{Yukiyoshi Sumi}, \bibinfo{person}{Fumi Masuda},
  {et~al\mbox{.}}} \bibinfo{year}{2018}\natexlab{}.
\newblock \showarticletitle{{Heart Rate Variability-Based Driver Drowsiness
  Detection and its Validation with EEG}}.
\newblock \bibinfo{journal}{\emph{IEEE Transactions on Biomedical Engineering}}
  \bibinfo{volume}{66}, \bibinfo{number}{6} (\bibinfo{year}{2018}),
  \bibinfo{pages}{1769--1778}.
\newblock


\bibitem[\protect\citeauthoryear{Giardino, Friedman, and Dager}{Giardino
  et~al\mbox{.}}{2007}]%
        {giardino2007anxiety}
\bibfield{author}{\bibinfo{person}{Nicholas~D Giardino},
  \bibinfo{person}{Seth~D Friedman}, {and} \bibinfo{person}{Stephen~R Dager}.}
  \bibinfo{year}{2007}\natexlab{}.
\newblock \showarticletitle{{Anxiety, Respiration, and Cerebral Blood Flow:
  Implications for Functional Brain Imaging}}.
\newblock \bibinfo{journal}{\emph{Comprehensive Psychiatry}}
  \bibinfo{volume}{48}, \bibinfo{number}{2} (\bibinfo{year}{2007}),
  \bibinfo{pages}{103--112}.
\newblock


\bibitem[\protect\citeauthoryear{Halperin, Hu, Sheth, and Wetherall}{Halperin
  et~al\mbox{.}}{2011}]%
        {b5:csitool}
\bibfield{author}{\bibinfo{person}{Daniel Halperin}, \bibinfo{person}{Wenjun
  Hu}, \bibinfo{person}{Anmol Sheth}, {and} \bibinfo{person}{David Wetherall}.}
  \bibinfo{year}{2011}\natexlab{}.
\newblock \showarticletitle{{Tool Release: Gathering 802.11n Traces with
  Channel State Information}}.
\newblock \bibinfo{journal}{\emph{ACM SIGCOMM Computer Communication Review}}
  \bibinfo{volume}{41}, \bibinfo{number}{1} (\bibinfo{year}{2011}),
  \bibinfo{pages}{53}.
\newblock


\bibitem[\protect\citeauthoryear{Huynh, Balan, Ko, and Lee}{Huynh
  et~al\mbox{.}}{2019}]%
        {Vitamon-Sensys}
\bibfield{author}{\bibinfo{person}{Sinh Huynh}, \bibinfo{person}{Rajesh~Krishna
  Balan}, \bibinfo{person}{JeongGil Ko}, {and} \bibinfo{person}{Youngki Lee}.}
  \bibinfo{year}{2019}\natexlab{}.
\newblock \showarticletitle{{VitaMon: Measuring Heart Rate Variability Using
  Smartphone Front Camera}}. In \bibinfo{booktitle}{\emph{Proc. of the 17th ACM
  Sensys}}. \bibinfo{pages}{1--14}.
\newblock


\bibitem[\protect\citeauthoryear{Jackson, Hilditch, Holmes, Reed, Merat, and
  Smith}{Jackson et~al\mbox{.}}{2011}]%
        {jackson2011fatigue}
\bibfield{author}{\bibinfo{person}{P Jackson}, \bibinfo{person}{C Hilditch},
  \bibinfo{person}{A Holmes}, \bibinfo{person}{N Reed}, \bibinfo{person}{N
  Merat}, {and} \bibinfo{person}{L Smith}.} \bibinfo{year}{2011}\natexlab{}.
\newblock \showarticletitle{{Fatigue and Road Safety: A Critical Analysis of
  Recent Evidence}}.
\newblock \bibinfo{journal}{\emph{Department for Transport, Road Safety Web
  Publication}}  \bibinfo{volume}{21} (\bibinfo{year}{2011}).
\newblock


\bibitem[\protect\citeauthoryear{Jia, Peng, Ruan, Tang, and Zhao}{Jia
  et~al\mbox{.}}{2018}]%
        {WiFind-IEEEToBD}
\bibfield{author}{\bibinfo{person}{Wenjia Jia}, \bibinfo{person}{Hongjian
  Peng}, \bibinfo{person}{Na Ruan}, \bibinfo{person}{Zhiping Tang}, {and}
  \bibinfo{person}{Wei Zhao}.} \bibinfo{year}{2018}\natexlab{}.
\newblock \showarticletitle{{WiFind: Driver Fatigue Detection with Fine-Grained
  Wi-Fi Signal Features}}.
\newblock \bibinfo{journal}{\emph{IEEE Transactions on Big Data}}
  (\bibinfo{year}{2018}), \bibinfo{pages}{1--14}.
\newblock


\bibitem[\protect\citeauthoryear{Jo, Kim, and Kim}{Jo et~al\mbox{.}}{2019}]%
        {korean2019hr}
\bibfield{author}{\bibinfo{person}{Sang-Ho Jo}, \bibinfo{person}{Jin-Myung
  Kim}, {and} \bibinfo{person}{Dong~Kyoo Kim}.}
  \bibinfo{year}{2019}\natexlab{}.
\newblock \showarticletitle{{Heart Rate Change While Drowsy Driving}}.
\newblock \bibinfo{journal}{\emph{Journal of Korean Medical Science}}
  \bibinfo{volume}{34}, \bibinfo{number}{8} (\bibinfo{year}{2019}).
\newblock


\bibitem[\protect\citeauthoryear{Leem, Khan, and Cho}{Leem
  et~al\mbox{.}}{2017}]%
        {leem2017vital}
\bibfield{author}{\bibinfo{person}{Seong~Kyu Leem}, \bibinfo{person}{Faheem
  Khan}, {and} \bibinfo{person}{Sung~Ho Cho}.} \bibinfo{year}{2017}\natexlab{}.
\newblock \showarticletitle{{Vital Sign Monitoring and Mobile Phone Usage
  Detection Using IR-UWB Radar for Intended Use in Car Crash Prevention}}.
\newblock \bibinfo{journal}{\emph{Sensors}} (\bibinfo{year}{2017}).
\newblock


\bibitem[\protect\citeauthoryear{Li, Lee, and Chung}{Li et~al\mbox{.}}{2015}]%
        {Smartwatch-IEEESens}
\bibfield{author}{\bibinfo{person}{Gang Li}, \bibinfo{person}{Boon-Leng Lee},
  {and} \bibinfo{person}{Wan-Young Chung}.} \bibinfo{year}{2015}\natexlab{}.
\newblock \showarticletitle{{Smartwatch-based Wearable EEG System for Driver
  Drowsiness Detection}}.
\newblock \bibinfo{journal}{\emph{IEEE Sensors Journal}} \bibinfo{volume}{15},
  \bibinfo{number}{12} (\bibinfo{year}{2015}), \bibinfo{pages}{7169--7180}.
\newblock


\bibitem[\protect\citeauthoryear{Li, Zhang, Lv, Xiong, Li, Zhang, and Mei}{Li
  et~al\mbox{.}}{2017}]%
        {Indotrack-UbiComp}
\bibfield{author}{\bibinfo{person}{Xiang Li}, \bibinfo{person}{Daqing Zhang},
  \bibinfo{person}{Qin Lv}, \bibinfo{person}{Jie Xiong},
  \bibinfo{person}{Shengjie Li}, \bibinfo{person}{Yue Zhang}, {and}
  \bibinfo{person}{Hong Mei}.} \bibinfo{year}{2017}\natexlab{}.
\newblock \showarticletitle{{IndoTrack: Device-Free Indoor Human Tracking with
  Commodity Wi-Fi}}.
\newblock \bibinfo{journal}{\emph{Proc. of the 17th ACM UbiComp}}
  \bibinfo{volume}{1}, \bibinfo{number}{3} (\bibinfo{year}{2017}),
  \bibinfo{pages}{72}.
\newblock


\bibitem[\protect\citeauthoryear{Lisper, Laurell, and Stening}{Lisper
  et~al\mbox{.}}{1973}]%
        {lisper1973effects}
\bibfield{author}{\bibinfo{person}{H-O Lisper}, \bibinfo{person}{Hans Laurell},
  {and} \bibinfo{person}{G Stening}.} \bibinfo{year}{1973}\natexlab{}.
\newblock \showarticletitle{{Effects of Experience of The Driver on Heart-Rate,
  Respiration-Rate, and Subsidiary Reaction Time in A Three Hours Continuous
  Driving Task}}.
\newblock \bibinfo{journal}{\emph{Ergonomics}} \bibinfo{volume}{16},
  \bibinfo{number}{4} (\bibinfo{year}{1973}), \bibinfo{pages}{501--506}.
\newblock


\bibitem[\protect\citeauthoryear{Lutfi}{Lutfi}{2015}]%
        {lutfi2015patterns}
\bibfield{author}{\bibinfo{person}{Mohamed~Faisal Lutfi}.}
  \bibinfo{year}{2015}\natexlab{}.
\newblock \showarticletitle{{Patterns of Heart Rate Variability and Cardiac
  Autonomic Modulations in Controlled and Uncontrolled Asthmatic Patients}}.
\newblock \bibinfo{journal}{\emph{BMC Pulmonary Medicine}}
  \bibinfo{volume}{15}, \bibinfo{number}{1} (\bibinfo{year}{2015}),
  \bibinfo{pages}{119}.
\newblock


\bibitem[\protect\citeauthoryear{Mardi, Ashtiani, and Mikaili}{Mardi
  et~al\mbox{.}}{2011}]%
        {mardi2011eeg}
\bibfield{author}{\bibinfo{person}{Zahra Mardi}, \bibinfo{person}{Seyedeh
  Naghmeh~Miri Ashtiani}, {and} \bibinfo{person}{Mohammad Mikaili}.}
  \bibinfo{year}{2011}\natexlab{}.
\newblock \showarticletitle{{EEG-Based Drowsiness Detection for Safe Driving
  Using Chaotic Features and Statistical Tests}}.
\newblock \bibinfo{journal}{\emph{Journal of Medical Signals and Sensors}}
  \bibinfo{volume}{1}, \bibinfo{number}{2} (\bibinfo{year}{2011}),
  \bibinfo{pages}{130}.
\newblock


\bibitem[\protect\citeauthoryear{Nabi, Hall, Koskenvuo, Singh-Manoux, Oksanen,
  Suominen, Kivim{\"a}ki, and Vahtera}{Nabi et~al\mbox{.}}{2010}]%
        {nabi2010psychological}
\bibfield{author}{\bibinfo{person}{Hermann Nabi}, \bibinfo{person}{Martica
  Hall}, \bibinfo{person}{Markku Koskenvuo}, \bibinfo{person}{Archana
  Singh-Manoux}, \bibinfo{person}{Tuula Oksanen}, \bibinfo{person}{Sakari
  Suominen}, \bibinfo{person}{Mika Kivim{\"a}ki}, {and} \bibinfo{person}{Jussi
  Vahtera}.} \bibinfo{year}{2010}\natexlab{}.
\newblock \showarticletitle{{Psychological and Somatic Symptoms of Anxiety and
  Risk of Coronary Heart Disease: the Health and Social Support Prospective
  Cohort Study}}.
\newblock \bibinfo{journal}{\emph{Biological Psychiatry}} \bibinfo{volume}{67},
  \bibinfo{number}{4} (\bibinfo{year}{2010}), \bibinfo{pages}{378--385}.
\newblock


\bibitem[\protect\citeauthoryear{{Novelda AS}}{{Novelda AS}}{2017a}]%
        {xethru}
\bibfield{author}{\bibinfo{person}{{Novelda AS}}.}
  \bibinfo{year}{2017}\natexlab{a}.
\newblock \bibinfo{title}{{Single-Chip Radar Sensors with Sub-mm Resolution -
  XETHRU}}.
\newblock \bibinfo{howpublished}{\url{https://www.xethru.com/}}.
\newblock
\newblock
\shownote{Accessed: 2019-10-03.}


\bibitem[\protect\citeauthoryear{{Novelda AS}}{{Novelda AS}}{2017b}]%
        {xethru_comparison}
\bibfield{author}{\bibinfo{person}{{Novelda AS}}.}
  \bibinfo{year}{2017}\natexlab{b}.
\newblock \bibinfo{title}{{XETHRU Sensor Emissions: An In-Depth Look at Radar
  Safety Regulations}}.
\newblock
  \bibinfo{howpublished}{\url{https://www.xethru.com/blog/posts/xethru-radar-emission-comparison}}.
\newblock
\newblock
\shownote{Accessed: 2019-10-03.}


\bibitem[\protect\citeauthoryear{Ogino and Mitsukura}{Ogino and
  Mitsukura}{2018}]%
        {ogino2018portable}
\bibfield{author}{\bibinfo{person}{Mikito Ogino} {and} \bibinfo{person}{Yasue
  Mitsukura}.} \bibinfo{year}{2018}\natexlab{}.
\newblock \showarticletitle{{Portable Drowsiness Detection Through Use of A
  Prefrontal Single-Channel Electroencephalogram}}.
\newblock \bibinfo{journal}{\emph{Sensors}} \bibinfo{volume}{18},
  \bibinfo{number}{12} (\bibinfo{year}{2018}), \bibinfo{pages}{4477}.
\newblock


\bibitem[\protect\citeauthoryear{Park, Hong, Lee, Jang, Yun, Lee, and
  Yook}{Park et~al\mbox{.}}{2019}]%
        {park2019noncontact}
\bibfield{author}{\bibinfo{person}{Jin-Kwan Park}, \bibinfo{person}{Yunseog
  Hong}, \bibinfo{person}{Hyunjae Lee}, \bibinfo{person}{Chorom Jang},
  \bibinfo{person}{Gi-Ho Yun}, \bibinfo{person}{Hee-Jo Lee}, {and}
  \bibinfo{person}{Jong-Gwan Yook}.} \bibinfo{year}{2019}\natexlab{}.
\newblock \showarticletitle{{Noncontact RF Vital Sign Sensor for Continuous
  Monitoring of Driver Status}}.
\newblock \bibinfo{journal}{\emph{IEEE Transactions on Biomedical Circuits and
  Systems}} \bibinfo{volume}{13}, \bibinfo{number}{3} (\bibinfo{year}{2019}),
  \bibinfo{pages}{493--502}.
\newblock


\bibitem[\protect\citeauthoryear{Pinheiro, Postolache, and Gir{\~a}o}{Pinheiro
  et~al\mbox{.}}{2010}]%
        {pinheiro2010theory}
\bibfield{author}{\bibinfo{person}{Eduardo Pinheiro}, \bibinfo{person}{Octavian
  Postolache}, {and} \bibinfo{person}{Pedro Gir{\~a}o}.}
  \bibinfo{year}{2010}\natexlab{}.
\newblock \showarticletitle{{Theory and Developments in An Unobtrusive
  Cardiovascular System Representation: Ballistocardiography}}.
\newblock \bibinfo{journal}{\emph{The Open Biomedical Engineering Journal}}
  \bibinfo{volume}{4} (\bibinfo{year}{2010}), \bibinfo{pages}{201}.
\newblock


\bibitem[\protect\citeauthoryear{Raemer}{Raemer}{1996}]%
        {raemer1996radar}
\bibfield{author}{\bibinfo{person}{Harold~R Raemer}.}
  \bibinfo{year}{1996}\natexlab{}.
\newblock \bibinfo{booktitle}{\emph{{Radar Systems Principles}}}.
\newblock \bibinfo{publisher}{CRC press}.
\newblock


\bibitem[\protect\citeauthoryear{{Raspberry Pi Foundation}}{{Raspberry Pi
  Foundation}}{2019}]%
        {rpi}
\bibfield{author}{\bibinfo{person}{{Raspberry Pi Foundation}}.}
  \bibinfo{year}{2019}\natexlab{}.
\newblock \bibinfo{title}{{Teach, Learn and Make with Raspberry Pi - Raspberry
  Pi}}.
\newblock \bibinfo{howpublished}{\url{https://https://www.raspberrypi.org/}}.
\newblock
\newblock
\shownote{Accessed: 2019-10-03.}


\bibitem[\protect\citeauthoryear{Sessa, Anna, Messina, Cibelli, Monda, Marsala,
  Ruberto, Biondi, Cascio, Bertozzi, et~al\mbox{.}}{Sessa
  et~al\mbox{.}}{2018}]%
        {sessa2018heart}
\bibfield{author}{\bibinfo{person}{Francesco Sessa}, \bibinfo{person}{Valenzano
  Anna}, \bibinfo{person}{Giovanni Messina}, \bibinfo{person}{Giuseppe
  Cibelli}, \bibinfo{person}{Vincenzo Monda}, \bibinfo{person}{Gabriella
  Marsala}, \bibinfo{person}{Maria Ruberto}, \bibinfo{person}{Antonio Biondi},
  \bibinfo{person}{Orazio Cascio}, \bibinfo{person}{Giuseppe Bertozzi},
  {et~al\mbox{.}}} \bibinfo{year}{2018}\natexlab{}.
\newblock \showarticletitle{{Heart Rate Variability As Predictive Factor for
  Sudden Cardiac Death}}.
\newblock \bibinfo{journal}{\emph{Aging (Albany NY)}} \bibinfo{volume}{10},
  \bibinfo{number}{2} (\bibinfo{year}{2018}), \bibinfo{pages}{166}.
\newblock


\bibitem[\protect\citeauthoryear{Shaffer and Ginsberg}{Shaffer and
  Ginsberg}{2017}]%
        {shaffer2017overview}
\bibfield{author}{\bibinfo{person}{Fred Shaffer} {and} \bibinfo{person}{JP
  Ginsberg}.} \bibinfo{year}{2017}\natexlab{}.
\newblock \showarticletitle{{An Overview of Heart Rate Variability Metrics and
  Norms}}.
\newblock \bibinfo{journal}{\emph{Frontiers in Public Health}}
  \bibinfo{volume}{5} (\bibinfo{year}{2017}), \bibinfo{pages}{258}.
\newblock


\bibitem[\protect\citeauthoryear{Shaffer, McCraty, and Zerr}{Shaffer
  et~al\mbox{.}}{2014}]%
        {shaffer2014healthy}
\bibfield{author}{\bibinfo{person}{Fred Shaffer}, \bibinfo{person}{Rollin
  McCraty}, {and} \bibinfo{person}{Christopher~L Zerr}.}
  \bibinfo{year}{2014}\natexlab{}.
\newblock \showarticletitle{{A Healthy Heart Is Not A Metronome: An Integrative
  Review of the Heart's Anatomy and Heart Rate Variability}}.
\newblock \bibinfo{journal}{\emph{Frontiers in Psychology}}
  \bibinfo{volume}{5} (\bibinfo{year}{2014}), \bibinfo{pages}{1040}.
\newblock


\bibitem[\protect\citeauthoryear{Sharma, Hirulkar, and Ranka}{Sharma
  et~al\mbox{.}}{2011}]%
        {sharma2011effect}
\bibfield{author}{\bibinfo{person}{Anupriya Sharma}, \bibinfo{person}{NB
  Hirulkar}, {and} \bibinfo{person}{Payal Ranka}.}
  \bibinfo{year}{2011}\natexlab{}.
\newblock \showarticletitle{{Effect of Hyperglycemia on Electrolytes
  Imbalance}}.
\newblock \bibinfo{journal}{\emph{Int J Pharm Biol Arch}}  \bibinfo{volume}{2}
  (\bibinfo{year}{2011}), \bibinfo{pages}{526--33}.
\newblock


\bibitem[\protect\citeauthoryear{Solaz, Laparra-Hern{\'a}ndez, Bande,
  Rodr{\'\i}guez, Veleff, Gerpe, and Medina}{Solaz et~al\mbox{.}}{2016}]%
        {solaz2016drowsiness}
\bibfield{author}{\bibinfo{person}{Jos{\'e} Solaz}, \bibinfo{person}{Jos{\'e}
  Laparra-Hern{\'a}ndez}, \bibinfo{person}{Daniel Bande},
  \bibinfo{person}{Noelia Rodr{\'\i}guez}, \bibinfo{person}{Sergio Veleff},
  \bibinfo{person}{Jos{\'e} Gerpe}, {and} \bibinfo{person}{Enrique Medina}.}
  \bibinfo{year}{2016}\natexlab{}.
\newblock \showarticletitle{{Drowsiness Detection Based on the Analysis of
  Breathing Rate Obtained from Real-Time Image Recognition}}.
\newblock \bibinfo{journal}{\emph{Transportation Research Procedia}}
  \bibinfo{volume}{14} (\bibinfo{year}{2016}), \bibinfo{pages}{3867--3876}.
\newblock


\bibitem[\protect\citeauthoryear{Strau{\ss}, Ewig, Richter, K{\"o}nig, Heller,
  and Bauer}{Strau{\ss} et~al\mbox{.}}{2014}]%
        {strauss2014prognostic}
\bibfield{author}{\bibinfo{person}{Richard Strau{\ss}},
  \bibinfo{person}{Santiago Ewig}, \bibinfo{person}{Klaus Richter},
  \bibinfo{person}{Thomas K{\"o}nig}, \bibinfo{person}{G{\"u}nther Heller},
  {and} \bibinfo{person}{Torsten~T Bauer}.} \bibinfo{year}{2014}\natexlab{}.
\newblock \showarticletitle{{The Prognostic Significance of Respiratory Rate in
  Patients with Pneumonia: A Retrospective Analysis of Data from 705 928
  Hospitalized Patients in Germany from 2010--2012}}.
\newblock \bibinfo{journal}{\emph{Deutsches {\"A}rzteblatt International}}
  \bibinfo{volume}{111}, \bibinfo{number}{29-30} (\bibinfo{year}{2014}),
  \bibinfo{pages}{503}.
\newblock


\bibitem[\protect\citeauthoryear{Sullivan, Deiss, and Cauwenberghs}{Sullivan
  et~al\mbox{.}}{2007}]%
        {sullivan2007low}
\bibfield{author}{\bibinfo{person}{Thomas~J Sullivan},
  \bibinfo{person}{Stephen~R Deiss}, {and} \bibinfo{person}{Gert
  Cauwenberghs}.} \bibinfo{year}{2007}\natexlab{}.
\newblock \showarticletitle{{A Low-Noise, Non-Contact EEG/ECG Sensor}}. In
  \bibinfo{booktitle}{\emph{2007 IEEE Biomedical Circuits and Systems
  Conference}}. \bibinfo{pages}{154--157}.
\newblock


\bibitem[\protect\citeauthoryear{Troutman}{Troutman}{2012}]%
        {troutman2012variational}
\bibfield{author}{\bibinfo{person}{John~L Troutman}.}
  \bibinfo{year}{2012}\natexlab{}.
\newblock \bibinfo{booktitle}{\emph{{Variational Calculus and Optimal Control:
  Optimization with Elementary Convexity}}}.
\newblock \bibinfo{publisher}{Springer Science \& Business Media}.
\newblock


\bibitem[\protect\citeauthoryear{Tsunoda, Endo, Hashimoto, Honma, and
  Honma}{Tsunoda et~al\mbox{.}}{2001}]%
        {tsunoda2001effects}
\bibfield{author}{\bibinfo{person}{Mihoko Tsunoda}, \bibinfo{person}{Takuro
  Endo}, \bibinfo{person}{Satoko Hashimoto}, \bibinfo{person}{Sato Honma},
  {and} \bibinfo{person}{Ken-Ichi Honma}.} \bibinfo{year}{2001}\natexlab{}.
\newblock \showarticletitle{{Effects of Light and Sleep Stages on Heart Rate
  Variability in Humans}}.
\newblock \bibinfo{journal}{\emph{Psychiatry and Clinical Neurosciences}}
  \bibinfo{volume}{55}, \bibinfo{number}{3} (\bibinfo{year}{2001}),
  \bibinfo{pages}{285--286}.
\newblock


\bibitem[\protect\citeauthoryear{{Volvo Group}}{{Volvo Group}}{2019}]%
        {volvo}
\bibfield{author}{\bibinfo{person}{{Volvo Group}}.}
  \bibinfo{year}{2019}\natexlab{}.
\newblock \bibinfo{title}{{Driver Alert System}}.
\newblock
  \bibinfo{howpublished}{\url{https://www.volvocars.com/en-th/support/manuals/s60/2014/driver-support/driver-alert-system/driver-alert-control-dac---operation}}.
\newblock
\newblock
\shownote{Accessed: 2019-10-09.}


\bibitem[\protect\citeauthoryear{Wang, Xie, Wang, Chen, Bu, and Lu}{Wang
  et~al\mbox{.}}{2018a}]%
        {RFECG-UbiComp}
\bibfield{author}{\bibinfo{person}{Chuyu Wang}, \bibinfo{person}{Lei Xie},
  \bibinfo{person}{Wei Wang}, \bibinfo{person}{Yingying Chen},
  \bibinfo{person}{Yanling Bu}, {and} \bibinfo{person}{Sanglu Lu}.}
  \bibinfo{year}{2018}\natexlab{a}.
\newblock \showarticletitle{{RF-ECG: Heart Rate Variability Assessment Based on
  COTS RFID Tag Array}}.
\newblock \bibinfo{journal}{\emph{Proc. of the 17th ACM UbiComp}}
  \bibinfo{volume}{2}, \bibinfo{number}{2} (\bibinfo{year}{2018}),
  \bibinfo{pages}{85}.
\newblock


\bibitem[\protect\citeauthoryear{Wang, Zhang, Zheng, Gu, Zhou, and
  Dorizzi}{Wang et~al\mbox{.}}{2018b}]%
        {AcousticResp-Ubicomp}
\bibfield{author}{\bibinfo{person}{Tianben Wang}, \bibinfo{person}{Daqing
  Zhang}, \bibinfo{person}{Yuanqing Zheng}, \bibinfo{person}{Tao Gu},
  \bibinfo{person}{Xingshe Zhou}, {and} \bibinfo{person}{Bernadette Dorizzi}.}
  \bibinfo{year}{2018}\natexlab{b}.
\newblock \showarticletitle{{C-FMCW Based Contactless Respiration Detection
  Using Acoustic Signal}}.
\newblock \bibinfo{journal}{\emph{Proc. of the 17th ACM UbiComp}}
  \bibinfo{volume}{1}, \bibinfo{number}{4} (\bibinfo{year}{2018}),
  \bibinfo{pages}{170}.
\newblock


\bibitem[\protect\citeauthoryear{Waring, Rhee, Bateman, Leggett, and
  Jamie}{Waring et~al\mbox{.}}{2008}]%
        {waring2008impaired}
\bibfield{author}{\bibinfo{person}{WS Waring}, \bibinfo{person}{JY Rhee},
  \bibinfo{person}{DN Bateman}, \bibinfo{person}{GE Leggett}, {and}
  \bibinfo{person}{H Jamie}.} \bibinfo{year}{2008}\natexlab{}.
\newblock \showarticletitle{{Impaired Heart Rate Variability and Altered
  Cardiac Sympathovagal Balance After Antidepressant Overdose}}.
\newblock \bibinfo{journal}{\emph{European Journal of Clinical Pharmacology}}
  \bibinfo{volume}{64}, \bibinfo{number}{11} (\bibinfo{year}{2008}),
  \bibinfo{pages}{1037--1041}.
\newblock


\bibitem[\protect\citeauthoryear{Wu, Zhang, Xu, Wang, and Li}{Wu
  et~al\mbox{.}}{2017}]%
        {wu2017device}
\bibfield{author}{\bibinfo{person}{Dan Wu}, \bibinfo{person}{Daqing Zhang},
  \bibinfo{person}{Chenren Xu}, \bibinfo{person}{Hao Wang}, {and}
  \bibinfo{person}{Xiang Li}.} \bibinfo{year}{2017}\natexlab{}.
\newblock \showarticletitle{{Device-Free WiFi Human Sensing: From Pattern-Based
  to Model-Based Approaches}}.
\newblock \bibinfo{journal}{\emph{IEEE Communications Magazine}}
  \bibinfo{volume}{55}, \bibinfo{number}{10} (\bibinfo{year}{2017}),
  \bibinfo{pages}{91--97}.
\newblock


\bibitem[\protect\citeauthoryear{Xu, Yu, Chen, Zhu, Kong, and Li}{Xu
  et~al\mbox{.}}{2019}]%
        {BreathListener-MobiSys19}
\bibfield{author}{\bibinfo{person}{Xiangyu Xu}, \bibinfo{person}{Jiadi Yu},
  \bibinfo{person}{Yingying Chen}, \bibinfo{person}{Yanmin Zhu},
  \bibinfo{person}{Linghe Kong}, {and} \bibinfo{person}{Minglu Li}.}
  \bibinfo{year}{2019}\natexlab{}.
\newblock \showarticletitle{{BreathListener: Fine-Grained Breathing Monitoring
  in Driving Environments Utilizing Acoustic Signals}}. In
  \bibinfo{booktitle}{\emph{Proc. of the 17th ACM MobiSys}}.
  \bibinfo{pages}{54--66}.
\newblock


\bibitem[\protect\citeauthoryear{Zeng, Wu, Gao, Gu, and Zhang}{Zeng
  et~al\mbox{.}}{2018}]%
        {Fullbreathe-UbiComp}
\bibfield{author}{\bibinfo{person}{Youwei Zeng}, \bibinfo{person}{Dan Wu},
  \bibinfo{person}{Ruiyang Gao}, \bibinfo{person}{Tao Gu}, {and}
  \bibinfo{person}{Daqing Zhang}.} \bibinfo{year}{2018}\natexlab{}.
\newblock \showarticletitle{{FullBreathe: Full Human Respiration Detection
  Exploiting Complementarity of CSI Phase and Amplitude of WiFi Signals}}.
\newblock \bibinfo{journal}{\emph{Proc. of the 18th ACM UbiComp}}
  \bibinfo{volume}{2}, \bibinfo{number}{3} (\bibinfo{year}{2018}),
  \bibinfo{pages}{148}.
\newblock


\bibitem[\protect\citeauthoryear{Zhang, Xu, Hu, and Kanhere}{Zhang
  et~al\mbox{.}}{2017}]%
        {WiCare-MobiQuitous}
\bibfield{author}{\bibinfo{person}{Jin Zhang}, \bibinfo{person}{Weitao Xu},
  \bibinfo{person}{Wen Hu}, {and} \bibinfo{person}{Salil~S Kanhere}.}
  \bibinfo{year}{2017}\natexlab{}.
\newblock \showarticletitle{{WiCare: Towards In-Situ Breath Monitoring}}. In
  \bibinfo{booktitle}{\emph{Proc. of the 14th ACM MobiQuitous}}. ACM,
  \bibinfo{pages}{126--135}.
\newblock


\bibitem[\protect\citeauthoryear{Zhao, Adib, and Katabi}{Zhao
  et~al\mbox{.}}{2016}]%
        {zhao2016emotion}
\bibfield{author}{\bibinfo{person}{Mingmin Zhao}, \bibinfo{person}{Fadel Adib},
  {and} \bibinfo{person}{Dina Katabi}.} \bibinfo{year}{2016}\natexlab{}.
\newblock \showarticletitle{{Emotion Recognition Using Wireless Signals}}. In
  \bibinfo{booktitle}{\emph{Proc. of the 22nd ACM MobiCom}}.
  \bibinfo{pages}{95--108}.
\newblock


\end{thebibliography}

\clearpage
\appendix

\end{document}